%% file: main.tex
\documentclass[fleqn,usenatbib]{mnras}

\usepackage{newtxtext,newtxmath}
\usepackage[T1]{fontenc}

\DeclareRobustCommand{\VAN}[3]{#2}
\let\VANthebibliography\thebibliography
\def\thebibliography{\DeclareRobustCommand{\VAN}[3]{##3}\VANthebibliography}

\usepackage{cleveref}
\usepackage{caption}

\usepackage{graphicx}	% Including figure files
\usepackage{amsmath}	% Advanced maths commands
\title[JWST predictions with \textsc{Flares} and \textsc{Shark}]{Unveiling the main sequence of galaxies at $z\geq5$ with the James Webb Space Telescope: predictions from simulations}

\author[J. C. J. D'Silva et al.]{
Jordan C. J. D'Silva,$^{1}$\thanks{E-mail: 22252335@student.uwa.edu.au}
Claudia D. P. Lagos,$^{1,2,3}$
Luke J. M. Davies,$^{1}$
Christopher C. Lovell,$^{4}$ \newauthor
Aswin P. Vijayan$^{3,5}$ 
\\
$^{1}$International Centre for Radio Astronomy Research (ICRAR), M468, University of Western Australia, 35 Stirling Hwy, Crawley, WA 6009, Australia\\
$^{2}$ARC Centre of Excellence for All Sky Astrophysics in 3 Dimensions (ASTRO 3D).\\
$^{3}$Cosmic Dawn Center (DAWN).\\
$^{4}$Centre for Astrophysics Research, School of Physics, Astronomy \& Mathematics, University of Hertfordshire, Hatfield AL10 9AB, UK.\\
$^{5}$DTU-Space, Technical University of Denmark, Elektrovej 327, DK-2800 Kgs. Lyngby, Denmark
}

\date{Accepted XXX. Received YYY; in original form ZZZ}

\pubyear{2022}

\begin{document}
\label{firstpage}
\pagerange{\pageref{firstpage}--\pageref{lastpage}}
\maketitle

\begin{abstract}
%The James Webb Space Telescope (JWST) is expected to revolutionise our understanding of extra-galactic astronomy. In preparation, 

We use two independent, galaxy formation simulations, \textsc{Flares}, a cosmological hydrodynamical simulation, and \textsc{Shark}, a semi-analytic model, to explore how well the James Webb Space Telescope (JWST) will be able to uncover the existence and parameters of the star-forming main sequence (SFS) at $z=5\to10$, i.e. shape, scatter, normalisation. Using two independent simulations allows us  to isolate predictions (e.g., stellar mass, star formation rate, SFR, luminosity functions) that are robust to or highly dependent on the implementation of the physics of galaxy formation. Both simulations predict that JWST can observe $\ge 70-90$\% (for \textsc{Shark} and \textsc{Flares} respectively) of galaxies up to ${z\sim10}$ (down to stellar masses  of $\approx 10^{8.3}\,\rm M_{\odot}$ and SFRs of $\approx 10^{0.5}\,\rm M_{\odot}\, yr^{-1}$) in modest integration times and given current proposed survey areas (e.g. the Web COSMOS $0.6\,\rm deg^2$) to accurately constrain the parameters of the SFS. Although both simulations predict qualitatively similar distributions of stellar mass and SFR, there are important quantitative differences, such as the abundance of massive, star-forming galaxies, with \textsc{Flares} predicting a higher abundance than {\sc Shark}; the early onset of quenching as a result of black hole growth in \textsc{Flares} (at $z\approx 8$), not seen in {\sc Shark} until much lower redshifts; and the implementation of synthetic photometry, with \textsc{Flares} predicting more JWST-detected galaxies ($\sim 90\%$) than \textsc{Shark} ($\sim 70\%$) at $z=10$. JWST observations will distinguish between these models, leading to a significant improvement upon our understanding of the formation of the very first galaxies. 

%; and the effect of chemical enrichment upon the observed light from galaxies {\bf(with attenuation scaling more so with SFR in \textsc{Shark} compared to \textsc{Flares} that we attribute to \textsc{Shark's} quick metal enrichment).}
%
%(with \textsc{Flares} predicting much less dust attenuation compared to {\sc Shark} that that we attribute to \textsc{Shark's} quick metal enrichment). 
% JWST observations will distinguish between these models, leading to a significant improvement upon our understanding of the formation of the very first galaxies. 

%We explicitly use \textsc{Flares}, a cosmological hydrodynamical zoom simulation, and \textsc{Shark}, a semi analytic model, to explore these predictions in two unique theoretical models, which while relying on the same principles of galaxy formation encapsulate them with different modelling recipes. As such, our predictions are model dependent indicating that the JWST should be able to distinguish between the intricacies of the two simulations and thus better improve our understanding of the processes that shaped the formation of the very first galaxies. 

%This is a simple template for authors to write new MNRAS papers.
%The abstract should briefly describe the aims, methods, and main results of the paper.
%It should be a single paragraph not more than 250 words (200 words for Letters).
%No references should appear in the abstract.
\end{abstract}


%%%%%%%%%%%%%%%%%%%%%%%%%%%%%%%%%%%%%%%%%%%%%%%%%%

%%%%%%%%%%%%%%%%% BODY OF PAPER %%%%%%%%%%%%%%%%%%

\input{Sections/Introduction}
\input{Sections/Simulations}
\input{Sections/SMF-SFRF}

\input{Sections/SFS}

\input{Sections/CSFH-CSMH}
\input{Sections/Discussion}
\input{Sections/Conclusion}
\section*{Acknowledgements}
We thank the anonymous reviewer for their time and constructive comments. JCJD is supported by an Australian Government Research Training Program (RTP) Scholarship. We thank the entire \textsc{Flares} team for their support and feedback.  CL has received funding from the ARC Centre of Excellence for All Sky Astrophysics in 3 Dimensions (ASTRO 3D), through project number CE170100013. LJMD acknowledges support from the Australian Research Councils Future Fellowship scheme (FT200100055). CCL acknowledges support from the Royal Society under grant RGF/EA/181016. Cosmic Dawn Centre is funded by the Danish National Research Foundation. This work was supported by resources provided by The Pawsey Supercomputing Centre with funding from the Australian Government and the Government of Western Australia. This work used the DiRAC@Durham facility man- aged by the Institute for Computational Cosmology on be- half of the STFC DiRAC HPC Facility (www.dirac.ac.uk). The equipment was funded by BEIS capital funding via STFC capital grants ST/K00042X/1, ST/P002293/1, ST/R002371/1 and ST/S002502/1, Durham University and STFC operations grant ST/R000832/1. DiRAC is part of the National e-Infrastructure.

% \clearpage

% The Acknowledgements section is not numbered. Here you can thank helpful
% colleagues, acknowledge funding agencies, telescopes and facilities used etc.

%%%%%%%%%%%%%%%%%%%%%%%%%%%%%%%%%%%%%%%%%%%%%%%%%%
\section*{Data Availability}
Figures, scripts and additional data is available upon reasonable request to the authors. \textsc{Flares} is hosted here \url{https://github.com/flaresimulations} and \textsc{Subfind}/photometry outputs are available here \url{https://flaresimulations.github.io/}. \textsc{Shark} is hosted here \url{https://github.com/ICRAR/shark}. \textsc{Shark} \textsc{Subfind}/photometry outputs are available upon reasonable request. 
 
% The inclusion of a Data Availability Statement is a requirement for articles published in MNRAS. Data Availability Statements provide a standardised format for readers to understand the availability of data underlying the research results described in the article. The statement may refer to original data generated in the course of the study or to third-party data analysed in the article. The statement should describe and provide means of access, where possible, by linking to the data or providing the required accession numbers for the relevant databases or DOIs.



\bibliographystyle{mnras}
\bibliography{refs} % if your bibtex file is called example.bib

%\begin{thebibliography}{99}
%\end{thebibliography}

%%%%%%%%%%%%%%%%%%%%%%%%%%%%%%%%%%%%%%%%%%%%%%%%%%

%%%%%%%%%%%%%%%%% APPENDICES %%%%%%%%%%%%%%%%%%%%%

\appendix

\section{Effect of uncertainties on the SFS fitting}
\label{apdx:uncertainties}
\begin{figure*}
    \includegraphics[width=\textwidth]{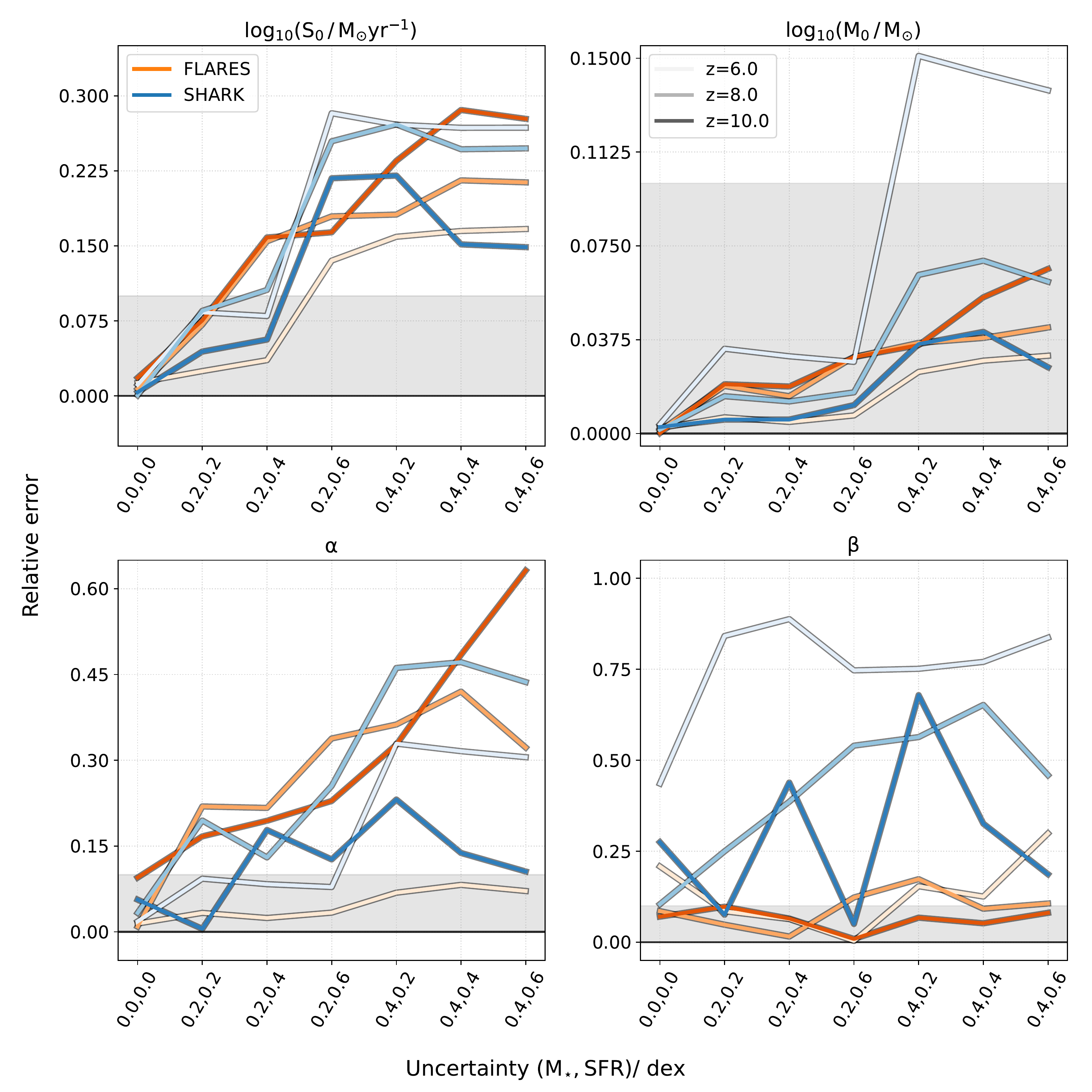}
     \caption{Relative error on each parameter, as labelled, used for fitting the SFS for the JWST population as a function of equal Gaussian uncertainty upon the stellar masses and star formation rates. Orange colours show the results for \textsc{Flares} and blue colours show the results for \textsc{Shark}. Each redshift, ${z=6,8,10}$, is shown by a shading gradient with ${z=10}$ being the darkest shade. The black line highlights no relative error, and the shaded region shows the 10\% level.} 
    \label{fig:relative}
\end{figure*}

To investigate the effect of measurement uncertainties on the ability to recover the SFS parameters of the true galaxy population we refit the SFS of the JWST population with varied Gaussian uncertainties added to the stellar masses and star formation rates. \Cref{fig:relative} shows the relative error of the parameters of the SFS as a function of the assumed uncertainty on the stellar masses and star formation rates of the JWST detected populations in the simulations. The relative error is 
\begin{equation}
    \mathrm{
    Relative \, error = \left | \frac{X_{JWST} - X_{All}}{X_{All}}\right |,
    }
    \label{eq:rel_err}
\end{equation}
where $\mathrm{X_{JWST}}$ are the parameters calculated for the JWST population with the absorbed uncertainties and $\mathrm{X_{All}}$ are the calculated parameters for the entire galaxy population in the simulations. When using the same prior distributions as those we used for the fitting in \Cref{fig:sfs_quantiles,fig:sfs_jwst_fit} we found that for varied uncertainty combinations the relative error was largest in different parameters, making it difficult to see just how additive normal noise affected the fitting. For this reason, we fit the parameters independently, keeping the remaining three parameters fixed, with each uncertainty combination. We achieve this by using a uniform prior on the desired parameter and narrow normal priors centred on the true population parameters for the remaining three parameters, cycling through this process until all parameters have been considered. So for 7 uncertainty combinations, 4 parameters and 3 selected redshifts we perform $\mathrm{7 \times 4 \times 3 = 84}$ MCMC fitting routines on each simulation. 

Both simulations predict a similar trend of increased relative error on all parameters of the SFS as the uncertainty on the stellar masses star formation rates increases, as expected. Even with zero uncertainty there is a non-zero relative error indicating systematic offsets between the JWST detected and total galaxy populations in the simulations. This is perhaps unsurprising as the JWST will miss faint galaxies, and so the relationship between the total and JWST populations is not one-to-one. There is, in general, a larger increase in the relative error when the uncertainty on the stellar mass is increased from 0.2 dex to 0.4 dex compared to when the uncertainty on the SFR is increased from 0.2 dex to 0.4 dex to 0.6 dex. We thus suggest that accurate constraints on the stellar mass are more necessary to describe the SFS than constraints on SFR. The grey shaded regions shows the 10\% margin. Beyond $\mathrm{\sim(0.2,0.4)}$ dex uncertainties on $\mathrm{(M_{\star}, SFR)}$ both simulations fail to recover the normalisation of the SFS to within 10\% at all redshifts shown. For almost any uncertainty combination, the simulations are able to recover the turn over mass to within $10$\% at all redshifts except ${z\sim6}$ in \textsc{Shark}. The intermediate and high stellar mass slopes require far more accurate constraints on the stellar mass and SFR. Beyond $\mathrm{\sim 0.2 \, dex}$ and $\mathrm{\lesssim 0.6 \, dex}$ uncertainty the intermediate stellar mass slope cannot be recovered to within 10\% at all redshifts except ${z=6}$ in \textsc{Flares}. For most uncertainty combinations used in \textsc{Flares} the high stellar mass slope can be recovered except beyond $\mathrm{\sim(0.2,0.6)}$ dex uncertainties on $\mathrm{(M_{\star}, SFR)}$. On the other hand, for almost every uncertainty combination \textsc{Shark} cannot recover the high stellar mass slope at all redshifts suggesting that the high stellar mass slope is subtle in \textsc{Shark} by default. Interestingly, the apparent redshift dependencies on the recovery of these parameters appear to be reversed between the simulations, with \textsc{Flares} showing decreasing relative error for the same uncertainty combination between ${z=10\to6}$, whereas this is true between ${z=6\to10}$ in \textsc{Shark}. We do not comment any further on this in this work. 

Some larger combinations of uncertainties result in a lower relative error than smaller uncertainties. For example, the $\mathrm{\sim(0.4,0.6)}$ dex combination is closer to the true population value of the turn over at all redshifts in \textsc{Shark} than the $\mathrm{\sim(0.4,0.2)}$ combination is. This is more than likely a statistical effect or a symptom of the fitting routine as, intuitively, a larger uncertainty should result in worse agreement; though we do not investigate this further here. Ultimately, these trends show that poorer constraints on $\mathrm{M_{\star}}$ and $\mathrm{SFR}$ result in a worse recovery of the true parameters of the SFS. These trends also provide an indication on the level of accuracy that would be needed to adequately determine the true SFS to within 10\%. 

\bsp	% typesetting comment
\label{lastpage}
\end{document}

%% file: Sections/Introduction.tex
\section{Introduction}
\label{sect:introduction}

A ubiquitous feature of galaxies is the tight coupling between star formation rates (SFR) and stellar masses that has been observed to exist out to high redshifts with very little scatter over the breadth of the $\mathrm{SFR-M_{\star}}$ plane \citep{Brichmann2004,Salim2007,Noeske2007, Noeske2007b,Whitaker_2012}; the so called star forming main sequence (SFS). To first order $\mathrm{SFR \propto M_\star^{\alpha}}$ where $\mathrm{\alpha}$ is a constant; though, differing opinions exist in the literature as to whether the SFS is best described as a single component power law model (e.g. \citealt{Wuyts_2011,Whitaker2012,Speagle2014,Pearson2018}) or a two component power law model (e.g. \citealt{Lee2015,Leslie2020,Thorne2021}). Two component functions model the low and high stellar mass end of the SFS separately to account for distinct slopes in these mass regimes, which accordingly trace different populations of galaxies \citep{Whitaker_2012}. The use of a double power law function to describe the SFS is motivated by a turn over in the SFS that has been shown to occur for massive galaxies on the $\rm{SFR-M_{\star}}$ plane that experience a downturn toward lower SFRs away from a linear trend \citep{Abramson2014,Cook2020}. The physical mechanism behind the SFS and its low scatter is believed to be that galaxies relax toward the locus of the SFS by the internal and self-consistent regulation of star formation. The inflow rate of gas that is needed for star formation is in equilibrium with a combination of the rate at which stars are being formed and the rate at which gas is diverted from the galaxy via outflows driven by feedback from stars, supernovae and active galactic nuclei (e.g. \citealt{Lilly2013, Tacchella2016}). This equilibrium needs to be quickly regained when galaxies are thrown out of it to explain the tightness of the SFS. Indeed, this has been demonstrated to be the case in simulations (e.g. \citealt{Schaye2010, Lagos2011, Lagos2014, Matthee2019}).
%CLAUDIA: I'm commenting the text below as I think you need to justify these statements much better - many in my opinion wouldn't be shared by many people, which may create issues with a potential referee.
%The SFS is essentially the parameter space of energy balance within galaxies. It is particularly useful for investigating the assembly of massive metal free stars in galaxies \citep{Bromm&Larson2004}, the transition from top heavy initial mass functions to distributed initial mass functions \citep{Bromm2009} and the effect of metal pollution via chemical enrichment upon that \citep{Bromm&Loeb2003,Grief2008,Grief2010}, the growth of metallicity in early galaxies \citep{Maio2010, Casey_2014} and the contributors to cosmic reionisation \citep{Loeb_2001,Omukai1999}.

Stellar masses and SFRs have been previously determined out to $z \sim 8$ \citep{Bouwens2015,Gonzalez2011,Katsianis2017} with instruments like the Advance Camera for Surveys on board the Hubble Space Telescope (HST; \citealt{Beckwith2006,Koekemoer2007,Grogin2011}). The trouble with these results is that they are generally short wavelength surveys that can be complicated by the effect of dust attenuation \citep{Draine2003}. The results of these surveys are sensitive to the methods that are used to correct for dust attenuation. This is particularly problematic as it is unclear whether dust obscured galaxy populations have been robustly accounted for, which have shown to become dominant at $z\sim 1-2$ \citep{Casey_2014}. At higher redshifts, it is unclear whether dust-obscured galaxies can make a significant contribution to the cosmic SFR density or not (e.g. \citealt{Casey_2018}.) Sub-millimetre (sub-mm) instruments, like the Atacama Large Millimetre Array (ALMA), have been able to constrain the sub-mm emission of hundreds of galaxies \citep{Fudamoto_2021} at $z>1$, and as sub-mm light is the result of the dust re-radiating absorbed short wavelength photons, these measurements can be used to constrain the SFRs of these galaxies. The problem with these observations is two fold. First, ALMA has a small field-of-view, and hence it could potentially be missing a number of intermediate-to-bright sub-mm bright galaxies, if their frequency is rare, leading to systematic uncertainties in the estimated cosmic SFR density of more than a factor of $2$ at $z\gtrsim 4$ \citep{Casey_2018}. Second, it is difficult to determine stellar masses from sub-mm observations \citep{Michalowski2014}.
%The problem of the latter observations is that they are limited only to the most star-forming galaxies, which heat the dust \citep{Casey_2014}, and it is difficult to determine stellar masses from sub-mm observations \citep{Michalowski2014}. 
While studies have attempted to confirm the existence of the SFS at $z\gtrsim 5$, the stellar mass range is too small ($<1$~dex) to obtain meaningful measurements of the slope and scatter of the SFS \citep{Pearson2018,Leslie2020,Thorne2021}, and the UV rest-frame wavelengths used to measure stellar masses can be extremely affected by dust attenuation.
%observations been successful at confirming the existence of the SFS at high redshifts, for example $\mathrm{z\sim 6}$ \citep{Pearson2018,Thorne2021}, these surveys are too severely affected by the extreme redshifts to adequately confirm the existence and constrain the parameters of the SFS during the epoch of reionisation at redshifts $\mathrm{z\geq5}$. 
This is about to dramatically change, thanks to the recently launched James Webb Space Telescope \citep[JWST;][]{Gardner2006}, which is observing in the infrared ($\mathrm{\lambda \sim 0.5 \to 30.0 \mu m}$) with exquisite resolution and sensitivity, a wavelength range that is less affected by dust than those traced by the HST, even at high redshift. The Near Infrared Camera (NIRCam) on board the JWST is an order of magnitude more sensitive and covers a greater area per pointing than the already existing infrared capabilities of the Wide Field Camera 3 on board the HST \citep{Gardner2006}. A key goal of the JWST is to provide a complete census of galaxies out to $z \sim 10$. 
%, with longer wavelength and spectroscopic follow ups on these sources with the Mid Infrared Camera (MIRI) and the Near Infrared Spectrograph (NIRSpec) also on the JWST.

A key outcome of new observations is to confirm or challenge the current understanding of astrophysical processes. It is thus important to produce tailored predictions for what the JWST will be able to uncover as it continues to observe the distant Universe. For this, we can turn our attention to physically motivated galaxy formation models. Galaxy formation simulations have a rich history \citep[e.g.,][]{Somerville&Dave2015}, and have proven to be a vital tool for interpreting and predicting physical observations of the Universe, especially those from large multi-wavelength galaxy surveys  \citep{Cole2000,Baugh2006,Benson2010,Vogelsberger2014,Schaye2015,Croton2016,Lagos2018}. In order to make meaningful predictions for the JWST, these simulations must include a description of galaxy formation in combination with a description of the spectral output of those simulated galaxies to directly test possible biases affecting the observations.
%Improving upon this theoretical understanding is accomplished by `recovering' the properties of galaxies, like their stellar masses and SFRs, from the light that those galaxies emit and then comparing those derived quantities against the same quantities predicted by galaxy formation models. The conversion of galaxy light to physical quantities relies on key assumptions, like the age and metallicity of the stellar populations, about how light translates to those physical quantities \citep{Taylor2011,Kennicutt2012,Davies2016,Bellstedt2020}. A powerful technique for deriving physical galaxy quantities comes in the form of Spectral Energy Density (SED) fitting, which model the broadband distribution of galaxy light and the physical processes that are most likely to have shaped them \citep{daCunha2012,Robotham2020}. Then, to interpret these results, the recovered galaxy quantities can then be compared with physically motivated models that realise galaxies under the consideration of what current theory permits \citep{Somerville&Dave2015}; simulations must include a description of galaxy formation in combination with a description of the spectral output of those simulated galaxies if they are to be used as apparatuses to test or predict observations. Galaxy formation simulations have a rich history, and have proven to be a vital tool interpreting and predicting physical observations of the Universe \citep{Cole2000,Vogelsberger2014,Schaye2015,Croton2016,Lagos2018}.

In this work we use state-of-the-art simulations to address whether the JWST will be able to predict the existence, and properties, of the SFS from $z=5\to10$.
% This work attempts to use galaxy formation simulations to address the question of what state-of-the-art simulations predict the SFS to be like at $\mathrm{z>5}$, and whether the JWST will be able to characterise it up to $\mathrm{z=10}$. 
%how the understanding of high-redshift galaxies, according to galaxy formation simulations, around the epoch of reionisation will be confirmed or challenged with the advent of the JWST. 
% Recent studies have explored JWST predictions \citep{Cowley2018, Vogelsberger2020,Shen2021, Curtis-Lake2021}, mostly focusing on luminosity functions and expected numbers of galaxies. In addition, these generally present predictions within an individual model, without exploring the dependence of their predictions on the details of galaxy formation modelling. 
Recent studies have explored JWST predictions \citep{Vogelsberger2020, Shen2021, Curtis-Lake2021,wilkinsFirstLightReionisation2022,wilkinsFirstLightReionisation2022a,wilkinsFirstLightReionisation2022b}, mostly focusing on luminosity functions (LF) and expected numbers of galaxies. In addition, these generally present predictions within an individual model, without exploring the dependence of their predictions on the details of galaxy formation modelling.
In this work instead, we make use of two independent galaxy formation models, using drastically different techniques, the cosmological hydrodynamical simulation suite {\sc Flares} \citep{Lovell2021,Vijayan2021}, and the semi-analytic model of galaxy formation {\sc Shark} \citep{Lagos2018}, with the aim of isolating predictions that appear robust to the details of the models, and those that are highly model dependent.
%meaning that the efficacy of these predictions will be reliant on the accuracy of the given model. 
%Singular models are not so effective at highlighting how the variation of particular parameters affects the predictions of the model. Which physical mechanisms are important or true to the details of galaxy formation around the epoch of reionsiation will be hard to pin down as it will be unclear how model parameters are sensitive to idiosyncratic biases when the actual JWST observations arrive and the individual predictions are compared to them. Essentially, two fundamentally distinct simulations, a hydrodynamical simulation and a semi analytic model, are used to explore which predictions are model dependent so that it will be clearer which precise parameters of the models are important for accurately simulating galaxies when they are compared to actual JWST observations together. It should be stressed that this work is not concerned with comparing one simulation against the other because, at the redshifts of interest, it is pointless to say that one simulation is more or less correct than the other. Specifically, we wish to predict what kinds of galaxies the JWST will observe and the distributions of stellar mass and SFRs of those galaxies from redshift $\mathrm{z=5}$ to redshift $\mathrm{z=10}$, and how each of these two things are model dependent. 

Broadly, we wish to predict what kinds of galaxies the JWST will observe and their distributions of stellar mass and SFRs of those galaxies from $z=5$ to $z=10$. The main question is whether those observations would be enough to establish the existence of a SFS out to $z=10$ and the parameters describing it. We choose to make predictions on the galaxies that will be observed in the $\mathrm{\sim 2\mu m}$, F200W filter with NIRCam on the JWST. As the F200W is the most sensitive imaging filter it will be an important component in the construction of high redshift sources \citep{rigbyCharacterizationJWSTScience2022a}. 

% As the initial census of high redshift sources will be taken with this observing instrument, in this filter, it is most natural to see how this population compares with the total population of galaxies throughout redshift. 

This paper is laid out as follows. In \Cref{sect:simulations} we describe the main differences between \textsc{Flares} and \textsc{Shark}, and we make note of the cosmological parameters and initial mass functions (IMFs) that are used in the two simulations. In \Cref{sect:smf-sfrf} we calculate predictions of stellar mass functions (SMF) and SFRs. In \Cref{sect:SFS} we calculate predictions on the parameters of the SFS and the stellar mass dependent scatter in the SFS. In \Cref{sect:cosmic} we predict the cosmic stellar mass and star formation history. In \Cref{sect:discussion} we discuss our results, and in \Cref{sect:conclusion} we present our final conclusions and main points. Unless explicitly specified, we have absorbed cosmological dependencies in our results. Furthermore, all magnitudes are quoted in AB, unless stated otherwise. 

%% file: Sections/Simulations.tex
\section{Simulations}
\label{sect:simulations}
We introduce the two galaxy formation simulations used in this work: \textsc{Flares}, a hydrodynamical zoom simulation (\Cref{subsect:Flares}), and \textsc{Shark}, a semi-analytic model of galaxy formation (\Cref{subsect:Shark}). We explore differences between the dust and stellar population synthesis (SPS) models between the two simulations in \Cref{sect:sps_and_dust}.

\subsection{\textsc{Flares}}
\label{subsect:Flares}
\textsc{Flares} (First Light And Epoch of Reionisation) was introduced in \citet{Lovell2021} and \citet{Vijayan2021}. Below we summarise the baryon model ($\S$~\ref{secflaredmod}) and the way the spectral energy distribution (SED) of galaxies is computed in {\sc Flares} ($\S$~\ref{secdustmodelflares}).

\subsubsection{Modelling and resimulation method}\label{secflaredmod}
\textsc{Flares} is a series of cosmological hydrodynamic zoom simulations built on the {\sc EAGLE} galaxy formation model \citep{Schaye2015,Crain_2015} and therefore utilises the modelling techniques and sub-grid recipes used in {\sc EAGLE}. {\sc EAGLE} is a suite of smoothed-particle-hydrodynamics simulations run on the P-GADGET3 N-body Tree-PM code that was last described in \citet{Springel2005} and uses the ANARCHY (see \citet{Schaller2015} for details) code to solve the coupled equations of hydrodynamics. {\sc EAGLE} includes a swathe of sub-grid recipes including radiative cooling and photo-heating, star formation, stellar evolution and chemical enrichment, black hole growth and feedback from active galactic nuclei (AGN) and massive stars (see \citet{Schaye2015,Crain_2015} for further details on the sub-grid models). {\sc EAGLE} was tuned to the $z\sim0$ galaxy SMF, stellar mass-black hole relation and galaxy sizes, but has also been shown to agree well with observations not explicitly used in the tuning of the sub-grid model free parameters. Relevant for this work, \citet{Furlong2015} showed that the SMF and SFS were reasonably well reproduced in {\sc EAGLE} up to $z=4$ and $z=3$, respectively; \citet{Katsianis2017} showed that the SFR function was reasonably well reproduced up to $z=4$; and both \citet{Katsianis2019} and \citet{Davies2019} showed that {\sc EAGLE} reproduced well the SFS's scatter at $z=0$.

Unfortunately, the dynamic range of stellar masses and star formation rates in the \textsc{EAGLE} suit is insufficient to probe the most massive and star forming galaxies that exist in the most overdense environments. Being a series of zoom simulations, \textsc{Flares} is able to sample these rare overdense environments from a larger dark matter-only simulation thereby providing a statistically complete set of environments. \textsc{Flares} concentrates on redshift snapshots $z=[5,6,7,8,9,10]$ and is therefore perfectly suited to provide predictions on what future surveys will be able to conclude about high-redshift galaxies and the epoch of reionisation. Previous studies have sampled the rarest cluster environments. \citet{Barnes2017,Bahe2017} for example sampled and resimulated 30 massive galaxy clusters, also with the {\sc EAGLE} model, with $z=0$ halo masses $\mathrm{10^{14} \leq M_{200}/M_{\odot} \leq 10^{15.4}}$. Attempting to use such a sample to parametrise the universal properties of high redshift galaxies is however troublesome because of the bias toward only massive cluster galaxies that is not representative of the diverse range of galaxy environments present in the Universe. A prediction of the properties of all galaxies in the early Universe thus requires an unbiased set of simulated galaxies existing in a range of environments and possessing a range of masses. \textsc{Flares} has generated a set of galaxies that are representative of the entire Universe by sampling from a range of 40 overdensities and applying statistical weights to each region \citep{Lovell2021}. The zoom regions are selected from the $\mathrm{3.2^{3} Gpc^{3}}$ dark matter-only {\sc EAGLE} box, which is also used in \cite{Barnes2017}. 40 spherical zoom regions with radius $\mathrm{14 \mathit{h}^{-1} cMpc}$ are resimulated down to $z=4.67$ at the same resolution as the fiducial {\sc EAGLE} simulation and with identical physics and parameters. The benefit of this is a sample representative of the early universe with sufficient number statistics on the most massive, clustered galaxies, but unbiased to those.  

The range of overdensities of the large dark-matter only box is partitioned into 50 bins of equal width in $\log_{10}(1+\delta)$, and the $i^{\rm th}$ bin is assigned a statistical weight, $w_{\rm true, i}$, according to the proportion of resimulated overdensities that exist in the bin against the total quantity of overdensities in the box, such that $\sum_{\rm i}{w_{\rm true, i}} =1$. Each of the resimulated regions are also distributed over the bin size, and are assigned weights, $w_{\rm i,j}$, according to the proportion of resimulated overdensities in the bin against the total quantity of the 40 resimulated overdensities, once again so that $\sum_{i}{w_{\rm ij}} =1$. The weight per bin associated with the resimulated regions is then $w_{\rm sample, i} = \sum_{\rm  j}{w_{\rm  ij}}$. Each density bin is then weighted by $r_{\rm i} =w_{\rm true, i}/w_{\rm sample,i}$. In order for each resimulation region, $\mathrm{j}$, to be statistically representative of the cosmic distribution, it is weighted by $f_{j} = \sum_{i}r_{i}w_{i,j}$ with $\sum_{j}f_{j} = 1$. 
%Each galaxy in the resimulation region has the same statistical weight that likely introduces variance from a truly representative sample; a correct weighting per local galaxy overdensity is deferred to future work. 
The weights do not change with redshift since the relative ordering of the overdensities is expected to be invariant through $z=10 \to 5$ as even the greatest overdensities would be a result of only mildly non-linear evolution. 

Structures are found in \textsc{Flares} first from a Friends-Of-Friends finder \citep{davis1985}, and bound substructure is further identified with the \textsc{SubFind} algorithm \citep{springel2001}. 

The dark matter particle mass resolution of the large $\mathrm{3.2^{3} Gpc^{3}}$ dark matter-only box is $\mathrm{m_{DM} = 8.01 \times 10^{10} M_{\odot}}$ and the gravitational softening length is $\mathrm{59\, ckpc}$ (note this is the same simulation used in \citealt{Barnes2017,Bah__2017} to select the regions to resimulate). The gas particle mass and dark matter particle mass in the fiducial {\sc EAGLE} simulation, and therefore \textsc{Flares}, is $\mathrm{m_{gas} = 1.8 \times 10^{6} M_{\odot}}$ and  $\mathrm{m_{DM} = 9.7 \times 10^{6} M_{\odot}}$ respectively; and the softening length is $\mathrm{\epsilon = 2.66\, ckpc}$. {\sc EAGLE}, and by extension \textsc{Flares}, use the \citet{Planck2014year1}  $\Lambda$ cold dark matter ($\Lambda$CDM) cosmology with $\mathrm{\mathit{H_{0}} = \mathit{h} \times 100 Mpc \, km s^{-1}}$, $\mathrm{\mathit{h}=0.6777}$; $\mathrm{\Omega_{m}=0.307}$ and $\mathrm{\Omega_{\Lambda} = 0.693}$. \textsc{Flares} adopts a universal Chabrier initial mass function \citep{Chabrier2003}. 

\subsubsection{Dust attenuation calculation in {\sc Flares}}\label{secdustmodelflares}

The light emitted by stars is attenuated by a two-phase dust medium, the birth clouds, which attenuate young stars, and the diffuse dust in the interstellar medium, which attenuates both young and older stars. Dust attenuation can be quantified by an examination of the optical depth in the V-band (550nm), $\mathrm{\tau_{V}}$, and seeing how this is affected by the dust. The emission from each simulated galaxy is computed using version 2.2.1 of the Binary Population and Spectral Synthesis (BPASS) stellar population synthesis (SPS) code \citep{Eldridge2017,stanwayReevaluatingOldStellar2018}. Stellar clusters form below sub-kpc scales meaning that dust attenuation from the birth clouds must be calculated using a sub-grid recipe, as:
\begin{equation}
\tau_{\rm BC, V} = 
    \begin{cases} 
      \kappa_{\rm BC}\times (Z_{\star}/0.01) & \mathrm{t\leq 10^{7} yr}, \\
      0 & \mathrm{t > 10^{7} yr},
   \end{cases}
   \label{eq:fl_bc_att}
\end{equation}
where $\kappa_{BC}$ is a normalisation that encodes information about the physical properties of the dust grains and the dust-to-metal ratio of the birth clouds, and $Z_{\star}$ is the metallicity of the star particle. The optical depth is sensitive to the age of the stellar populations and the piecewise function is obtained from \citet{Charlot&Fall2000} who find that birth clouds disperse by $\mathrm{10^{7}}$ years, beyond which there are few attenuating particles around the stellar cluster and hence the optical depth of the birth cloud tends to zero.

In the case of the diffuse dust component, the metal content of the ISM is used as a proxy of dust content  \citep{Vijayan2021}. The metal content of the interstellar medium is directly determined from the SPH particles. 

The equation to calculate the V-band optical depth is 
\begin{equation}
    \tau_{\rm ISM, V} = \mathrm{DTM} \, \kappa_{\mathrm{ISM}} \, \Sigma(x, y)
    \label{eq:fl_ism_att}
\end{equation}
where $\mathrm{DTM}$ is the dust-to-metal mass ratio of the galaxy, $\kappa_{\mathrm{ISM}}$ is a normalisation that encodes information about the physical properties of the dust grains and $\Sigma(x,y)$ is the metal column density integrated over a line of sight. The $\mathrm{DTM}$ ratio adopted in {\sc Flares} is a fitting function \citet[Equation 15]{Vijayan2019} that connects the dust-to-metal mass ratio to the age and metallicity of the galaxy. This function fits the relation between the three latter quantities in the semi-analytic model of galaxy formation {\sc L-galaxies}, and specifically the version introduced in \citet{Vijayan2019} which includes a model for dust formation, growth and destruction. $\Sigma(x,y)$ is evaluated by integrating the density field of SPH particles and linking this to the metallicity and mass of each particle along the line of sight. The viewing angle is fixed such that the line of sight is along the z-axis of the simulation. This prescription yields a sufficient calculation of interstellar medium attenuation without having to simulate the distinct geometry and properties of the dust. 

Attenuation from both physical components, described with \Cref{eq:fl_bc_att} and \Cref{eq:fl_ism_att}, is combined with a general wavelength dependence as 
\begin{equation}
    \label{eq:flares_attenuation_lambda}
    \tau_{\rm \lambda} = (\tau_{\mathrm{ISM}} + \tau_{\mathrm{BC}}) 
    \times (\lambda/550\mathrm{nm)}^{-1}.
\end{equation}
\noindent The form of \Cref{eq:flares_attenuation_lambda} is reminiscent of the \citet{Charlot&Fall2000} attenuation curve. This is the attenuation curve for each star particle along the line of sight. Lower metallicities at high redshift motivates the use of an attenuation curve that is similar to the Small Magellanic Cloud; \citet{Vijayan2021} remark that the slope of the attenuation curve in \textsc{Flares} is flatter in the UV region when compared to the same curve of the Small Magellanic Cloud, but not as flat as the attenuation curve of \citet{calzettiDustContentOpacity2000}. Furthermore, the clumpiness of the ISM can affect the resultant attenuation curve. The dust model is calibrated using the UV light function, UV continuum slope and the [OIII]+H$\beta$ EW distribution \citep[for further details see the appendix of ][]{vijayan2021light}. At all redshifts, $\kappa_{\mathrm{BC}} = 1$ and $\kappa_{\mathrm{ISM}} = 0.0795$. \textsc{Flares} does not include a redshift evolution of the dust grain sizes, masses or composition. 

\subsection{\textsc{Shark}}
\label{subsect:Shark}
The second simulation used in this work is the semi-analytic model of galaxy formation \textsc{Shark}, first introduced in \citet{Lagos2018}. 
\subsubsection{N-body skeleton and semi-analytic method}
The cosmic structure in \textsc{Shark} is provided by the N-body dark matter only simulation Synthetic UniveRses For Surveys: SURFS \citep{Elahi2018}. SURFS was designed to provide theoretical test beds to compliment ongoing and upcoming galaxy surveys. SURFS was run using a memory-lean version of the GADGET-2 code \citep{Springel_2005}, producing a variety of box sizes between $\mathrm{40 \mathit{h}^{-1} cMpc}$ and $\mathrm{210 \mathit{h}^{-1} cMpc}$, each including between $\mathrm{512^{3}}$ to $\mathrm{1563^{3}}$ dark matter particles. Different simulation boxes also cover a variety of particle masses and softening lengths. The SURFS box that underpins the \textsc{Shark} simulation used in this work is the L210N1563 simulation that has box size  $\mathrm{L_{box} = 210 \mathit{h}^{-1} cMpc}$, number of dark matter particles $\mathrm{N_{p} = 1563^{3}}$, dark matter particle mass $\mathrm{m_{DM} = 2.21 \times 10^{8} \mathit{h}^{-1} M_{\odot}}$ and softening length $\mathrm{\epsilon = 4.5 \mathit{h}^{-1} ckpc}$. In total, 200 snapshots spaced in logarithmic intervals of growth factor from redshift $z=24$ to $z=0$ were produced using the \citet{Planck2025XIII} $\Lambda$CDM cosmology with cosmological parameters: $\mathrm{\mathit{H_{0}} = \mathit{h} \times 100 Mpc \, km s^{-1}}$, $\mathrm{\mathit{h}=0.6751}$; $\mathrm{\Omega_{m}=0.3121}$, $\mathrm{\Omega_{b} = 0.0491}$ and $\mathrm{\Omega_{\Lambda} = 0.6879}$ being the Hubble's constant, matter density, baryon density and $\mathrm{\Lambda}$ density respectively.

Dark matter halos and their substructure are identified using the halo finding code, VELOCIraptor \citep{Canas2019,Elahi2019a}, and merger trees of dark matter halos are constructed using the halo merger tree builder, TreeFrog \citep{Elahi2019b}. 

The dark matter halo catalogues from SURFS provide the static skeletons that \textsc{Shark} uses to simulate galaxies. Since links between halo descendants are found for only up to 4 snapshots into the future, it can be the case that discontinuities exist in the merger trees across snapshots. To smooth over these discontinuities and enforce continuity over the equations of galaxy formation, \textsc{Shark} places subhalos between the snapshots of the current subhalo and its descendent, if snapshots were skipped by TreeFrog. The dark matter reservoir in which galaxies are embedded are assumed to have an NFW profile \citep{NFW1997}. 

% Central subhalos are defined as the most massive subhalos within a host halo at $\mathrm{z=0}$. It is in these central subhalos that central galaxies exist. Once this definition is first made at redshift $\mathrm{z=0}$, an iterative process back through time is done that makes the progenitors of the defined centrals the centrals of their own halos at the earlier time step. For galaxies that merged together at each snapshot but are not the main progenitors, the central is designated using the same process described above. There are three different types of \textsc{Shark} galaxies: $\mathrm{type=0}$ are the central galaxies of the central subhalo, $\mathrm{type=1}$ are the central galaxies of the satellite subhalos. In the event that a subhalo merges onto another one but is not the main progenitor, it is called defunct. The galaxies in these defunct subhalos are $\mathrm{type=2}$ and are transferred to the central subhalo of their descendant host halo  \citep{Lagos2018}. For the remainder of this work, central galaxies will be $\mathrm{type=0}$ while every other galaxy will be a satellite. 

Galaxies are evolved by first finding the halos that first appear in the snapshot and do not have a progenitor. The most massive subhalo of these first generation halos are assigned a gas supply of mass
\begin{equation}
    \mathrm{M_{gas} = \Omega_{b}/\Omega_{m} \times M_{halo}},
    \label{eq:sh_mgas}
\end{equation}
and these artificial galaxies are then evolved forward in time. The differential equations that determine the mean mass and metallicity of the baryonic components that are simulated in \textsc{Shark} are given by Equations 49-58 in \citet{Lagos2018}. In \textsc{Shark}, the physical modelling includes: baryonic matter accretion onto halos, radiative gas cooling, star formation in discs and bulges, stellar feedback, reincorporation of ejected gas, chemical enrichment, galaxy mergers, disc instabilities, photoionisation feedback, black hole growth and AGN feedback and environmental effects. As \textsc{Shark} lacks the ability to model complex morphologies of galaxies, the basic morphological description that is used is a bulge-disc distinction. This distinction is important because the assembly of stellar mass, and subsequent chemical enrichment, in either the disc or the bulge is not necessarily identical. This distinction is reflected in the star formation law that is a function of the molecular-to-atomic gas fraction, surface density of gas and pressure in either the disc or the bulge, depending on where the star formation is occurring. In addition, it is assumed that star formation proceeding in the bulge in the form of starbursts converts the molecular gas into stars more efficiently (by a factor of 10) than star formation occurring in the disc.

For extended details on this modelling, the reader is referred to \citet{Lagos2018}. \textsc{Shark} adopts a universal Chabrier IMF \citep{Chabrier2003}. 

\begin{figure*}
    \includegraphics[width=\textwidth]{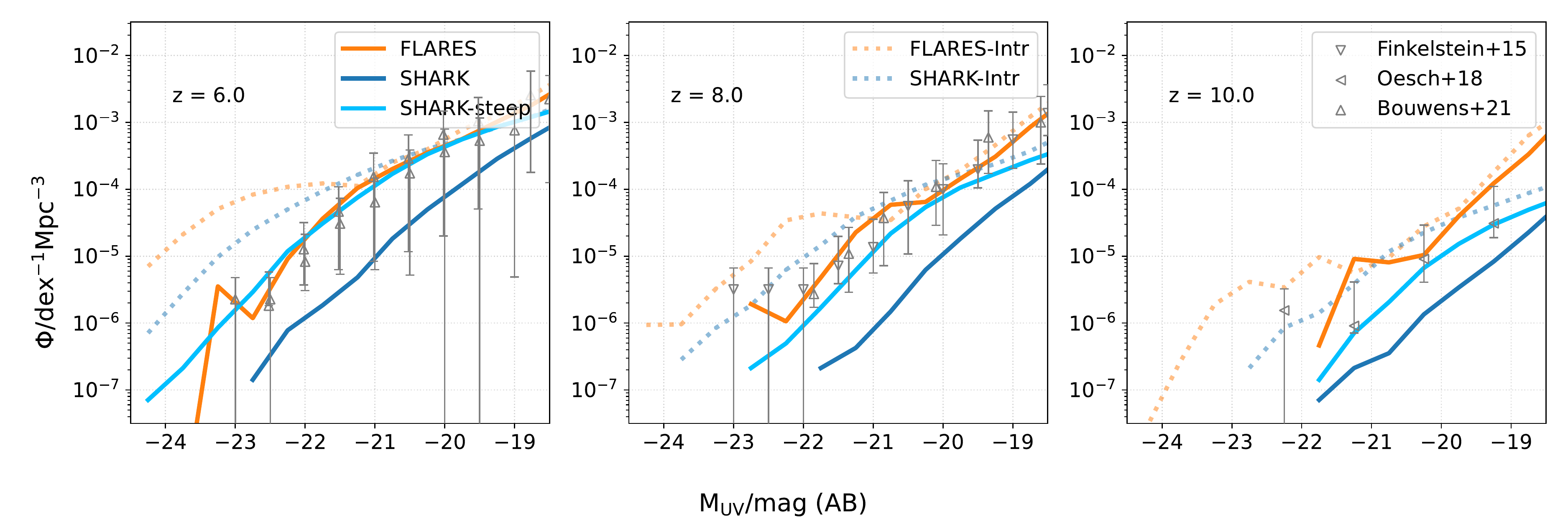}
    \caption{Obscured {rest-frame FUV LFs} for \textsc{Flares} (orange), \textsc{Shark}-default (blue) and \textsc{Shark}-steep (light blue) at redshifts $z=6,8,10$, as labelled. The dotted lines show the intrinsic LFs for {\sc Flares} and {\sc Shark}. Grey symbols with error bars show observational constraints of the obscured rest-frame FUV LFs from \citet{Finkelstein2015, Oesch_2018, Bouwens2021}. It is apparent that {\sc Flares} and {\sc Shark}-steep offer the best fits to the observations, while \textsc{Shark}-default predicts number densities that are too low.}  
    \label{fig:fuv_z6-10}
\end{figure*}

\subsubsection{Dust attenuation calculation in {\sc Shark}}
The work horse behind the lighting of \textsc{Shark} galaxies is \textsc{ProSpect} \citep{Robotham2020}, a multiwavelength spectral energy distribution package capable of generating and fitting spectral energy distributions under the consideration of many different astrophysical mechanisms. \textsc{ProSpect} by default utilises the SPS libraries of \citet{Bruzual&Charlot2003}, though the highly flexible nature of \textsc{\textsc{ProSpect}} allows for the use of many different SPS libraries (such as e-MILES, \citealt{Vazdekis2016}, and BPASS, \citealt{Eldridge2017}). Dust attenuation is described using the model of \citet{Charlot&Fall2000}. Star formation and metallicity histories from \textsc{Shark} galaxies are fed to \textsc{ProSpect} with sensible assumptions about how dust attenuates galaxy emission and how energy is then reradiated in the infrared. Much like \textsc{Flares}, the attenuation is multicomponent in nature being related to the optical depth of both birth clouds and the interstellar medium.

%Much like \textsc{Flares}, the attenuation is multicomponent in nature being related to the V-band optical depth of both birth clouds and the interstellar medium. The relevant equations are 
% \begin{equation}
%     \mathrm{\tau_{BC} = \tau_{ISM} + {\tau}_{BC, V}
%     \times (\lambda/550nm)^{\eta_{BC}}},
%     \label{eq:sh_att_bc}
% \end{equation}
% \begin{equation}
%     \mathrm{\tau_{ISM} = \tau_{ISM, V}
%     \times (\lambda/550nm)^{\eta_{ISM}}},
%     \label{eq:sh_att_ism}
% \end{equation}

% \noindent respectively. 

There are four distinct methods of dust attenuation that are included in \textsc{Shark}, and the reader is referred to \citet{Lagos2019} for further details on all of them. Here we only concentrate on methods 3 and 4 as referred to in \citet{Lagos2019}, and we briefly outline them below. 

The optical depth of the diffuse ISM is modelled as 
\begin{equation}
    \tau_{\mathrm{ISM}} = \tau_{\mathrm{ISM, V}}
    \times (\lambda/550\mathrm{nm})^{\eta_{\mathrm{ISM}}},
    \label{eq:sh_att_ism}
\end{equation}
\citet{Trayford2020} use the radiative transfer code SKIRT \citep{Camps2015} to derive scaling relations between $\tau_{\mathrm{ISM, V}}$, $\eta_{\mathrm{ISM}}$ and the surface density of dust $\Sigma_{\mathrm{dust}}$ for galaxies between $z=0 \to 2$ in the {\sc EAGLE} simulation, finding the relationships to be mostly independent of redshift. In \textsc{Shark} the dust surface density is computed for each galactic component (disc and bulge), and the attenuation parameters of \Cref{eq:sh_att_ism} are determined by sampling the scaling relations of \citet{Trayford2020}. Note that these parameters are found independently for the bulge and disc. 

% , attenuation from the interstellar medium is done by obtaining the attenuation parameters in \Cref{eq:sh_att_ism} with the dust mass surface density from the scaling relations presented by \citet{Trayford2020}. In \citet{Trayford2020}, the attenuation of {\sc EAGLE} galaxies up to redshift $\mathrm{z=4}$ is modelled using a \citet{Charlot&Fall2000} attenuation scheme whose shape experiences little variation with redshift. They find that the surface density of dust, $\mathrm{\Sigma_{dust}}$ scales with $\mathrm{\tau_{ISM, V}}$ and $\mathrm{\eta_{ISM}}$ with a low scatter across the breadth of the plane. \textsc{Shark} galaxies with an associated dust surface density are assigned with the ISM attenuation parameters used in \Cref{eq:sh_att_ism} from the scaling relations and then perturbed by a Gaussian with width, $\mathrm{\sigma}$, being the 16-84 percentiles. 
The opacity of light from the birth clouds is calculated with 
\begin{equation}
    \tau_{\rm BC} = \tau_{\mathrm{ISM}} + {\tau}_{\mathrm{BC, V}}
    \times (\lambda/550 \mathrm{nm})^{\eta_{\mathrm{BC}}}.
    \label{eq:sh_att_bc}
\end{equation}
\noindent The parameter $\tau_{\mathrm{ISM}}$ appears in \Cref{eq:sh_att_bc} because light from stars inside the birth clouds is attenuated by both the birth clouds themselves and the ISM.
The V-band optical depth of the birth cloud is derived from the following equation:
\begin{equation}
    \tau_{\mathrm{BC, V}} = \tau_{\mathrm{BC, 0}}\frac{\mathrm{DTM}  \, Z_{\mathrm{gas}}  \, \Sigma_{\mathrm{gas, cl}}}{\mathrm{DTM_{MW}}  \, Z_{\odot} \, \Sigma_{\mathrm{MW, cl}}}
    \label{eq:sh_bc_vband}
\end{equation}
$\tau_{\mathrm{BC, 0}}=1$, DTM is the dust-to-metal ratio, $Z_{\mathrm{gas}}$ is the gas metallicity, $\Sigma_{\mathrm{gas, cl}}$ is the surface density of gas, $\mathrm{DTM_{MW}}=0.33$ is the dust-to-metal ratio of the Milky Way, $Z_{\odot}=0.0189$ is the solar metallicity and $\Sigma_{\mathrm{MW, cl}}=85 \rm \,M_{\odot}\mathrm{pc}^{-2}$ is the typical surface density of molecular clouds in the Milky Way \citep{Krumholz_2009}. The value of $\eta_{\mathrm{BC}}$ is taken to be the default for the \citet{Charlot&Fall2000} attenuation model: $\eta_{\mathrm{BC}} = -0.7$.

The dust masses, which are used to compute $\Sigma_{\mathrm{dust}}$, are by default determined from the metallicity of the gas in the galaxy according to the best fitting $\mathrm{M_{dust}/M_{Z} - Z_{gas}}$ relation computed by \citet{Remy-Ruyer2014}. An alternative method for calculating dust masses uses a steeper fit to the $\mathrm{M_{dust}/M_{Z} - Z_{gas}}$ within the errors of the best-fitting relation, which is more consistent with the more recent data of \citet{DeVis2019}. \textsc{Shark} uses these as the dust-to-metal ratio. The surface density of the dust is computed for discs and bulges separately according to the following equations;
\begin{equation}
    \mathrm{\Sigma_{dust, disc} = \frac{0.5 \, M_{dust, disc}}{\pi r_{50, disc}, \, l_{50}}},
\end{equation}
where $\mathrm{M_{dust, disc}}$ is the dust mass in the disc, $\mathrm{r_{50, disc}}$ is the half-gas mass radius of the disc and $\mathrm{l_{50} = sin(i) \times (r_{50, disc} - r_{50, disc}/7.3) + r_{50, disc}/7.3)}$ is the projected minor axis with inclination $\mathrm{i}$. The factor $7.3$ originates from the average ratio between the scale height and scale length observed in local galaxy discs \citep{Kregel2002}, and the inclination is determined from the angular momentum vector of the host sub halo or chosen randomly for orphan galaxies. For bulges,
\begin{equation}
    \mathrm{\Sigma_{dust, bulge} = \frac{0.5 \, M_{dust, bulge}}{\pi r_{50, bulge}^{2}}},
\end{equation}
where $\mathrm{M_{dust, bulge}}$ is the dust mass of the bulge and $\mathrm{r_{50, bulge}}$ is the half-gas mass radius of the bulge. Bulges are considered to be spherically symmetric so the inclination is unimportant.

The two models are thus referred to as 

\begin{itemize}
    \item {\sc EAGLE}-$\mathrm{\tau}$ RR14, which is the default for \textsc{Shark} and uses the {\sc EAGLE} parametrisation of the \citet{Charlot&Fall2000} model presented in \citet{Trayford2020} and the best-fitting relation of the $\mathrm{M_{dust}/M_{Z} - Z_{gas}}$ from \citet{Remy-Ruyer2014}.
    \item {\sc EAGLE}-$\mathrm{\tau}$ RR14-steep that is the same as above but uses a steeper relation for the $\mathrm{M_{dust}/M_{Z} - Z_{gas}}$. This indicates that this model will have lower dust masses for fixed metallicities than the previous model. 
\end{itemize}
We choose to concentrate on only two of the possible four dust models included in \textsc{Shark} here for the purpose of demonstrating the varying effects of dust models upon a single simulation.

\subsection{SPS and Dust Models: comparing {\sc Flares} and {\sc Shark}}
\label{sect:sps_and_dust}

We investigate how different parameters used in the generation of the artificial spectral energy distributions affect the emission of our simulated galaxies. Later we use synthetic photometry to apply magnitude cuts in  galaxies to isolate those that would be considered JWST detected.

% For the intrinsic and obscured light functions we show the FLARES results, with the default parameters explained in \Cref{sect:simulations}, in orange. 
% 
% Because of the highly flexible nature of SHARK and \textsc{ProSpect} we are able to show the affect of using the BPASS \citep{Eldridge2017} and BC03 \citep{Bruzual&Charlot2003} stellar population synthesis (SPS) libraries on the intrinsic UV emission, and we can see that the differences between using these two libraries is minimal; the main difference being that BPASS is able to extend down to a few brighter UV magnitudes. Meanwhile, we show the UV emission using only the default parameters for FLARES, in this case BPASS. Nevertheless, the minute differences in using two different SPS models in SHARK reflects that the difference between FLARES and SHARK is more so driven by each independent method of stellar mass assembly as opposed to differences in the synthetic stellar populations. 
% 
% 
% 
\Cref{fig:fuv_z6-10} shows the obscured rest-frame far ultraviolet (FUV, $\mathrm{\sim1500}$\AA) LFs at $z=6,8,10$. We show the effect of using both the default and steep dust models in \textsc{Shark} as described in  \Cref{sect:simulations}. 
% As a consequence of the way that dust attenuation is treated in \textsc{Flares} i.e. it using the SPH particles directly, we do not have the flexibility to adopt different methods for determining the parameters in 
% \Cref{eq:fl_bc_att,eq:fl_ism_att}. 
We do not have the flexibility to adopt different methods for determining the parameters in \Cref{eq:fl_bc_att,eq:fl_ism_att} for \textsc{Flares} in an efficient way as the use of a range of possible recipes that treat the dust is computationally expensive to run. We also do not perform any stellar mass cut in these plots. Observational constraints of the rest-frame UV LFs at these redshifts from \citet{Finkelstein2015, Oesch_2018, Bouwens2021} are shown. These use a combination of optical and near infrared images from the Hubble Space Telescope. We also show the unobscured LFs of each simulation for reference. The unobscured LFs are similar to the obscured ones at the faint end but diverge at the bright end, where there is significant difference between the intrinsic and obscured brightness per unit volume, showing that the effect of dust upon the FUV LF is significant, even at $z=10$ in both \textsc{Flares} and \textsc{Shark}. The kinks in the intrinsic LFs of \textsc{Flares} for $\mathrm{M_{UV} \lesssim -21}$ are due to the overdense regions that \textsc{Flares} samples at fixed magnitude, which \textsc{Shark} does not due to its limited volume. Because of the highly flexible nature of \textsc{Shark} and \textsc{ProSpect}, we are able to investigate the effect of using the BPASS \citep[][version 2.2.1]{Eldridge2017,stanwayReevaluatingOldStellar2018} and BC03 \citep{Bruzual&Charlot2003} SPS libraries on the intrinsic rest-frame UV and optical emission, and we found that the differences between using these two libraries is minimal. Therefore, although \textsc{Flares} uses BPASS, we continue to use the default BC03 library with \textsc{Shark} for consistency with previous {\sc Shark} papers. 

The default \textsc{Shark} dust model causes a greater deficit in the FUV brightness than the steep model over nearly all magnitudes at each redshift, as expected from the higher dust mass at fixed metallicity in the this model than in the steep model. 
%There is a slight turnover at the bright end where the default model predicts more FUV brightness suggesting that the steep dust model predicts more dust for a fixed metallicity than the default model, but only for the already intrinsically FUV bright galaxies. 
% which boils down to the default model causing fewer UV photons to escape from the galaxies that can possibly enter an observing line of sight than the steep model. The culprit behind these differences in UV brightness, and therefore the escape fractions, is differing distributions of intervening dust because UV photons can be absorbed or scattered by dust grains \citep{Mathis1977,Draine2003}; and so, redden galaxies. As described in \Cref{sect:simulations} dust masses in \textsc{Shark}, for these two dust models, are derived from a scaling between the gas metallicity and the dust masses obtained from observations. For a fixed metallicity the steep dust model coincides with a lower dust mass than for the default dust model; therefore, this explains why the steep model shows a higher normalisation in the UV light function than the default model. 
While both simulations agree reasonably well with the observational results, such as the UV-FIR emission of galaxies \citep[e.g.,][]{Lagos2018,Lagos2019,Vijayan2021,vijayan2021light}, it is important to highlight the large uncertainties that characterise current observational measurements. Due to the \textsc{Shark}-steep dust model matching these observations far better than the default model, we will use the steep dust model wherever we make use of \textsc{Shark} photometry within the subsequent sections. \citet{Lagos2019} showed that the normalisation of the $z=0$ FUV LF predicted by steep \textsc{Shark} model was slightly higher around the break, $\mathrm{L^{*}}$, than the default model by a fraction of a dex. The fact that the steep model provides a better fit to the observed UV LFs at $z >6$ means that either we require a redshift dependent dust-to-metal ratio relation or a chemical evolution model that enriches galaxies more slowly than the current instantaneous recycling model in {\sc Shark}. We come back to this discussion later in the context of the predicted cosmic stellar mass and SFR history.
%  Additionally, the latter two models are most similar to \textsc{Flares} as they are explicitly derived from the {\sc EAGLE} simulations \citep{Lagos2019}. Therefore, by fixing \textsc{Flares} while changing \textsc{Shark}, we are able to see that the steep dust model better agrees with \textsc{Flares}, even though both simulations model this dust attenuation uniquely. It will thus be a task for higher clarity, statistically complete, near infrared observations done with the JWST to disentangle the uncertainties between two independent models, in this way further comprehending the implications of dust upon galaxy formation. 
\begin{figure}
    \includegraphics[width=\linewidth]{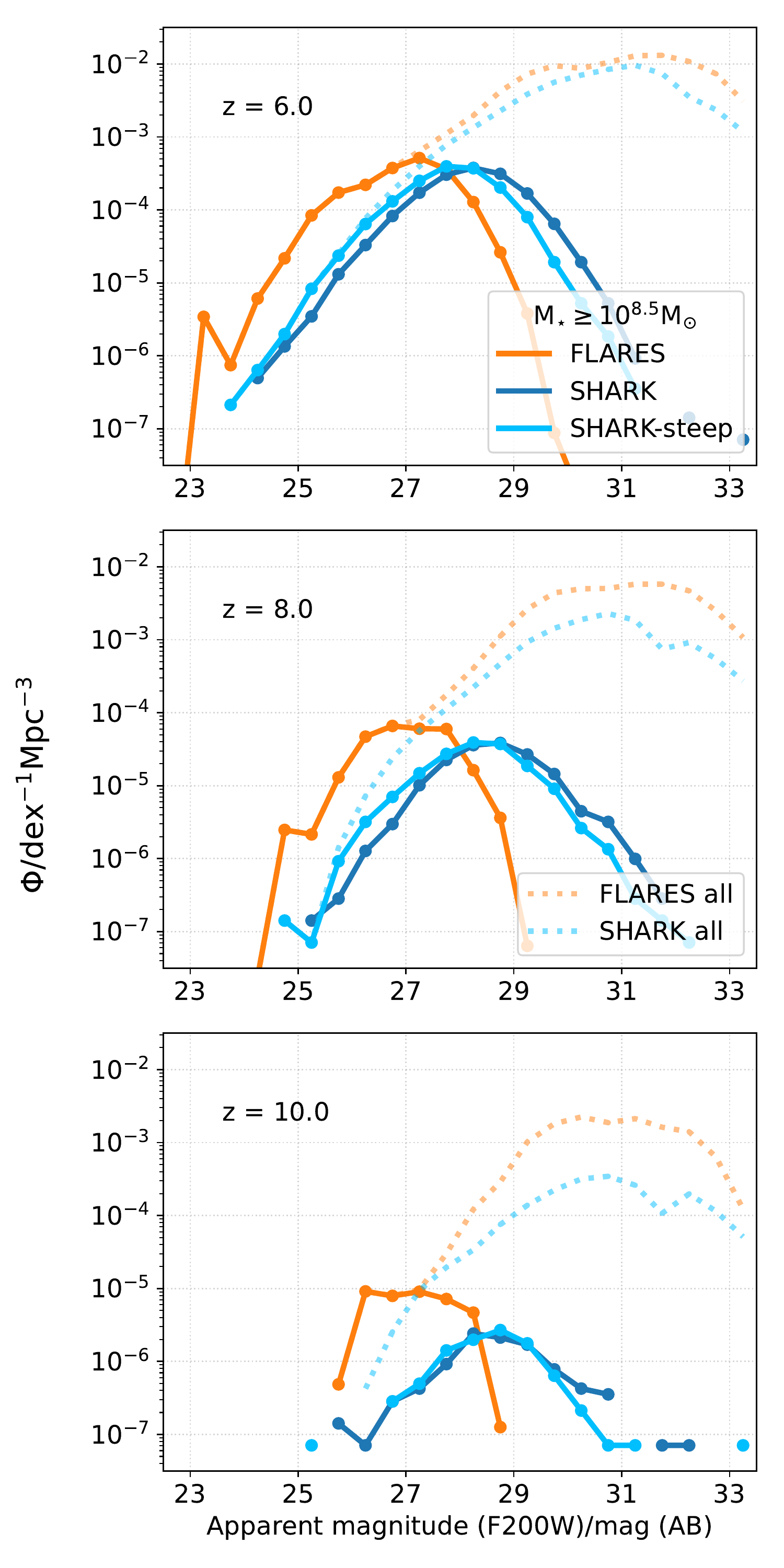}
    \caption{The JWST NIRCAM F200W apparent magnitude volume weighted distribution in \textsc{Flares} (orange), \textsc{Shark}-default (blue) and \textsc{Shark}-steep (light blue) at redshifts $z=6,8,10$. We have performed a stellar mass cut of $\mathrm{M_{\star} \geq 10^{8.5}M_{\odot}}$. The faint dotted lines show the LFs for all galaxies in \textsc{Flares} and \textsc{Shark}-steep.}  
    \label{fig:f200w_z6-10}
\end{figure}

The distribution of galaxies that will be recoverable with the JWST, using NIRCam and the F200W filter for example, depends on the fraction of $\mathrm{\lambda \lesssim 0.5\mu m}$ photons that escape from the galaxy at $z \gtrsim 5$.  \Cref{fig:f200w_z6-10} shows the apparent magnitude distribution per unit volume in the F200W band for \textsc{Flares} and \textsc{Shark} galaxies. As with the FUV LFs, both simulations predict similar shapes to the F200W distributions over the breadth of the magnitude domain and throughout redshift. The \textsc{Shark} steep model predicts more bright galaxies than the default model, and the cause of this difference is the same as it is for the UV LFs in \Cref{fig:fuv_z6-10}. These differences are, however, lesser than those seen in \Cref{fig:fuv_z6-10} as the rest wavelength traced by the JWST is less affected by dust than that traced by the Hubble Space Telescope, for example, at $z\gtrsim 5$ (we elaborate on this further in \Cref{sect:smf-sfrf}). Also note that both simulations have a peak at around $\sim 30$~mag. This peak is driven by resolution effects becoming significant at higher magnitudes, as galaxies with stellar masses $\lesssim 10^{8.5}\,\rm M_{\odot}$ start to dominate the number density fainter than $30$~mag as can be seen from the dotted lines that show the distribution for these galaxies in the simulations. The sensitivity of the JWST in the F200W filter is predicted to be 29 mag, as per the exposure calculator tool \footnote{The 29mag cut comes from using an integration time of $10,000$~seconds and imposing detections to be above $5\sigma$. The exposure calculator for JWST can be found here \url{https://jwst.etc.stsci.edu}}. Hence, throughout the text we will refer to JWST detected galaxies as those brighter than 29 mag in F200W. 
%CONFIRM THIS.
%15/04/22 updated plot to show < 10^8 galaxies

The FUV and Nircam F200W LFs highlight clear differences between \textsc{Flares} and \textsc{Shark}. Although variations in the dust model used in simulations can alleviate this tension, fundamental differences in the way that either simulation assembles stellar mass affect the predicted emission of galaxies \citep[e.g.,][]{bellstedtGalaxyMassAssembly2020,koushanGAMADEVILSConstraining2021,wilkinsFirstLightReionisation2022a}. These processes are interconnected as dust tracks metal enrichment that is itself connected to star formation and stellar evolution. As such, it is useful to investigate how each simulation models the distribution of stellar mass and SFR in galaxies out to redshift $z=10$ and whether the JWST will be able improve our understanding of them.   

%% file: Sections/SMF-SFRF.tex
\section{Stellar mass functions and star formation rate functions}
\label{sect:smf-sfrf}

%In this section we present predictions on the stellar mass and SFR functions.

\begin{figure*}
    \includegraphics[width=\textwidth]{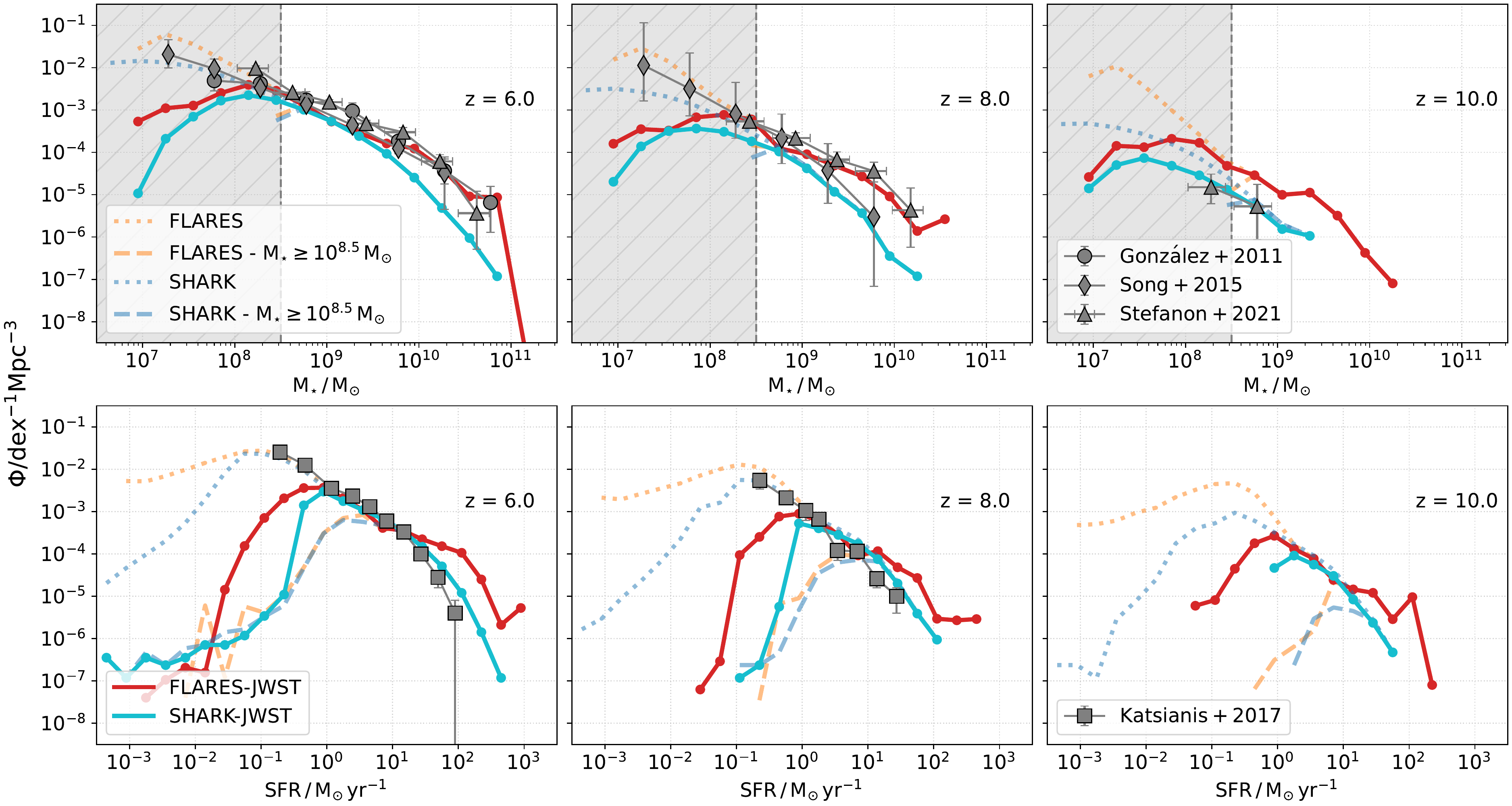}
    \caption{SMFs (top) and SFRFs (bottom) at redshifts $z = 6,8,10$. The light coloured, dotted lines show these quantities for the total population of galaxies in the \textsc{Flares} (orange) and \textsc{Shark} (blue). The light coloured dashed lines show these quantities for the galaxies with stellar masses above $\mathrm{10^{8.5}M_{\odot}}$ in \textsc{Flares} (orange) and \textsc{Shark} (blue). The solid lines show these quantities for the JWST detected populations in \textsc{Flares} (red) and \textsc{Shark} (cyan). Points with error bars show observational constraints of the SMF from \citet{Gonzalez2011, Song2016, Stefanon2021} and SFRF from \citet{Katsianis2017}.}  
    \label{fig:smf_sfrf_z6-10}
\end{figure*}
\begin{figure*}
    \centering
    \includegraphics[width  = \linewidth]{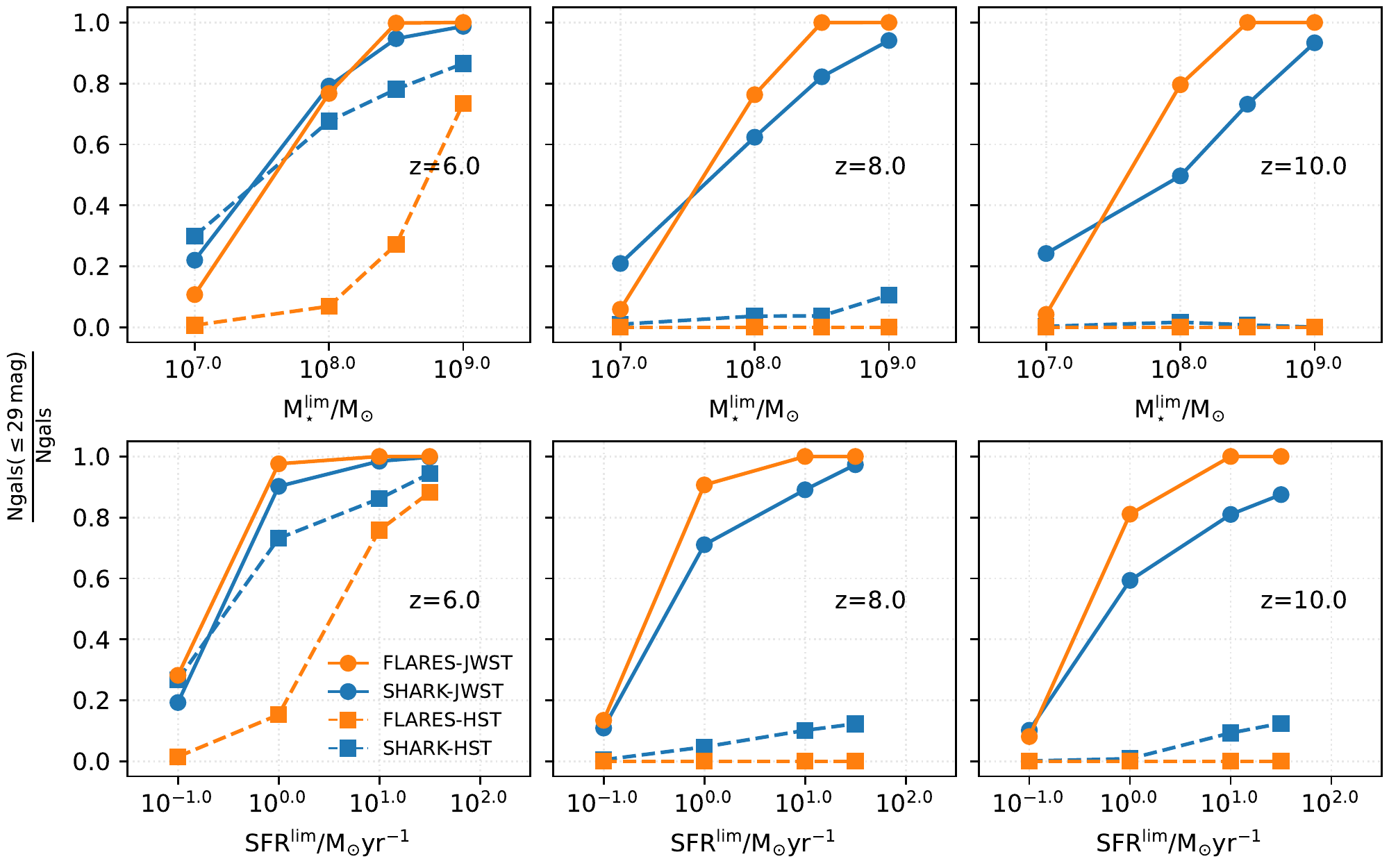}
    \caption{Fraction of galaxies that are brighter than 29 magnitudes in a population selected to have stellar masses (top panels) or SFRs (bottom panels) above a limit, as a function of that limit. This is shown for \textsc{Flares} (orange) and \textsc{Shark} (blue) at $z=6,\,8,\,10$, as labelled. The solid lines with connecting dots show the fraction of galaxies detected in the F200W filter of NIRCam on JWST, while the dashed lines with connecting squares shows this for  galaxies detected in the F775W filter of ACS on HST. At $z\ge 8$, HST is detecting $<10$\% of even the most massive, or highest SFR galaxies in both models.}
    \label{fig:dark_frac_mass_sfr}
\end{figure*}

\Cref{fig:smf_sfrf_z6-10} shows the stellar mass functions (SMF) and SFR functions (SFRF) for galaxies at redshifts $z=6,8,10$ in both \textsc{Flares} and \textsc{Shark}. Note that here we present the predictions assuming no errors in stellar mass or SFRs. It is common, however, for predictions to be presented assuming random errors for stellar mass and SFR, which due to the Eddington bias, tend to shift the high-mass end towards higher stellar masses or SFRs (e.g. \citealt{Lagos2018}). We use the instantaneous SFRs of the galaxies in the results presented for both simulations. For \textsc{Flares} galaxies we follow \citet{Lovell2021} and define the stellar mass as the total mass of star particles within a 30 kpc aperture centred on the potential minimum of the subhalo, as such the SFR of \textsc{Flares} galaxies is taken to be within the same aperture. We use these definitions throughout this work.

An important effect for which we must account that motivates our particular selection is the stellar mass resolution of simulated galaxies. \citet{Schaye2015} and \citet{Furlong2015} suggest that galaxies are sufficiently well sampled provided that they consist of at least 100 star particles, which for the star mass resolution of EAGLE and therefore \textsc{Flares}, results in a mass resolution of $\mathrm{M_{\star}^{lim} \approx 10^{8.2577}M_{\odot}}$ in the simulation. In a similar fashion, reasonably sampled dark matter halos include 100 dark matter particles in the SURFs simulations \citep{Elahi2018} that precipitates into a galaxy stellar mass resolution of $\mathrm{M_{\star}^{lim} \approx 10^{8}M_{\odot}}$ when considering the gas seeding and star formation recipes in \textsc{Shark} \citep{Lagos2018,Elahi2018}. We thus show the distributions of galaxies with  $\mathrm{M_{\star} \geq 10^{8.5} M_{\odot}}$ with the dashed, coloured lines. 

Below $\mathrm{M_{\star}\sim 10^{8.5}M_{\odot}}$, and therefore $\mathrm{SFR\sim1\, M_{\odot}yr^{-1}}$, the simulations diverge at all redshifts. This divergence is below the mass resolution of the simulations, so it is likely artificial. At $z=6$, the simulations agree up to $\mathrm{M_{\star}\sim 10^{10}M_{\odot}}$ and $\mathrm{SFR\sim10^{1}\to10^{2}M_{\odot}yr^{-1}}$. For higher stellar masses and SFRs the simulations are in tension, with \textsc{Flares} predicting an excess in the number densities compared to \textsc{Shark}. At $z>6$, the maximum stellar masses and SFRs at which the simulations agree is generally lower than at later redshifts. At $z=8$ the simulations agree up to $\mathrm{M_{\star}\sim 10^{9.0}M_{\odot}}$ and $\mathrm{SFR\sim10^{1}M_{\odot}yr^{-1}}$, an order of magnitude lower than the agreement threshold at $z=6$. By $z=10$ the stellar mass distributions of the simulations are in tension over all stellar masses, with \textsc{Flares} predicting a higher number density than \textsc{Shark} across the whole stellar mass range probed. The extension to higher SFRs and stellar masses in \textsc{Flares} is due to the larger effective volume of \textsc{Flares} compared to \textsc{Shark}. Interestingly, at $z=10$ the distribution of SFRs agree over a similar SFR range as that at $z=8$. This indicates that \textsc{Shark} assembles less stellar mass than \textsc{Flares} for a SFR in the range where they agree. In \cref{sect:discussion} we discuss the causes behind the differences at the high mass end of the SMF between the two simulations.
%{\bf The difference between {\sc Flares} and {\sc Shark} at the high-mass end of the stellar mass function is due to {\sc Flares} sampling better the rare, large density peaks. This is clear from the 
%Above $\mathrm{M_{\star}\sim 10^{9} \to 10^{10}M_{\odot}}$ and $\mathrm{SFR\sim10^{0}\to10^{1}M_{\odot}yr^-1}$ \textsc{Flares} exhibits a clear excess in the number density of stellar masses and SFRs at all redshifts when compared to \textsc{Shark}. 
%Both \textsc{Flares} and \textsc{Shark} are in agreement with each other for the stellar mass selected galaxies. At the largest stellar masses and by association the greatest star formation rates \textsc{Flares} exhibits a greater number density per unit volume compared to \textsc{Shark}. As \textsc{Flares} galaxies are more sensitive to the most clustered, overdense environments in the universe this enhancement is likely due to clustered galaxies that are contributing large stellar masses and star formation rates.
The solid cyan and red lines show the SMF and SFRF for the galaxies that will be detected by JWST at this wavelength. In \textsc{Shark} and \textsc{Flares} there is remarkable agreement between the distributions calculated for the total population and the JWST detected population above the resolution limit in the simulations; this means that both simulations predict that the JWST will be able to observe all galaxies of $\mathrm{M_{\star} \geq 10^{8.5}M_{\odot}}$ and $\mathrm{SFR \geq 10^{0} M_{\odot}yr^{-1}}$ out to at least $z = 10$.

The top panels of \Cref{fig:smf_sfrf_z6-10} show observational inferences of the SMF. In particular, the $z=6$ SMF of \citep{Gonzalez2011} who use a combination of rest-frame optical, ultraviolet and infrared fluxes obtained from the HST and Spitzer Telescope to derive stellar masses; the SMFs at redshifts $z=6$ and $z=8$ from \citet{Song2016} who also use a combination of HST and Spitzer observations, which cover a greater area than \citet{Gonzalez2011}; the redshift $z=6,8,10$ SMFs of \citet{Stefanon2021} use a combination of HST and deep IRAC observations. Both simulations generally agree with the available observations within the  uncertainties. At the highest redshifts, the observations exhibit large uncertainties meaning that with current observations, the high redshift SMF cannot be well constrained. A good example of the poor constraining power of these observations, is that even though {\sc Shark} and {\sc Flares} predict different number densities of massive galaxies at $z=8$, different sets of observations appear to prefer one or the other model.
Critically, these observations at these extreme redshifts are almost entirely derived from short wavelength observations that are sensitive to the methods that are used to correct for dust attenuation. Additionally, some estimates require near infrared measurements to make use of the full galaxy SED to determine the stellar mass for the highest redshift objects, which is troublesome with previous observatories, such as Spitzer, as a result of poor NIR sensitivity. This type of systematic uncertainty is not accounted for in the error bars of the observations in \Cref{fig:smf_sfrf_z6-10}, nor in the simulations.

A similar story is also true for the observational results shown for the SFRF in the lower panels of \Cref{fig:smf_sfrf_z6-10}. We show results from \citet{Katsianis2017} who use the UV LFs of \citet{Bouwens2015} to derive dust-corrected SFRs. \citet{Katsianis2017} use the $\mathrm{IRX-\beta}$ of \citet{Meurer1999} and the linear $\mathrm{\beta-M_{UV}}$ of \citet{Bouwens2012} to determine the absorption at $1600$\AA, and thus the optical depth, to correct the UV luminosities for dust attenuation following the method of \citet{Hao2011}. It can be seen that there is good agreement between these results and the predictions of \textsc{Flares} and \textsc{Shark} across the range of SFRs probed. Where possible, we have tried to standardise the cosmologies utilised by the simulations and observations as differences in cosmologies can introduce tensions \citep{Croton2013}; it should however be noted that discrepancies precipitated by different cosmologies are likely dwarfed by differences induced by modelling details, light-to-mass derivations and dust corrections \citep{Speagle2014}. It can also be seen that the agreement with these observational results persist for the galaxies below our chosen stellar mass selection. However, this agreement is sensitive to the dust corrections used to derive SFRs.

\begin{figure*}
    \includegraphics[width=\textwidth]{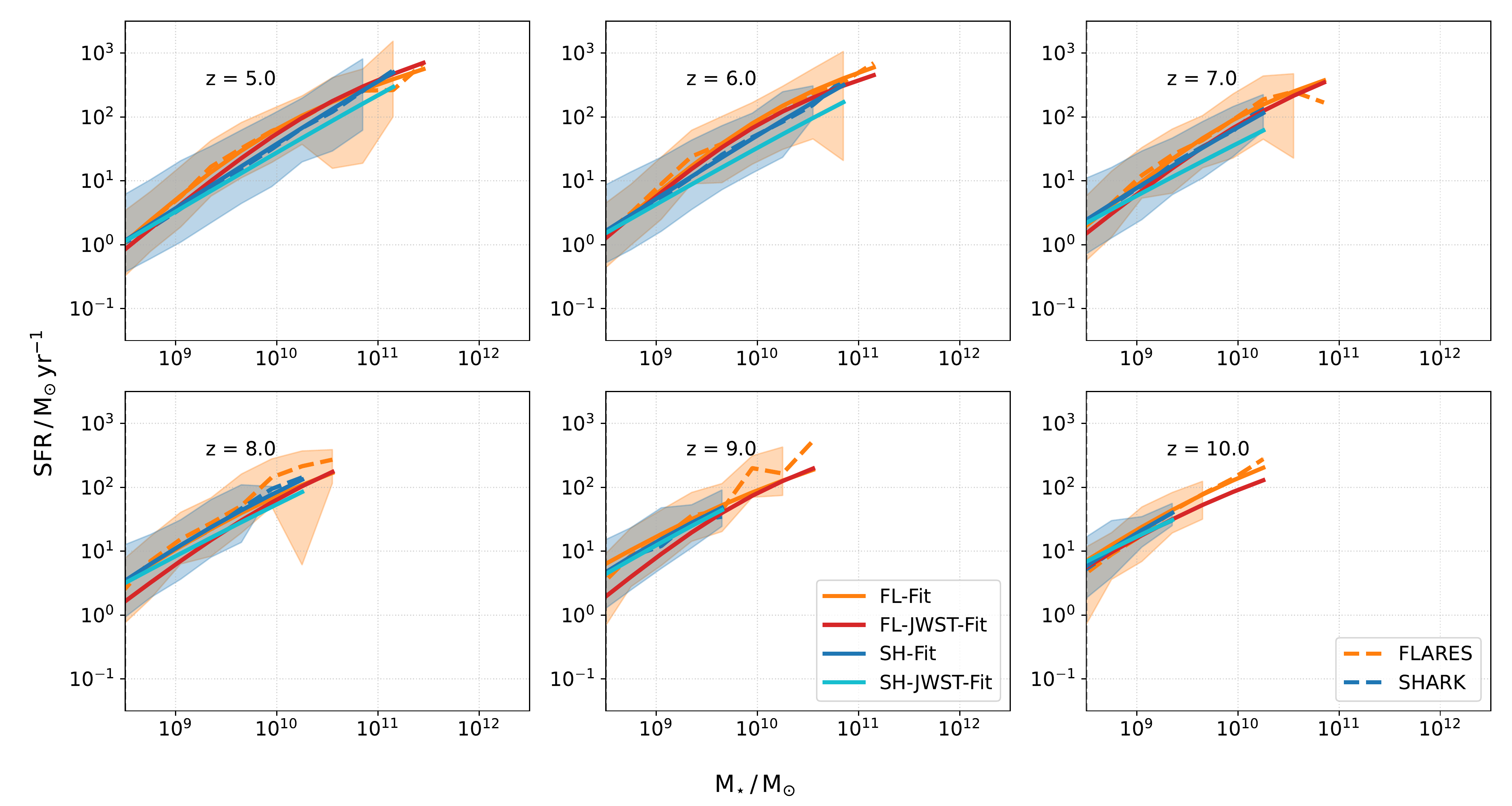}
     \caption{$\mathrm{SFR-M_{\star}}$ between redshift $z=5\to10$ for entire population in \textsc{Flares} (orange) and \textsc{Shark} (blue). The dashed lines with markers show the median trend and the coloured shaded regions show the $5^{\rm th}-95^{\rm th}$ percentiles. The solid orange and blue lines show the fit to the SFS as per \Cref{eq:SFS-functional-form} for the population with masses above $\mathrm{10^{8.5}M_{\odot}}$ in \textsc{Flares} and \textsc{Shark} respectively. The red and light blue solid lines show the fit for the JWST detected population in \textsc{Flares} and \textsc{Shark} respectively. The grey shaded regions highlights the stellar mass resolution limit of $\mathrm{10^{8.5}M_{\odot}}$.} 
    \label{fig:sfs_quantiles}
\end{figure*}

\Cref{fig:dark_frac_mass_sfr} shows the fraction of galaxies brighter than 29 magnitudes above a certain mass or SFR. We calculate this fraction for galaxies observed with the 2$\mu$m F200W filter of NIRCam on the JWST and the 0.75$\mu$m F775W filter of the Advanced Camera for Surveys on the HST. Our choice of HST filter for comparison is motivated by its use as the detection band of HST deep fields \citep[e.g.,][]{beckwithHubbleUltraDeep2006a}. 

We see that the fraction of JWST detected galaxies is greater than HST detected galaxies for almost all stellar mass and SFR limits and at all redshifts, showing that the rest-wavelengths traced by JWST are much less affected by dust. At $z=10$ there are virtually zero HST-detected galaxies predicted by \textsc{Flares} at all stellar masses and SFRs, while only $\sim 10\%$ of star-forming galaxies (SFR $\mathrm{\gtrsim 10 M_{\odot}yr^{-1}}$) are detected in \textsc{Shark}. 

\textsc{Flares} predicts that JWST can detect $\gtrsim 90\%$ of galaxies with $\mathrm{M_{\star}\gtrsim10^{8.3}M_{\odot}}$ and $\mathrm{SFR \gtrsim 10^{0.5}M_{\odot}yr^{-1}}$ at $z=10$, whereas \textsc{Shark} predicts that JWST can detect $\sim 70\%$ above those limits. So, in both simulations a significant population of undetected JWST galaxies do indeed exist. Though the fraction of galaxies missed by JWST is not significant enough to hinder a thorough exploration of the $\mathrm{SFR-M_{\star}}$ as is the case with the HST. The fact that these fractions are varied between the simulations boils down to different prescriptions of stellar mass assembly and the generation of synthetic photometry. The JWST should be able to detect sufficiently many galaxies to distinguish between these model dependent prescriptions.

%% %May be better for discussion %%%

% The key results of this plot is that, throughout redshifts $\mathrm z=6 \to 10$, the JWST will be able to detect all intrinsically bright galaxies up to redshift $\mathrm z=10$, and thus will be able to parametrise the distributions of stellar mass and star formation rates of the first galaxies in the Universe. This is especially useful as JWST observations should be able to explain the slight differences between the two simulations, and so further disentangle  

%%%

%% file: Sections/SFS.tex
\section{Star forming main sequence}
\label{sect:SFS}
In this section, we present our predictions on the shape and stellar mass-dependent scatter of the SFS at $z\geq 5$. 
\begin{figure*}
    \includegraphics[width=\textwidth]{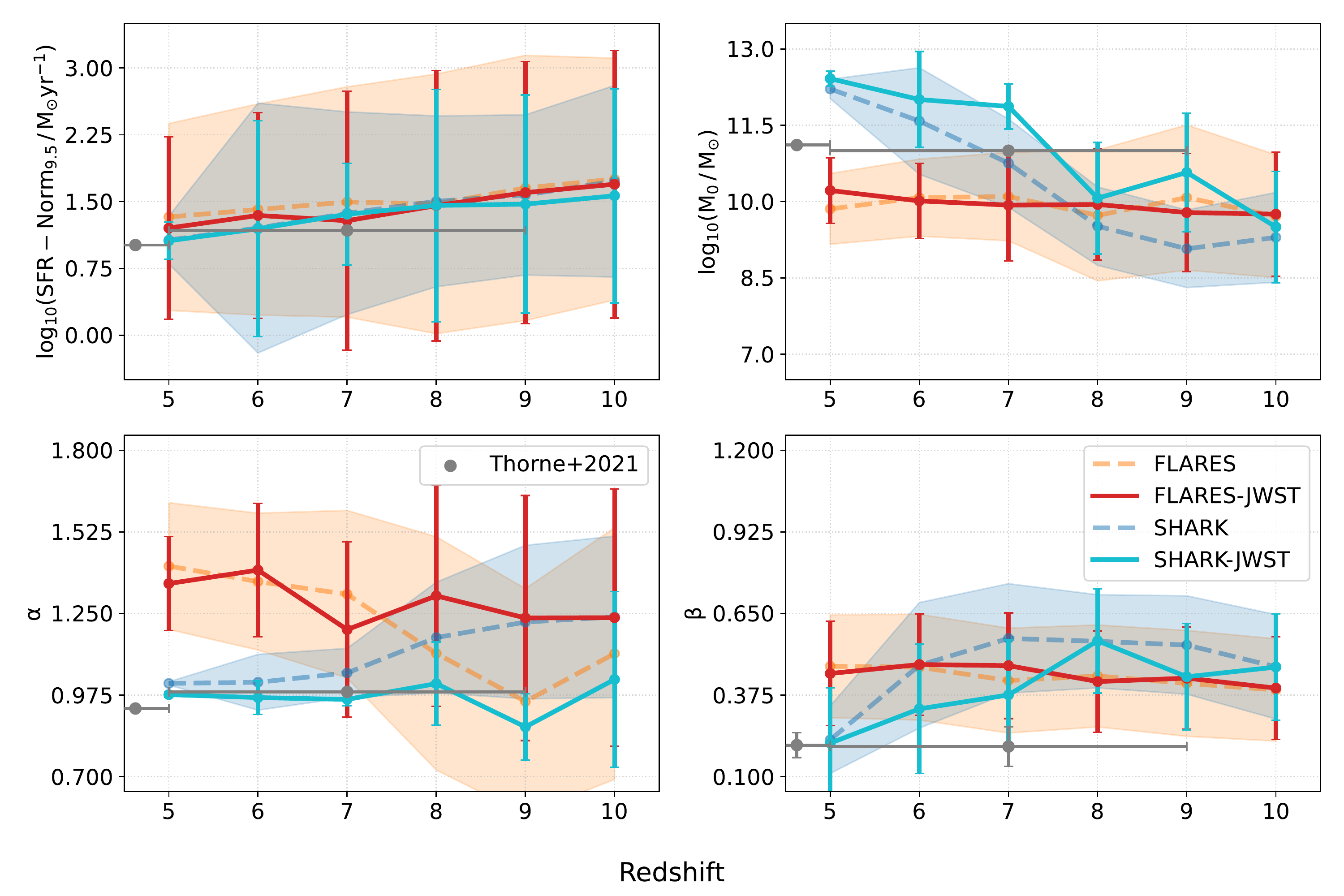}
     \caption{Redshift evolution of the parameters of the SFS as per \Cref{eq:SFS-functional-form}; the normalisation, $\mathrm{S_{0}}$, is calculated at $\mathrm{10^{9.5}M_{\odot}}$ instead of the maximum value to which \Cref{eq:SFS-functional-form} approaches at high stellar masses. The dashed lines with shaded regions show the results and $\mathrm{1 \sigma}$ uncertainties for the total population of galaxies in \textsc{Flares} (orange) and \textsc{Shark} (blue). The solid lines with error bars show the results and $\mathrm{1 \sigma}$ uncertainties for the JWST detected galaxies in \textsc{Flares} (red) and \textsc{Shark} (cyan). The grey points with error bars show the observational results and uncertainties obtained from \citet{Thorne2021}.} 
    \label{fig:sfs_jwst_fit}
\end{figure*}
% \subsection{Fitting the SFS}
% \label{sect:fittosfs}

% Whether the SFS is best described as a single component power law model \citep{Wuyts_2011,Whitaker2012,Speagle2014,Pearson2018} or a two component power law model \citep{Lee2015} is unclear. Two component models model the low stellar mass and high stellar mass shape of the SFS separately to account for distinct slopes in these mass regimes. Generally, the SFS is described as having a slope close to unity (citations) in the $\log_{10}(SFR)-\log_10(M_{\star})$ plane; accordingly, this is the value often quoted in single component power laws or the low stellar mass value of the slope in two component models. The high stellar mass slopes are frequently measured to be far shallower than the lower stellar mass slopes (citations). For this reason, the SFS is thought to bend downward beyond a certain turn-over mass; though the physical reasons responsible for this turn over are still as of yet not fully understood \citep{Abramson2014,Cook2020}

\subsection{Fitting the SFS}
To fit our simulation's SFS, we follow the same procedures outlined in \citet{Thorne2021} who use a two-component power law.
The latter allows to fit a possible turn over in the SFS, which, when present, appears in massive galaxies. \citet{Lovell2021} presented evidence of a turn over in the SFS in \textsc{Flares}. The functional form of the SFS is thus 

\begin{equation}
    \mathrm{\mathnormal{f}(\mathcal{M}) = S_{0} - \log_{10} \left[ \left(\frac{10^{\mathcal{M}}}{10^{M_{0}}}\right)^{-\alpha} + \left(\frac{10^{\mathcal{M}}}{10^{M_{0}}}\right)^{-\beta} \right ]},
    \label{eq:SFS-functional-form}
\end{equation}
where $\mathrm{\mathcal{M} = \log_{10}(M_{\star}/M_{\odot})}$, $\mathrm{S_{0}}$ is the limiting SFR of the function at high stellar mass in units of $\mathrm{\log_{10}(M_{\odot}yr^{-1})}$, $\mathrm{M_{0}}$ is the stellar mass in units of $\mathrm{\log_{10}(M_{\odot})}$ at which the SFS turns over and $\mathrm{\alpha}$ and $\mathrm{\beta}$ are the respective low and high stellar mass slopes. \Cref{eq:SFS-functional-form} from \citet{Thorne2021} is itself an adaptation of the two component model used in \citet{Lee2015} that, instead of assuming that the SFS flattens to a slope of zero at high stellar masses as is the case in \citet{Lee2015}, allows for an additional degree of freedom, $\beta$, to describe the high stellar mass slope. We use \Cref{eq:SFS-functional-form} as the foundation of a Bayesian fitting routine. The likelihood function that we maximise is a student-t distribution with a fixed scale of $0.3$. We use broad uniform priors on the parameters of the SFS: $\mathrm{S_{0} \sim U(0.01, 4.01)}$; $\mathrm{M_{0} \sim U  (8.0, 19.0)}$; $\mathrm{\alpha \sim U(0.5, 2.0) }$; $\mathrm{\beta \sim U(0.1, 0.61) }$. We are thus hypothesising that the SFS does in fact exist out to $z=10$ and that its shape is consistent with our function. We estimate the parameters and $\mathrm{1 \sigma}$ uncertainties by performing a Markov-Chain-Monte-Carlo (MCMC) analysis using the Python package EMCEE \citep{emcee}. We have elected to not perform any kind of SFR or specific SFR ($\mathrm{\log_{10}\left[ \frac{SFR}{M_{\star}} yr^{-1} \right]}$) selection in defining the SFS here, as is the case in many works that focus on the SFS at lower redshifts in the literature \citep{Davies2016, Davies2019, Davies2021, Katsianis2019}. Our choice is motivated by the fact that we do not expect a significantly quenched and off-SFS galaxy population contained in our high redshift sample. We note, however, that there are passive galaxies in \textsc{Flares} up to $z\sim8$ though they constitute $\mathrm{<3\%}$ of the total population (Lovell et al. in prep).

\Cref{fig:sfs_quantiles} shows the $5^{\rm th}-95^{\rm th}$ percentiles and medians between redshifts $z=5 \to 10$ for the total galaxy populations in both simulations. It is clear that both simulations predict the existence of a tight SFS up to $z=10$, indicating that the process of self-regulation in galaxy growth is present and efficient in the very early universe. 
%The agreement between the two simulations is evidence that both \textsc{Flares} and \textsc{Shark} are assembling stellar mass similarly.  
Despite differences in the stellar mass assembly, the two simulations predict similar SFSs.
\textsc{Flares} exhibits a slightly clearer high stellar mass turn over than \textsc{Shark}. \textsc{Shark} exhibits approximately uniform variance in the SFS with stellar mass, whereas \textsc{Flares} shows more puffed up variance at low and high stellar masses and a minimum in the variance around $\mathrm{\sim10^{9}M_{\odot}}$ particularly at $z \lesssim 7$. Note, however, that the increase in the variance at low stellar masses in {\sc Flares} happens mostly at the regime where resolution is expected to significantly affect our results. Therefore, the increased scatter is likely dominated by star formation stochasticity and requires higher resolution simulations to probe robustly. We return to the subject of variance about the SFS in \Cref{sect:sfssigma}. The skew in the SFS at high stellar masses seen in the $5^{\rm th}-95^{\rm th}$ percentiles is due to the population of massive galaxies in {\sc Flares} undergoing quenching, the cause of which is likely energy injection from their AGN driving a suppression of star formation in massive galaxies.

We also show the fitted relations as solid lines, and it can be seen that the fits agree well with the predicted relations. We only fit galaxies with $\mathrm{M_{\star} \geq 10^{8.5}M_{\odot}}$. We do see that the lower stellar mass slope, $\alpha$, is steeper in \textsc{Flares} than it is in \textsc{Shark} at $z \lesssim 7$, which may originate from the highly overdense, starburst galaxy populations that \textsc{Flares} samples but \textsc{Shark} does not; this is likely to be strongly related to the enhancement in stellar mass and SFR seen in \Cref{fig:smf_sfrf_z6-10}. 

Most notably at $z \lesssim 7$ the \textsc{Flares} fits are able to better capture the full extent of the turn over in the SFS observed in the median trend, with the median showing a slightly higher normalisation around $\mathrm{\sim 10^{9.5} M_{\odot}}$ than the fit. This is likely caused by quenching that is driving a cessation of star formation in massive galaxies. The fact that this turn over is not seen in \textsc{Shark} galaxies below $\mathrm{M_{\star}\sim 10^{10}M_{\odot}}$ indicates that massive galaxy quenching occurs earlier in \textsc{Flares} (Lovell et al. in prep). Above $z \sim 7$ the SFSs of both simulations are remarkably similar, indicating that self-regulation is occurring in a similar fashion between these two simulations.
%astrophysical processes, as they are modelled in each simulation, become more significant at driving the differences between the simulations as their synthetic galaxies continue to evolve. 

We have calculated the same parameters of the SFS for the population that will be detected by the JWST in both simulations. We have included Gaussian uncertainties of $0.2$~dex and $0.4$~dex on the stellar masses and SFRs of the JWST populations respectively to account for potential uncertainties in deriving those quantities from actual observations. We investigate the effect of uncertainties more closely in \Cref{apdx:uncertainties}. Remarkably, the JWST population is very closely congruent with the total population in the simulations over the entire $\mathrm{SFR-M_{\star}}$ plane at all redshifts shown in \Cref{fig:sfs_quantiles}. This indicates that planned JWST observations are deep enough to recover the SFS.

% A possible reason for this is the fact that we have assumed that the likelihood function is symmetric about the functional form of the SFS, i.e. along all stellar masses. Mathematically, this is 

% \begin{equation}
%      \mathrm{\log_{10}(SFR) \sim t(\mu = \mathnormal{f}(\mathcal{M}), \sigma = 0.3, \theta)},
%     \label{eq:sfrs_prob}
% \end{equation}

% where $\mathrm{t}$ is the student-t distribution, $\mathrm{\mu = \mathnormal{f}(\mathcal{M})}$ is from \Cref{eq:SFS-functional-form} and is the mean of the distribution, $\mathrm{\sigma = 0.3}$ is the fixed scale of the PDF and $\mathrm{\theta}$ are additional parameters. We see that at each stellar mass the variance is not symmetric about the median line meaning that per stellar mass, at least in \textsc{Flares}, the 3D number distribution along the SFS may be skewed at different stellar masses, and most certainly not always a student-t distribution. 

%As this slight discrepancy between the fit and the median is only seen in \textsc{Flares} and not \textsc{Shark}, we can conclude that this reduces to the slight different, systematic methods of star formation and stellar mass assembly in the simulations, which in consequence produce systematically distinct 3D number distributions along the SFS. In the case of \textsc{Flares} the skew in the 3D number distribution around $\mathrm{\sim 10^{9.5} M_{\odot}}$ at $\mathrm{z \lesssim 7}$ favours more star forming systems compared to \textsc{Shark}.  

\begin{table}
    \centering
    \resizebox{\columnwidth}{!}{%
    \begin{tabular}{|c | c | c | c | c | c|}
    \hline \hline
        Redshift & $\mathrm{S_0}$ & $\mathrm{M_{0}}$ & $\alpha$ & $\beta$\\
        \hline
        \textsc{Flares} &&&&\\
         5.0  &$2.40\pm0.77$&$10.38\pm0.75$&$1.31\pm0.16$&$0.46\pm0.17$ \\
         6.0  &$2.18\pm0.77$&$10.01\pm0.77$&$1.36\pm0.23$&$0.44\pm0.17$ \\
         7.0  &$2.32\pm0.92$&$10.17\pm0.97$&$1.28\pm0.29$&$0.47\pm0.17$ \\
         8.0  &$2.18\pm1.00$&$10.09\pm1.14$&$1.22\pm0.36$&$0.47\pm0.17$ \\
         9.0  &$2.11\pm1.04$&$9.88\pm1.23$&$1.30\pm0.42$&$0.42\pm0.17$ \\
         10.0 &$1.98\pm0.98$&$9.58\pm1.33$&$1.06\pm0.42$&$0.40\pm0.17$ \\
         \hline
         \textsc{Flares}-JWST &&&&\\
         5.0  & $2.02 \pm 0.71$ & $ 9.90 \pm 0.72$ & $ 1.40 \pm 0.21 $ & $ 0.48 \pm 0.18$\\
         6.0  & $2.35 \pm 0.79$ & $10.11\pm 0.76$ & $1.37 \pm 0.23$ & $0.45 \pm 0.17 $ \\
         7.0  & $2.28 \pm 0.85$ & $9.98 \pm 0.89$ & $1.33 \pm 0.29$ & $0.41 \pm 0.17 $ \\
         8.0  & $1.93 \pm 0.96$ & $9.69 \pm 1.27$ & $1.12 \pm 0.40$ & $0.46 \pm 0.17$ & \\
         9.0  & $2.07 \pm 0.98$ & $9.73 \pm 1.49$ & $0.97 \pm 0.41$ & $0.39 \pm 0.17$ & \\
         10.0 & $2.22 \pm 0.95$ & $9.69 \pm 1.27$ & $1.09 \pm 0.43$ & $0.43 \pm 0.17$ & \\
         \hline
         \textsc{Shark}-JWST &&&&\\
         5.0   &$3.90\pm0.32$&$12.62\pm0.40$&$0.93\pm0.02$&$0.37\pm0.24$\\
         6.0  &$3.36\pm0.88$&$11.93\pm1.11$&$0.92\pm0.05$&$0.50\pm0.21$& \\
         7.0  &$2.35\pm0.94$&$10.58\pm1.20$&$0.92\pm0.07$&$0.62\pm0.17$& \\
         8.0  &$2.67\pm1.03$&$10.83\pm1.34$&$0.90\pm0.10$&$0.59\pm0.17$& \\
         9.0  &$2.42\pm0.91$&$10.25\pm1.12$&$0.98\pm0.17$&$0.50\pm0.18$& \\
         10.0 &$2.05\pm0.97$&$9.72\pm1.38$&$0.89\pm0.27$&$0.51\pm0.17$& \\
         \hline
         \textsc{Shark} &&&&\\
         5.0  &$3.84\pm0.19$&$12.22\pm0.19$&$1.01\pm0.01$&$0.23\pm0.14$\\
         6.0  &$3.51\pm0.94$&$11.75\pm1.05$&$1.01\pm0.09$&$0.37\pm0.21$ \\
         7.0  &$2.77\pm0.85$&$10.73\pm0.95$&$1.05\pm0.11$&$0.56\pm0.19$ \\
         8.0  &$1.76\pm0.73$&$9.44\pm0.80$&$1.18\pm0.17$&$0.61\pm0.16$ \\
         9.0  &$1.70\pm0.84$&$9.30\pm0.98$&$1.14\pm0.24$&$0.53\pm0.17$ \\
         10.0 &$1.86\pm0.82$&$9.30\pm0.91$&$1.24\pm0.29$&$0.49\pm0.17$ \\
         \hline \hline
    \end{tabular}%
    }
    \caption{Redshift evolution of the SFS fit parameters and associated $1-\sigma$ uncertainties as described in \Cref{sect:SFS} and shown in \Cref{fig:sfs_jwst_fit}.}
    \label{tab:sfs_parms}
\end{table}

\Cref{fig:sfs_jwst_fit} shows the redshift evolution of the parameters of the SFS for \textsc{Shark} and \textsc{Flares}, for both the total and JWST galaxy detected populations. Tabular data of these fits is recorded in \Cref{tab:sfs_parms}. The shaded regions and error bars show $\mathrm{1\sigma}$ uncertainties on the parameter estimates from the MCMC sampling for the total and JWST detected populations respectively. It is interesting that above $z \sim 7$ the simulations predict very similar SFS considering the different modelling processes, and that only $z=0$ observations were used for the tuning of free parameters. The normalisation evolves similarly in the simulations for all considered redshifts. Below $z \sim 7$, the fitting to the SFS in \textsc{Shark} indicates a turn over mass that is an order of magnitude higher mass than that in \textsc{Flares} and a shallower low stellar mass slope. The fitted turn over mass in \textsc{Shark} is actually higher than the most massive galaxies that the simulations predict, indicating that there is no turn over in the SFS in {\sc Shark} at $z\gtrsim 5$. This, in combination with the low stellar mass slope tending to unity, suggests that the SFS in \textsc{Shark} is better described by a single power law at $z \gtrsim 5$. The main difference between the SFS in the simulations is thus the turn over as a result of quenching, and the redshift at which that quenching becomes prevalent. We elaborate on this in \Cref{sect:discussion}. 

Shown as well are observationally derived estimates on SFS parameters from \citet{Thorne2021} who fit the SFS using the same two-component power law in 20 redshift bins with an equal width of $\mathrm{\sim 0.75 Gyrs}$. SFRs and stellar masses are derived by SED fitting galaxies between $\mathrm{\lambda \sim 0.154 \to 504 \mu m}$ in the D10-COSMOS field of DEVILS \citep{Davies2018,Davies2021b}. These observations have a coarse temporal resolution at high redshift (with the highest redshift bin being $z=5 \to 9$) and a high stellar mass cut of $\mathrm{\gtrsim 10^{10.4}M_{\odot}}$ for $z > 5$ meaning the range of stellar masses probed is small. Despite these caveats, the simulations agree somewhat with the observations. The SFS normalisation is consistent with the simulations at redshift $z\lesssim 9$. The turn over mass is more consistent with \textsc{Shark} than \text{Flares}, with \textsc{Flares} being lower, however, the turnover mass in \textsc{Shark} happens above the range of simulated stellar masses and hence is not well constrained. This suggests that the observations favour a single power law SFS at $z\gtrsim 5$, which was a similar conclusion reached in \citet{Thorne2021} who found negligible turn overs beyond $z\sim2$. To assert that the observations favour one model or the other is disingenuous, however, because the lack of observations here hinder a rigorous exploration of this parameter space. Furthermore, mass completeness limit used in \citet{Thorne2021} is $\mathrm{\gtrsim 10^{10.4}M_{\odot}}$ for $z > 5$ meaning the range of stellar masses probed is small.

We thus predict that the JWST will be able to sufficiently detect all intrinsically bright galaxies and offer for the first time reliable fits to the SFS for the first galaxies to have formed in the Universe. It is encouraging then that we predict that the JWST will be able to recover the SFS for redshifts up to $z=10$ as these observations will be necessary to decouple slight differences in the simulations. This will elucidate the origins of these kinds of fine distinctions between the two simulations (and potentially other simulations), which are embodied in galaxy formation theory.

\subsection{Stellar mass dependence of the SFS scatter}
\label{sect:sfssigma}
The usefulness of the SFS is that it encodes information about astrophysical process that may be driving the shape and normalisation. In particular, the stellar mass dependent scatter is an indicator of how effective different feedback mechanisms are in regulating star formation in galaxies. At lower redshifts, $z \lesssim 0.7$, the $\mathrm{\sigma_{SFS}-M_{\star}}$ has been observed to exhibit a minimum vertex parabolic shape, with its turning point in the neighbourhood of $\mathrm{\sim 10^{9} M_{\odot}}$ \citep{Davies2019,Davies2021}. The minimum point of the stellar mass dependent scatter is interpreted as the galaxy phase at which gas inflow is balanced by both star formation and feedback. Galaxies in this phase experience little variation about the SFS, and so minimal scatter, as gas compaction and subsequent depletion events are highly self regulated for these galaxies \citep{Tacchella2016}.  
%{\bf ADD HERE SENTENCE OF HOW A MINIMUM POINT IS INTERPRETED.}, added 15/04/22

\begin{figure*}
    \includegraphics[width=\textwidth]{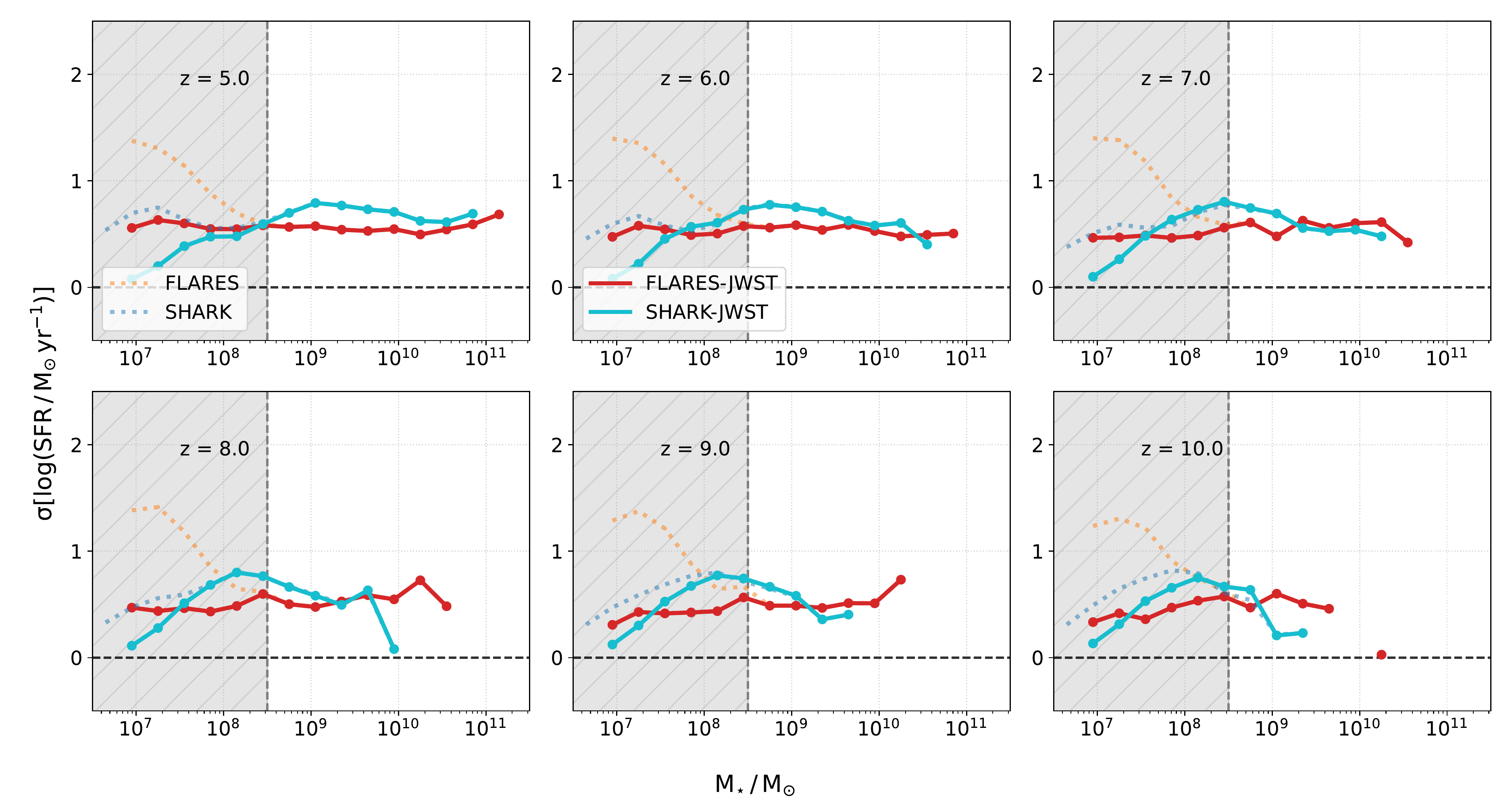}
     \caption{$\mathrm{\sigma_{SFS}-M_{\star}}$ relation for the total galaxy populations in \textsc{Flares} (orange) and \textsc{Shark} (blue) at redshifts $z=5\to10$. Also shown is the same relation but for the JWST detected galaxies in \textsc{Flares} (red) and \textsc{Shark} (cyan). The grey shaded regions shows the stellar mass resolution limit, $\mathrm{M_{\star} =  10^{8.5} M_{\odot}}$.}
    \label{fig:sigma_fl_sh}
\end{figure*}

\Cref{fig:sigma_fl_sh} shows the $\mathrm{\sigma_{SFS}-M_{\star}}$ for \textsc{Shark} and \textsc{Flares}, where $\mathrm{\sigma_{SFS}}$ is the $16^{\rm th}-84^{\rm th}$ percentile range. It can be seen that both simulations are fairly consistent about the shape of the relation with stellar mass over most stellar masses and redshifts probed. Below $\mathrm{M_{\star} = 10^{8.5}M_{\odot}}$ the simulations disagree, with \textsc{Shark} predicting a decrease in the scatter at $z\gtrsim9$ and a uniform scatter at $z\lesssim8$, while \textsc{Flares} predicts a significant uptick of the scatter at all redshifts. The decreased scatter in low stellar mass \textsc{Shark} galaxies at $z\gtrsim9$ may suggest that they are better self regulated as most are not far off from the SFS. Stochastic star formation then becomes more important at $z\lesssim8$ to drive up the scatter. Conversely the uptick seen in \textsc{Flares} throughout all redshifts is likely an effect of stochastic star formation due to the poor resolution at these masses. For both simulations, these effects occur below the stellar mass resolution limit of the simulations however, and so these results cannot be considered robust.

A striking feature of the $\mathrm{\sigma_{SFS}-M_{\star}}$ is the lack of significant scatter at the massive end, $\mathrm{M_{\star} \gtrsim 10^{10}M_{\odot}}$, that both simulations predict at redshifts $z=5\to10$. The likely reason for this is that the massive galaxies at these redshifts in the simulations have not yet had enough time to sufficiently grow their supermassive black holes. AGN feedback is thought to increase the asymmetric variance from the SFS as feedback continuously inhibits star formation and drives massive galaxies below the SFS \citep{Davies2021}. This is consistent with other studies that investigate this effect in simulations \citep{Katsianis2019}. Although there is a general lack of significant scatter at these large stellar masses, with decreasing redshift, {\sc Flares} starts to show an increase in $\mathrm{\sigma_{SFS}-M_{\star}}$ at stellar masses $\gtrsim 10^{10.3}\rm M_{\odot}$ at $z \sim 5$, which shows the initiation of quenching in these galaxies. This is consistent with lower redshift results from simulations that show that the scatter in the SFS for galaxies $\mathrm{M_{\star} \gtrsim 10^{10}M_{\odot}}$ increases by 0.05 dex from z=5 to z=0 \citep{Matthee2019}. This is not seen in {\sc Shark}. This difference between the simulations together with the JWST-like samples following the same SFS as all the simulated galaxies, indicates that the JWST will be able to place strong constraints on the onset of quenching in massive galaxies from the characterisation of the SFS at high redshift.
%\textsc{Flares} exhibits an increased asymmetric scatter at the high-mass end at $\mathrm{z \lesssim 7}$ than \textsc{Shark} that, in combination with the prevalence of bending in the SFS for \textsc{Flares} occurring before \textsc{Shark} as was seen in \Cref{fig:sfs_quantiles}, hints at the earlier systematic quenching of massive \textsc{Flares} galaxies compared to \textsc{Shark}. 

While there is the issue of stellar mass resolution we can conclude that faint dwarf galaxies below the detection sensitivity of the JWST are likely, if not completely, responsible for the disparity between the total and JWST detected populations observed for stellar masses $\mathrm{M_{\star} \lesssim 10^{8.5}M_{\odot}}$. JWST will not detect many galaxies with low star formation rates, as was confirmed in \Cref{fig:smf_sfrf_z6-10}. This means that the contribution to the scatter about the SFS for low stellar mass objects is biased to those on or above the SFS, which would lead to an underestimation of the scatter. This trend is seen in \Cref{fig:sigma_fl_sh} for both simulations, although again this is the mass range that falls below the resolution of the simulations.

Most importantly, above stellar masses of $\mathrm{10^{8.5} M_{\odot}}$ the JWST detected population in each simulation is in agreement with the total population up to the highest stellar masses probed and for all redshifts shown up to $z=10$. We thus predict that the JWST should be able to sufficiently detect all intrinsically bright galaxies to understand the $\mathrm{\sigma_{SFS}-M_{\star}}$ of the Universe up to redshift $z = 10$.

%% file: Sections/CSFH-CSMH.tex
\section{Cosmic stellar mass and star formation rate history}
\label{sect:cosmic}

\begin{figure*}
    \includegraphics[width=\textwidth]{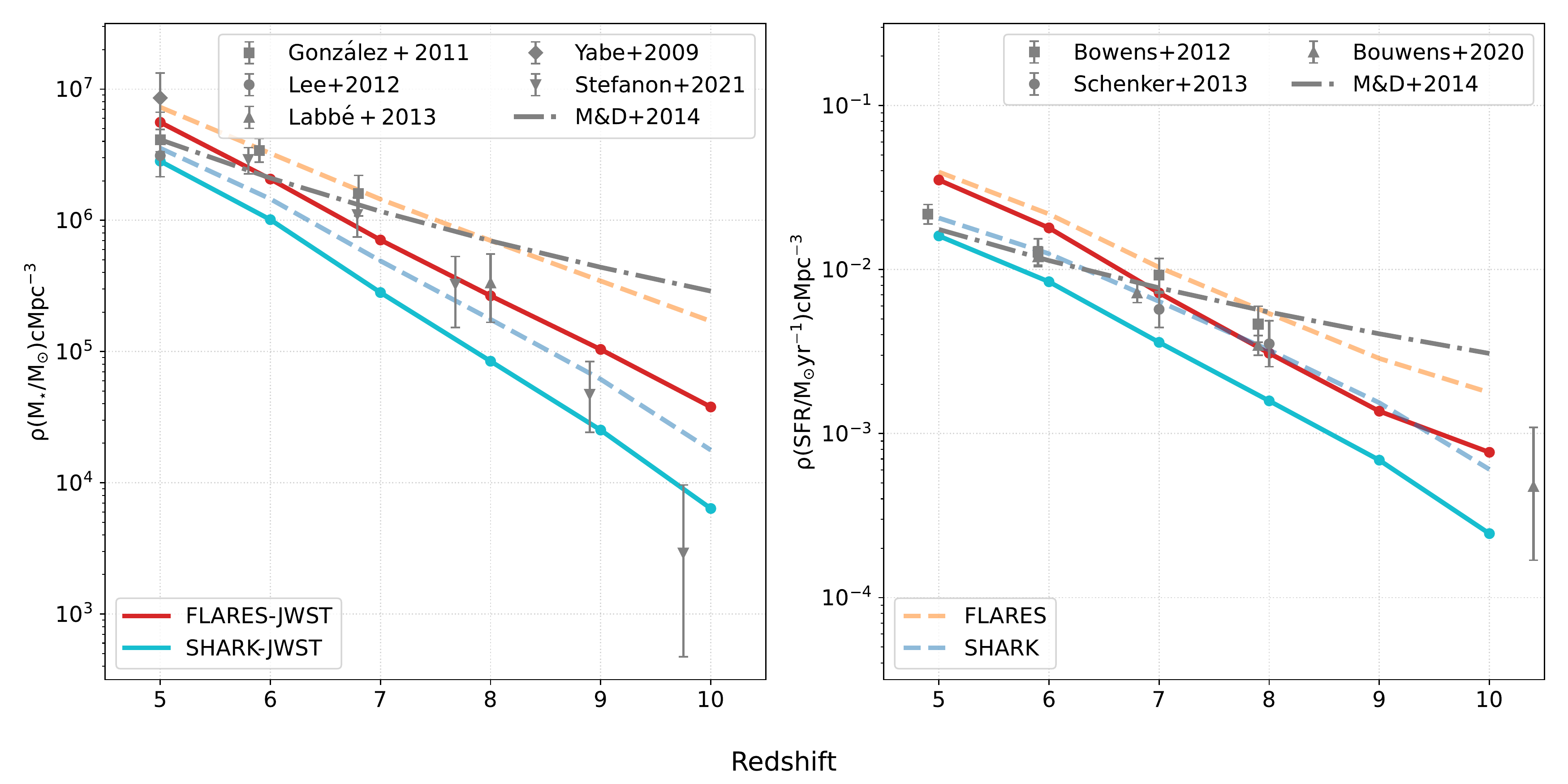}
    \caption{The CSMH (left) and CSFRH (right) at redshifts $z = 5 \to 10$. The light coloured, dashed lines show these quantities for the entire population of galaxies in \textsc{Flares} (orange) and \textsc{Shark} (blue). The solid lines show the results for the population that will be detected by the JWST in \textsc{Flares} (red) and \textsc{Shark} (cyan). The dot dashed grey lines show the fits to the CSFRD and CSMD from \citet{Madau&Dickinson2014}. Data points with error bars in the left panel show observational results of the CSMH \citep{Yabe2009,Gonzalez2011,Lee2012,Labbe2013,Stefanon2021}. In the right panel observational estimates on the CSFRH \citep{Bouwens2012, Schenker2013, bouwensALMASpectroscopicSurvey2020} are shown as data points with error bars.}
    \label{fig:cosmic_densities_sh_fl}
\end{figure*}

\Cref{fig:cosmic_densities_sh_fl} shows the cosmic stellar mass and cosmic star formation rate densities (CSMD/CSFRD) as a function of redshift, or cosmic stellar mass/cosmic star formation rate history (CSMH/CSFRH). The dashed lines show the results for all galaxies in the simulations. Both simulations predict consistent shapes of both the CSMH and CSFRH over all redshifts shown.
%, suggesting that the distributions of stellar mass and SFR are reasonably similar between them. 
\textsc{Flares}, however, exhibits a noticeably higher normalisation of both the CSMH and CSFRH compared to \textsc{Shark}. \textsc{Flares} is $\mathrm{\sim 4.4}$ ($\mathrm{\sim 2.0}$) times higher in CSMD (CSFRD) than \textsc{SHARK} averaged over redshift, but up to $\mathrm{\sim 9.6}$ ($\mathrm{\sim 2.9}$) times larger at $z=10$. The difference in CSMD becomes systematically smaller with decreasing redshift, with \textsc{Flares} being $\mathrm{\sim 2}$ times larger than \textsc{Shark} by $z=5$. The difference in CSFRD is lowest at $z=7$ where \textsc{Flares} is $\mathrm{\sim 1.6}$ times larger than \textsc{Shark} but becomes $\mathrm{\sim 1.9}$ times larger by $z=5$. 
%{\bf ADD NUMBERS. HOW MUCH HIGHER IS THE NORMALISATION IN FLARES VS SHARK}. added 15/04
This is potentially a result of the stellar mass and SFR enhancement captured by \textsc{Flares} as a result of the contribution from overdense environments sampled in \textsc{Flares} that are not present in \textsc{Shark}. It can also be seen that the difference in the normalisation between the two simulations is larger in the CSMH than in the CSFRH, which appears to be the result of the different chemical enrichment models used in the simulations. \textsc{Shark} instantaneously recycles and enriches the interstellar medium with metals following star formation \citep{Lagos2019}, while there is a delay between star formation and enrichment in \textsc{Flares} to account for the evolution of type Ia and II supernovae and AGB stars \citep{Schaye2015,Crain_2015}. We also inspected the gas phase metallicities in the simulations and found that \textsc{Shark} predicted higher metallicities per unit stellar mass and SFR at $z=5$ and $z=10$ than \textsc{Flares}. This can explain the larger offset in the CSMH, as more mass is diverted from stars toward metal pollution of the ISM in \textsc{Shark} than \textsc{Flares}, particularly at earlier times. The fact that the differences in the CSMH become smaller at later times is the result of the instantaneous recycling approximation becoming a better approximation as the dynamical time of halos becomes longer.

To ensure that resolution effects are not biasing these results, we also examined the CSFRH and CSMH only for galaxies that satisfy $\mathrm{10^{8.5}M_{\odot} < M_{\star} \leq M^{9.0}M_{\odot}}$. The reason for this particular selection is to be above the resolution limit but below the stellar mass range where \textsc{Flares} predicts a much greater number density of stellar mass and SFR compared to \textsc{Shark}, which would bias our results. We found that the same trends as the total population persist in this case except at $z=10$ where the difference in normalisation between the CSMD and CSFRD are similar. As such, while chemical enrichment is an important factor in influencing the normalisation of the CSMH and CSFRH, we cannot rule out different star formation histories also somewhat driving this difference.

We have also included some observational estimates on both the CSMH and CSFRH. The observations on the right panel of \Cref{fig:cosmic_densities_sh_fl} show the CSFRH derived by \citet{Madau&Dickinson2014} who curated UV LFs of \citet{Bouwens2012} and \citet{Schenker2013}, and then converted them into a  CSFRH. The observational data points shown in the left panel are UV-derived estimates of the CSMD \citep{Yabe2009, Gonzalez2011, Lee2012, Labbe2013, Stefanon2021}. Where possible, we have standardised the cosmologies and IMFs used for these observational results to be most comparable with our simulations. The observations for both the CSMH and CSFRH are consistent with the shapes of the predicted relations from both simulations up to $z\sim10$ and $z\sim8$ for the CSMH and CSFRH respectively. The \textsc{Shark} predictions of the CSMH and CSFRH are in better agreement with the observations, with \textsc{Flares} generally predicting a higher normalisation than the observations in both the CSMH and CSFRH over the relevant redshift range. We must again stress, however, that all of these observational estimates are entirely from UV sources that have been corrected for dust attenuation using the infrared-excess technique, and are thus subject to the uncertainties of the method \citep[e.g.,][]{shivaeiMOSDEFSurveyVariation2020b}. We showed in \Cref{fig:dark_frac_mass_sfr} that HST-dark systems are significant in number in the simulations up to $z\sim10$, implying that dust must be an important factor when determining SFR from rest-frame UV observations.

The solid lines show the cosmic densities for the population that will be detectable with the NIRCam F200W filter of the JWST. We only show the result with a photometric cut, truncating at $\mathrm{29 \, ABmag}$, and not a cut in stellar mass. Noticeably, both simulations predict that the JWST detected populations do not account for the entire distribution of stellar mass and SFR in the universe over all these redshifts. Perhaps unsurprisingly, this indicates that there are faint galaxies that fall below the detection sensitivity of the JWST and are not accounted for in the calculation of the CSMH and CSFRH. Fortunately, this is well understood as the Malmquist bias for which there are numerous correction methods \citep{Weigel2016}. As only the normalisation, and not the shape, of the CSMH and CSFRH for the JWST detected population is different from that of the total population it shows that the JWST should only be sensitivity limited, and not miss a significant population of heavily obscured systems compared to other shorter wavelength instruments such as the HST.

%be able to detect \textit{every} intrinsically bright galaxy out to redshift $z = 10$, just not \textit{all} galaxies in the universe.  

%The fit is 
%\begin{equation}
%    \mathrm{\psi(z)/M_{\odot}yr^{-1}Mpc^{-3} = 0.015 \frac{(1+z)^{2.7}}{1 + [(1+z)/2.9]^{5/6}}}.
%    \label{eq:md2014csfrd}
%\end{equation}
%

%\begin{equation}
%    \mathrm{\rho(z)/M_{\odot}Mpc^{-3} = (1-R)\int^{\infty}_{z} dz' \frac{\psi(z')}{H(z') (1+z')}}
%    \label{eq:md2014csmd}
%\end{equation}

%where $\mathrm{R=0.41}$ is the return fraction of mass into the ISM with each episode of star formation for a Chabrier initial mass function and $\mathrm{H(z')=H_{0}[\Omega_{M}(1+z')^{3} + \Omega_{\Lambda}]^{1/2}}$ is the Hubble parameter for a flat cosmology.  

%% file: Sections/Discussion.tex
\section{Discussion}
\label{sect:discussion}

%The take home message of this work is that galaxy formation simulations predict that the James Webb Space Telescope will be able to observe statistically complete samples of galaxies out to redshift $\mathrm{z=10}$ needed to parametrise the distributions of stellar masses and star formation rates for the first galaxies to have formed in the Universe.

\label{subsect:abundance_massive}
\begin{figure*}
    \includegraphics[width=\textwidth]{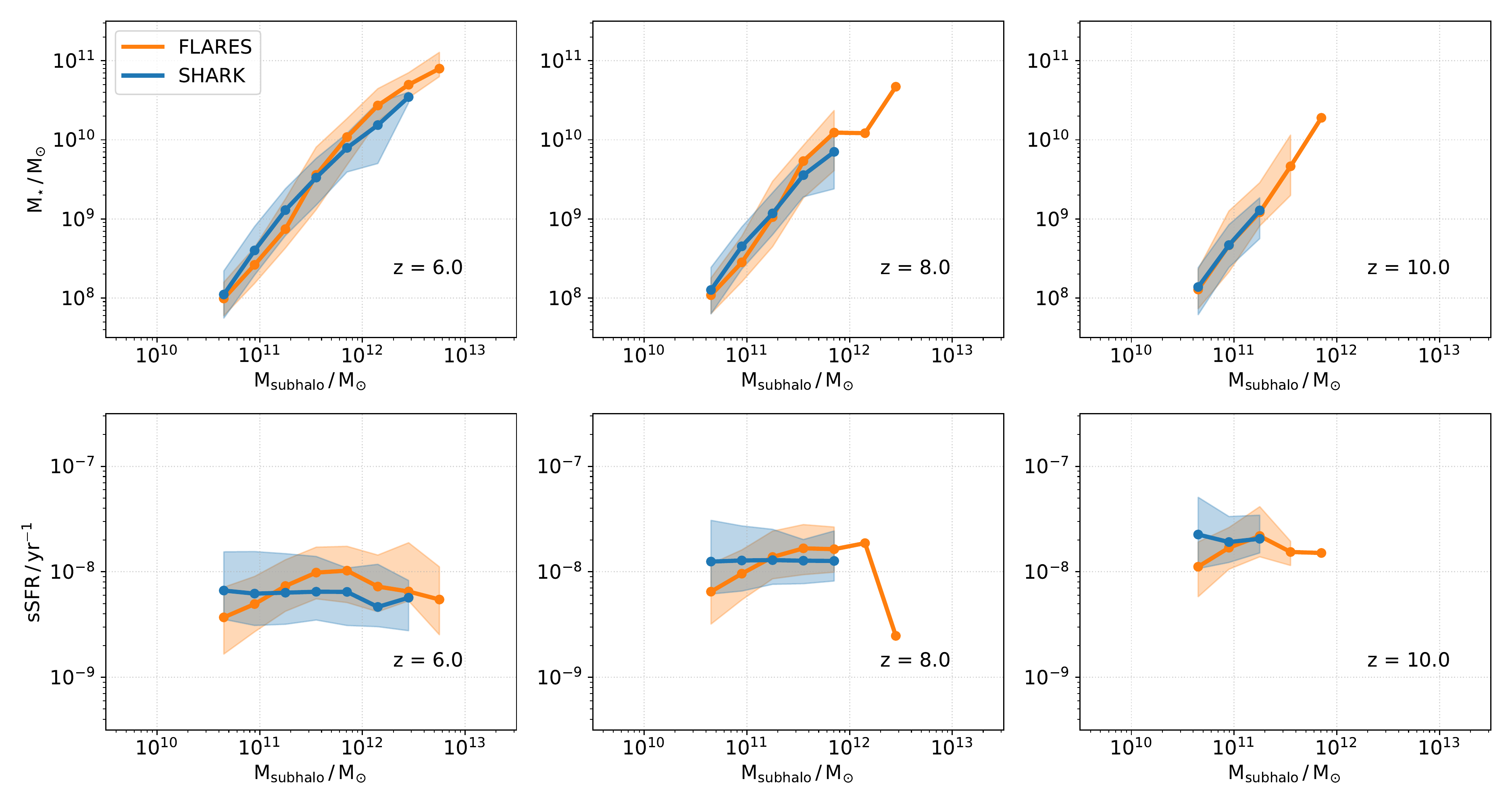}
    \caption{\textit{Top}: $\mathrm{M_{subhalo}-M_{\star}}$ at $z=6,8,10$, as labelled, for \textsc{Flares} (orange) and \textsc{Shark} (blue). The dotted lines show the median quantity while the shaded regions show the 5-95 quantiles. We only show central galaxies here. \textit{Bottom}: $\mathrm{sSFR-M_{subhalo}}$. The colours and labels are the same as the top row. We only show central galaxies here.}
    \label{fig:ssfr_mahlo}
\end{figure*}

%While high redshift measurements have been successful at determining, to some extent, stellar mass and SFR functions \citep[e.g.,][]{Katsianis2017,Stefanon2021}, the myriad assumptions needed to account for the effect of dust attenuation and the lack of large, deep near-infrared catalogues, severely complicate studies of the $\mathrm{SFR-M_{\star}}$ plane. The shape and variance of the $\mathrm{SFR-M_{\star}}$ relation encode information about astrophysical processes shaping it, which is especially important to understand the formation of first galaxies and how those drive the epoch of reionisation. 
We have shown that both simulations, \textsc{Flares} and \textsc{Shark}, predict that the JWST will be able to observe galaxies with a large enough range of stellar masses and SFRs to allow a thorough study of SFS, including the onset of quenching in massive galaxies, and thus stellar mass assembly during the early Universe. Additionally, we predict that the JWST will be a vital tool to better understand cosmic reionisation as a large census of galaxies that span a high dynamic range of stellar masses and star formation rates will will likely encompass ionising sources.
%, we predict that the JWST should be able to identify early, galactic sources of cosmic reionisation, and how the evolution of these sources is connected with stellar mass assembly. 
Prior observational studies \citep[e.g.,][]{Thorne2021} indicate the prevalence of a SFS around $z\sim5$, and here we show that both the simulations analysed here predict the existence of a SFS up to at least $z=10$. The existence of the SFS is a product of self-regulation of star formation in \textsc{Flares} and \textsc{Shark} where galaxies relax to the locus of the SFS through a balance of star formation and feedback. Each of these simulations implement these prescriptions differently so it is an interesting result that self-regulation appears in both independently. \textsc{Flares} for example uses a gas density floor above which stars form, whereas this is not the case for \textsc{Shark}. Furthermore, \textsc{Flares} implements stellar feedback stochastically to neighbouring SPH gas particles directly \citep{DallaVecchiaSchaye2012}, whereas stellar feedback is implemented with the calculation of the mass-ejection rate in \textsc{Shark}, which depends on the maximum circular velocity of the galaxy \citep[e.g.,][]{Lagos2018}. We show in \Cref{subsbect:onset_quenching} that the implementation of feedback from AGN between the simulations can explain the different shapes of the SFS. In this work, we quantified the predicted SFS's shape and scatter as a function of stellar mass in each simulation. We expect the JWST to be able to test these predictions in the near future.
%and as such we predict the shape of the $\mathrm{SFR-M_{\star}}$ under the assumption that a SFS persists out to redshift $\mathrm{z=10}$. We hypothesise the existence of a SFS and its shape and scatter so that when JWST observations arrive we can test this hypothesis.
%Once the JWST provides its first observations a comparison to each simulation will make it possible to constrain which details of galaxy formation theory are most critical for describing galaxies as we have explicitly used two unique galaxy formation simulations, each with their own independent methods. 
\Cref{fig:smf_sfrf_z6-10} shows that both \textsc{Flares} and \textsc{Shark} predict similar intrinsic distributions of stellar mass and SFR which indicates that, although these simulations differ in many details, the sub-grid methods of stellar mass assembly used in each must act  similarly. The relative agreement with the existing observations indicate that our understanding of stellar mass assembly must not be so far removed from how the process functions in nature; though, it will be a task for the JWST to confirm this suspicion with updated, near infrared photometry. 

Despite this overall agreement, there are important differences between the two simulations that we discuss below. 

\subsection{The abundance of massive galaxies}

A glaring difference between the simulations is the excess number density of high stellar mass galaxies with high star formation rates at fixed stellar mass in \textsc{Flares} compared to \textsc{Shark}.
To get to the bottom of this difference, we show in the top panels of \Cref{fig:ssfr_mahlo} the subhalo-stellar mass relation for both \textsc{Flares} and \textsc{Shark} at $z=6,8,10$ for central galaxies. We only use centrals in these calculations to simplify the comparison; including satellite galaxies only slightly increases the higher mass, $\mathrm{M_{\star} \gtrsim 10^{10}M_{\odot}}$, scatter of these trends. \textsc{Flares} extends the $\mathrm{M_{subhalo}-M_{\star}}$ 0.2~dex and 1~dex beyond \textsc{Shark} at $z=6$ and $z=10$ respectively. By the fact that the simulations broadly agree on the shape of the relation over all redshifts, it is possible to conclude that the difference in the predicted SMFs is due to {\sc Flares} covering a wider dynamic range, extending to rare, large over-densities (i.e. cosmic variance). 
%This is particularly important going forward as the JWST, with its enhanced sensitivity, should be able to investigate the impact of overdensity on the distribution of stellar masses and SFRs. 

Despite the broad agreement between the simulations in the top panel of \Cref{fig:ssfr_mahlo}, it is worth highlighting that \textsc{Flares} predicts slightly more stellar mass for a fixed subhalo mass than \textsc{Shark}, at $\mathrm{M_{subhalo}\gtrsim 10^{12}\,\rm M_{\odot}}$ at $z=6$ and $z=8$.  
%This could either mean that \textsc{Flares} contains more mass in halos than \textsc{Shark}, or \textsc{Flares} produces more stellar mass for a fixed halo mass compared to \textsc{Shark}; or a combination of the two. 
% {\bf CHANGE LOWER PANEL FOR SSFR VS MSUBHALO TO MAKE MORE SENSE OF THE COMPARISON. REWRITE BELOW BASED ON NEW PLOT}
The lower panels of \Cref{fig:ssfr_mahlo} show the specific SFR as a function of subhalo mass for the simulations at $z=6,8,10$. Especially at $z=6,8$, \textsc{Flares} predicts a higher sSFR than \textsc{Shark} for $\mathrm{M_{subhalo}\gtrsim10^{11}M_{\odot}}$, indicating that star formation is more efficient in massive halos in \textsc{Flares} than in \textsc{Shark}. Therefore it is likely a combination of both \textsc{Flares} sampling greater overdensities than \textsc{Shark} and the star formation prescription that are driving differences in the resulting stellar mass and SFR distributions.

\subsection{The onset of quenching in massive galaxies}
\label{subsbect:onset_quenching}
Another important difference between the simulations is their predictions on the onset of quenching in massive galaxies, $\mathrm{\geq 10^{10}M_{\odot}}$ as indicated by the prevalence of the turn-over and increased scatter in the SFS shown in \Cref{fig:sfs_quantiles,fig:sfs_jwst_fit,fig:sigma_fl_sh}. \citet{Davies2021} interpreted the increased scatter at the massive end of the SFS for the DEVILS sample as being caused by AGN feedback inhibiting star formation and driving galaxies below the SFS. This is corroborated by earlier {\sc EAGLE} simulations' results of \citet{Katsianis2019}, who also attributed the increased scatter at the massive end to AGN feedback. 
%While we can suggest a similar explanation for the features we observe in our results, 
\begin{figure*}
    \includegraphics[width=\textwidth]{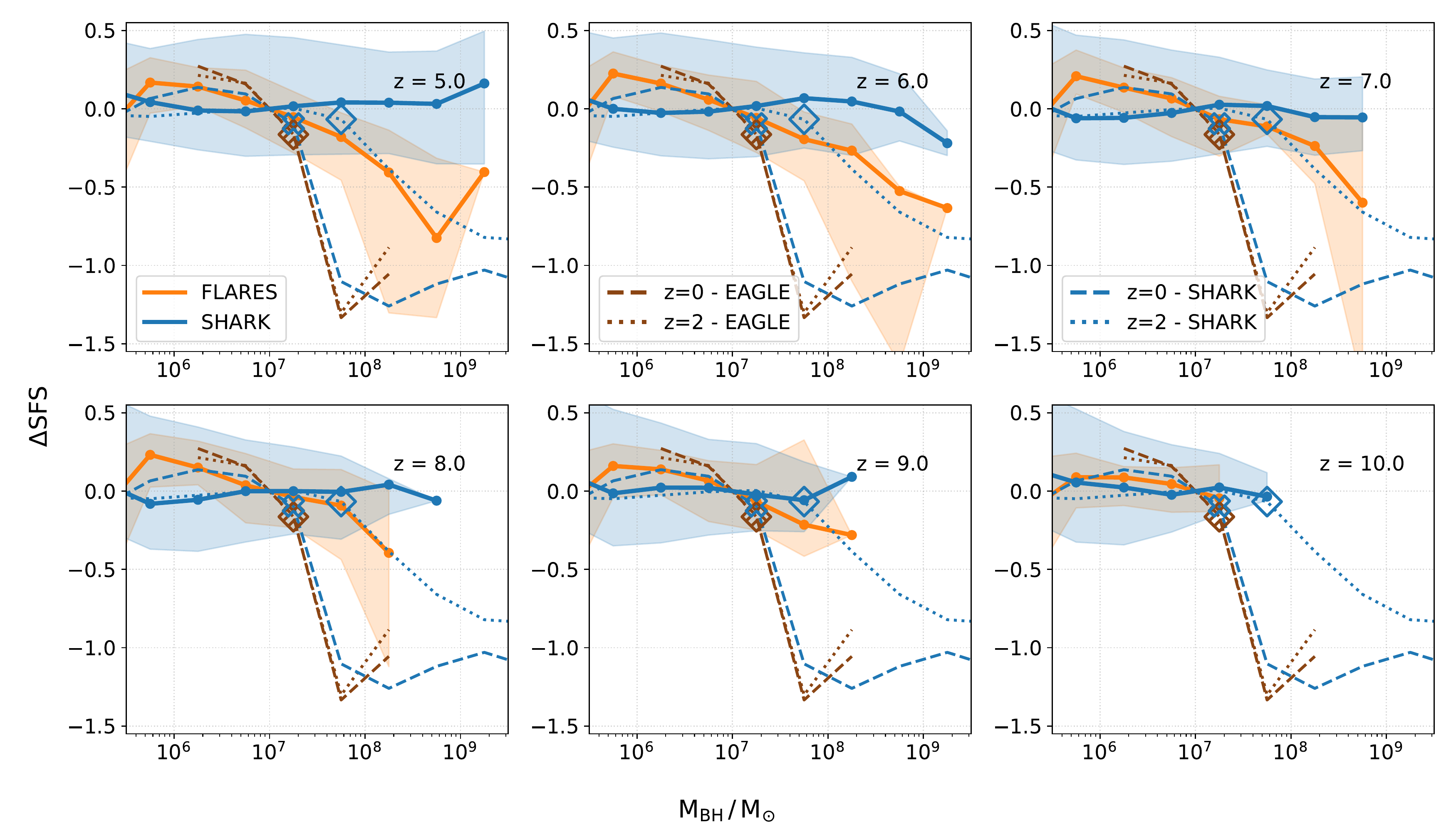}
    \caption{$\mathrm{\Delta SFS-M_{BH}}$ at redshifts $z=5,6,7,8,9,10$ as labelled for \textsc{Flares} (orange) and \textsc{Shark} (blue). We use the sSFR per unit black hole mass shifted up by the median sSFR between $\mathrm{M_{BH}=10^{6.75} \to 10^{7.25} M_{\odot}}$ as a proxy measure for distance to the SFS. The solid lines indicate the median and the shaded regions indicate 16-84 quantiles. The dashed and dotted lines show the results at $z=0$ and $z=2$. We show $z=0$ and $z=2$ results from \textsc{EAGLE}-AGNdT9 \citep{Schaye2015,Crain_2015,McAlpine2016} with brown lines in place of \textsc{Flares} as \textsc{Flares} stops at $z \approx 5$. Because \textsc{Flares} uses the same code and model as \textsc{EAGLE}-AGNdT9, this is an appropriate comparison to make. The X symbols indicate the black hole mass at which $\mathrm{\Delta SFS}$ experiences its greatest decline at $z=2$, while the diamond symbols indicate the same thing but for $z=0$. Note that for \textsc{EAGLE}, the Xs and diamonds positions are identical.}
    \label{fig:ssfr-mbh}
\end{figure*}

Here, we explore much higher redshifts than \citet{Davies2021} ($0.1 < z< 0.85$) and \citet{Katsianis2019} ($z<4$). To investigate the effect of AGN feedback on quenching and the increased scatter in the SFS at the high-mass end we show the distance from the SFS as a function of super massive black hole (SMBH) mass for the two simulations in \Cref{fig:ssfr-mbh}. We use the sSFR per unit mass shifted up by the median sSFR between $\mathrm{M_{BH}=10^{6.75} \to 10^{7.25} M_{\odot}}$ as a proxy measure for distance to the SFS. We use this measure instead of our fits to the SFS to promote comparisons with lower redshift results for which we do not have SFS fits. We see that the $\mathrm{\Delta SFS \sim 0}$ is approximately constant, in \textsc{Shark} over all black hole masses at all redshifts. We see at $z \lesssim 9$ that the $\mathrm{\Delta SFS}$ between \textsc{Shark} and \textsc{Flares} begins to diverge with \textsc{Flares} experiencing a downturn in $\mathrm{\Delta SFS}$ for SMBH masses $\mathrm{M_{BH} \gtrsim 10^{7}M_{\odot}}$. The divergence for $\mathrm{M_{BH} \gtrsim 10^{7}M_{\odot}}$ is greatest at $z\lesssim7$ that lines up well with the similar redshifts where the turn-over and increased scatter become more prevalent in \textsc{Flares} compared to \textsc{Shark}, indicating a connection between forming very massive SMBHs and quenching, congruous with the lower redshift conclusions in the literature. \citet{Bower17} showed that in {\sc EAGLE}, galaxies being quenched by AGN feedback are characterised by a strongly non-linear SMBH growth phase. The SMBH mass is thus a good predictor of the regime in which AGN feedback is efficient. \citet{Terrazas20} showed that in another cosmological hydrodynamic simulation, {\sc Illustris-TNG}, a similar behaviour is seen, and galaxies above a given SMBH mass experience AGN feedback quenching. 
%ACA VOY
 
The dashed and dotted lines in \Cref{fig:ssfr-mbh} show the $z=0$ and ${z=2}$ relations respectively for {\sc EAGLE} AGNdT9-50 \citep{Schaye2015,Crain_2015,McAlpine2016} and \textsc{Shark}. We remind the reader that {\sc Flares} adopted the same model as {\sc EAGLE} meaning the latter is the right comparison dataset for {\sc Flares} at ${z<5}$. The \textsc{EAGLE} Ref-100 simulation uses a different set of parameters meaning that it is unsuitable for comparison despite being a larger volume. At ${z=0}$, both {\sc EAGLE} and {\sc Shark} predict a sharp downturn in sSFR around $\mathrm{M_{BH}\gtrsim10^{7}M_{\odot}}$ indicating that above this SMBH mass AGN feedback becomes efficient at inhibiting star formation in galaxies. 
%
%
%\begin{figure}
%    \centering
%    \includegraphics[width = %\linewidth]{Figures/eagle_shark_mbh_ssfr.pdf}
%    \caption{$\mathrm{z=0, 0.5, 1, 2 \; sSFR-M_{BH}}$ in EAGLE (above) and \textsc{Shark} below. The dotted lines indicate the median. Colours indicate the different redshifts as labelled. Crosses indicate the black hole mass before the greatest decrease in sSFR occurs; turn-over mass in sSFR. The black hole masses are binned by 0.3 dex between $\mathrm{10^{4.5} \to 10^{10.5} M_{\odot}}$.}
%    \label{fig:eagle-shark-mbh-ssfr}
%\end{figure}
%
This SMBH mass threshold for quenching does not evolve in {\sc EAGLE} and {\sc Flares}, staying around $\mathrm{M_{BH}\gtrsim10^{7}M_{\odot}}$ at all times; i.e., as soon as a galaxy hits that SMBH, its star formation activity starts to quench due to AGN feedback. This is highlighted by the orange X and diamond symbols that correspond to the black hole mass at which the $\mathrm{\Delta SFS}$ experiences its sharpest down turn at ${z=2}$ and ${z=0}$ respectively. Note, that the position of these symbols are identical. However, {\sc Shark} behaves very differently, with the SMBH threshold mass above which quenching driven by AGN feedback happens evolves towards higher SMBH masses as the redshift increases. By ${z=2}$, the SMBH mass threshold for quenching in {\sc Shark} is $\sim$1~dex higher than at ${z=0}$ as indicated by the blue X and diamond symbols respectively.

%.\Cref{fig:eagle-shark-mbh-ssfr} shows the $\mathrm{z=0, 0.5, 1, 2 \; sSFR-M_{BH}}$ relations for EAGLE and \textsc{Shark}. We mark with crosses the black hole mass corresponding to the point before which the sSFR experiences its sharpest downfall, i.e. the turn-over black hole mass. It can be seen in EAGLE that the turn-over mass remains fairly constant, $\mathrm{\sim 10^{7} M_{\odot}}$, over $\mathrm{\sim 10.4 \, Gyr}$ of time, while there is almost an order of magnitude variation, $\sim \mathrm{10^{7} \to 10^{8} M_{\odot}}$, in \textsc{Shark} for the same amount of time. 
In \textsc{Shark} the SMBH accretion rate due to hot halo cooling is what regulates the efficiency of star formation quenching in central galaxies \citep{Lagos2018}. We observe a greater variation in turn-over mass over time in \text{Shark} that is a reflection of the AGN entering the appropriate feedback mode at different SMBH masses at different redshifts. 
This is not the case in {\sc EAGLE} and {\sc Flares}, and quenching mostly occurs at a fixed mass. This implies that the AGN in \textsc{Shark} for ${z \geq 5}$  are mostly growing in the QSO mode, and are yet to have significant accretion rates from the hot-halo cooling mode. This means they cannot quench galaxies, while all that matters in \textsc{Flares} is that the AGN have grown to sufficient masses, which happens at ${z \lesssim 7}$. 
Therefore, the cause of the difference seen in the SFS at the massive end between the simulations is a result of different models of AGN feedback. The JWST will be able to probe the SFS up to $\mathrm{M_{\star}\sim 10 ^{10}M_{\odot}}$, providing stringent constraints on theoretical models of AGN feedback.

\subsection{The volume probed by JWST surveys}

So far we have not accounted for the survey volume required to ascertain a sample representative of the Universe. Uncertainties in pencil beam surveys such as the Hubble Ultra Deep Field \citep{Beckwith2006} are dominated by cosmic variance where the probability of obtaining a survey sample representative of the entire Universe scales with area coverage \citep{DriverRobotham2010}. 

\begin{figure*}
    \includegraphics[width=0.8\textwidth]{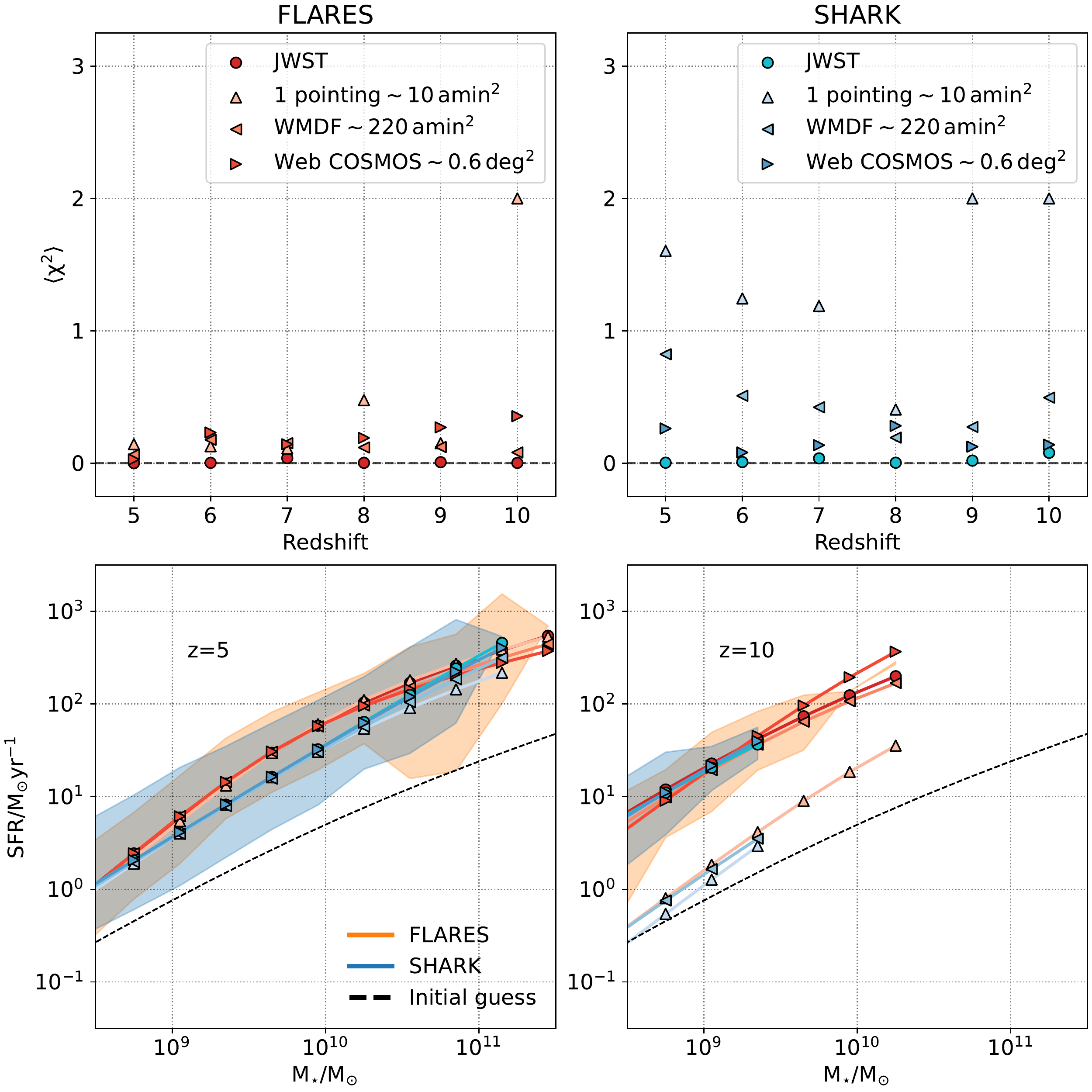}
    \caption{\textit{Top: } Weighted-averaged $\mathrm{\chi^{2}}$ values for each of the SFS fits where we have resampled galaxy stellar masses and SFRs according to different survey volumes 20 times for \textsc{Flares} (left) and \textsc{Shark} (right). $\mathrm{\chi^{2} = \sum^{4}_{i=0}\frac{(x_{i}-x^{true}_{i})^{2}}{x^{true}_{i}}}$, where $\mathrm{x_{i}}$ represents the derived parameters of the resampled SFS and $\mathrm{x_{i}^{true}}$ are the derived parameters for the entire population in the simulations. When calculating the average $\mathrm{\chi^{2}}$, we weight each of the $\mathrm{\chi^{2}}$ values by the inverse of the sum of squared uncertainties of the four parameters. The Filled circles show the  $\mathrm{\chi^{2}}$ for the JWST-detected galaxies; upright, left-pointing and right-pointing triangles show the $\mathrm{\chi^{2}}$ for a JWST surveys of areas $\mathrm{10arcmin^{2}}$ (i.e. a single pointing), $\mathrm{220arcmin^{2}}$ (i.e. a medium deep field) and $\mathrm{0.6deg^{2}}$ (i.e. the Web COSMOS survey), respectively, as labelled. In cases where the number of data points from the sample is less than the number of fit parameters, e.g., at $z=10$ for single-pointing surveys, we set the averaged $\mathrm{\chi^{2}}$ to an upper limit of 2. \textit{Bottom: } SFS plots at ${z=5}$ (left) and ${z=10}$ (right), for \textsc{Flares} (red and orange colours) and \textsc{Shark} (blue and cyan colours). We use the same symbols and colour gradient as the top panels to reflect different fits for each sample. Orange and blue shaded regions show the $5^{\rm th}-95^{\rm th}$ percentile ranges for all galaxies in the simulations. We find the average set of parameters for each of our 20 samples, weighting by the inverse of the error of each of the four SFS parameters. The dashed, black line shows the SFS for our initial guess of parameters ($\mathrm{[1.0, 10.0, 1.0, 0.5] = S_{0}, M_{0}, \alpha, \beta}$). The offsets of some of the fitted relations, e.g., single-pointing surveys at $z=10$, are due to the fits defaulting to the initial guess because of insufficient number statistics, which biases the average SFS relation toward the initial guess.}
    \label{fig:resampled}
\end{figure*}

To better assess the impact of survey area on the ability to constrain the shape of the SFS, we refit the SFS after resampling the stellar mass and star formation rate distributions assuming some survey volume. Specifically, we calculate the 2D histogram of SFR and stellar mass per unit volume and for a given redshift for JWST-selected galaxies in both simulations, and calculate the cumulative distribution function per unit volume at ${z=5,\,6,\,7,\,8,\,9\,\mathrm{and}~10}$. Multiplying this by a given survey volume gives the expected cumulative number distribution of galaxies in the survey at a given redshift. Here, we do not assume any random errors on stellar masses or star formation rates to aid in comparison between fits from the same simulation. We repeat this resampling and fitting process twenty times and average the derived SFS parameters to mitigate randomness caused by sampling the joint stellar mass and SFR distributions. We calculate $\mathrm{\chi^{2} = \sum^{4}_{i=0}\frac{(x_{i}-x^{true}_{i})^{2}}{x^{true}_{i}}}$, where $\mathrm{x_{i}}$ represents the derived parameters of the resampled SFS and $\mathrm{x_{i}^{true}}$ are the derived parameters for the entire population in the simulations, to assess how well each survey can reconstruct the SFS of the true population. We average this over the 20 samples. When calculating the average, we weight each sample by the inverse of the sum of squared uncertainties of each of the four parameters. We also impose a narrower prior on the turn-over mass ($\mathrm{M_{0}\sim U(8, 13)}$) when fitting our samples to prevent highly unphysical solutions to the high-mass end, especially in cases where high mass data points are missing as is the case for surveys that cover small areas. We note that the upper bound of this prior straddles the derived turn-over masses shown in \Cref{tab:sfs_parms}.

We show the results in \Cref{fig:resampled} for the entire population of galaxies in both simulations, the JWST-detected galaxies (determined from their F200W apparent magnitude), galaxies detected in a single pointing of $\mathrm{10\,arcmin^{2}}$, galaxies detected in $\mathrm{220\,arcmin^{2}}$ and galaxies detected in $\mathrm{0.6\,deg^{2}}$. The choice of $\mathrm{220\,arcmin^{2}}$ and $\mathrm{0.6\,deg^{2}}$ survey areas comes from the JWST Medium Deep Field (WMDF) Survey (Proposal ID: 1176, PI: R. Windhorst) and the Web COSMOS survey (Proposal ID: 1727, PI: J. Kartaltepe) \footnote{\url{https://www.stsci.edu/jwst/science-execution/program-information.html}}. 

% We see that even with a single NIRCam pointing that the {\bf SFS's parameters can be well recovered at $\mathrm{z=5}$} as can be seen from the low $\mathrm{\chi^{2}}$ values for both simulations {\bf and the SFS fits in the bottom panels of \Cref{fig:resampled}}. The agreement with the true population is better in \textsc{Flares} than in \textsc{Shark} at $\mathrm{z=5}$.

We see that in general it becomes more difficult to constrain the shape of the SFS with smaller area surveys. This is true in \textsc{Shark} over all redshifts shown, while only at ${z>7}$ in \textsc{Flares} do single pointings struggle to constrain the SFS. The fact that all survey areas can constrain the SFS in \textsc{Flares} at ${z\sim5}$ while \textsc{Shark} can not is a reflection of the two component SFS function being unsuitable for \textsc{Shark} at these lower redshifts. The wide offsets of the SFS fits for the average set of parameters of the narrowest surveys at $z=10$ seen in the bottom right panel of \Cref{fig:resampled} are due to the fits essentially failing and defaulting to the initial guess of the parameters, which biases the SFS to the initial guess. We do see that both simulations predict that a Web COSMOS-like survey without any assumed random errors will be able constrain the true SFS parameters at ${z\geq5}$. We further explore the effect of random errors on $\mathrm{M_{\star}}$ and SFR in \cref{apdx:uncertainties}. Nevertheless, we advocate for wider area surveys that will be able to get accurate SFRs and stellar masses needed to thoroughly study the SFS. 
%highlighting the model dependence when it comes to fitting the SFS. 
% We see that it becomes much harder to constrain the shape of the SFS at $\mathrm{z=10}$ with even the {\bf so far} widest planned survey of $\mathrm{0.6deg^{2}}$; therefore, we advocate for even wider JWST surveys to {\bf reliably measure the SFS} at $\mathrm{z=10}$. 

Interestingly, at ${z=6,10}$ in \textsc{Flares} the average $\chi^{2}$ for the Web COSMOS-like survey is higher than the narrower WMDF-like survey. This is likely a result of the fitting technique. At ${z=10}$ for example, the turn-over mass is $\mathrm{10^{10}M_{\odot}}$ for the WMDF-like survey, which goes beyond the range of resampled galaxies implying that the turn-over is unphysical for this sample, despite the new imposed prior. For the Web COSMOS-like survey, we predict more galaxies at higher stellar mass, but not so many that the turn-over in the SFS can be constrained. In this case, the SFS favours a single power law fit, and the average $\chi^{2}$ increases. This just further shows why it is necessary to have accurate SFRs and stellar masses over the breadth of the $\mathrm{SFR-M_{\star}}$. 
% Interestingly, we observe instances when the SFS is better constrained, with a lower $\mathrm{\chi^{2}}$, for smaller area surveys. For example, the result at $\mathrm{z=9,10}$ for \textsc{Flares} shows that the constraint from {\bf a Webb COSMOS-like survey} is predicted to be worse than the WMDF-like survey. Similarly, at $\mathrm{z=10}$ in \textsc{Shark} the SFS is better constrain by a single JWST pointing than the wider WMDF-like survey. This is an effect of the resampling and fitting process, which further shows why it is necessary to have accurate stellar masses and SFRs over a broad range of the $\mathrm{SFR-M_{\star}}$. 

% {\bf This is due to the uncertainties we introduce in stellar mass and SFR in some cases skewing the SFS-derived parameters quite significantly, by chance. Thus, getting accurate stellar masses and SFRs is as important as pushing for as deep and wide as possible surveys.}
%This highlights that fitting the SFS is not so trivial at any redshift, especially when there is level of uncertainty attached to stellar masses and star formation rates that can influence the shape of the SFS. 

Our resampling method does not take into account the underlying structure of the Universe that will affect our predicted constraints on the SFS. As \textsc{Flares} is comprised of 40 unique subvolumes that have each been accordingly weighted, a smooth mass distribution cannot be reconstructed, and thus cannot be used to test cosmic variance directly. Therefore, to test the effect of cosmic variance, we limit our attention to \textsc{Shark}, where we can partition the box into smaller sub volumes of equal weighting and refit the SFS in each. We divide the whole \textsc{Shark} box in 3 dimensions into 8 independent, contiguous and equal size subvolumes. We, once again, choose not to use random errors to emphasise the effect of cosmic variance only.

\begin{figure*}
    \includegraphics[width=\textwidth]{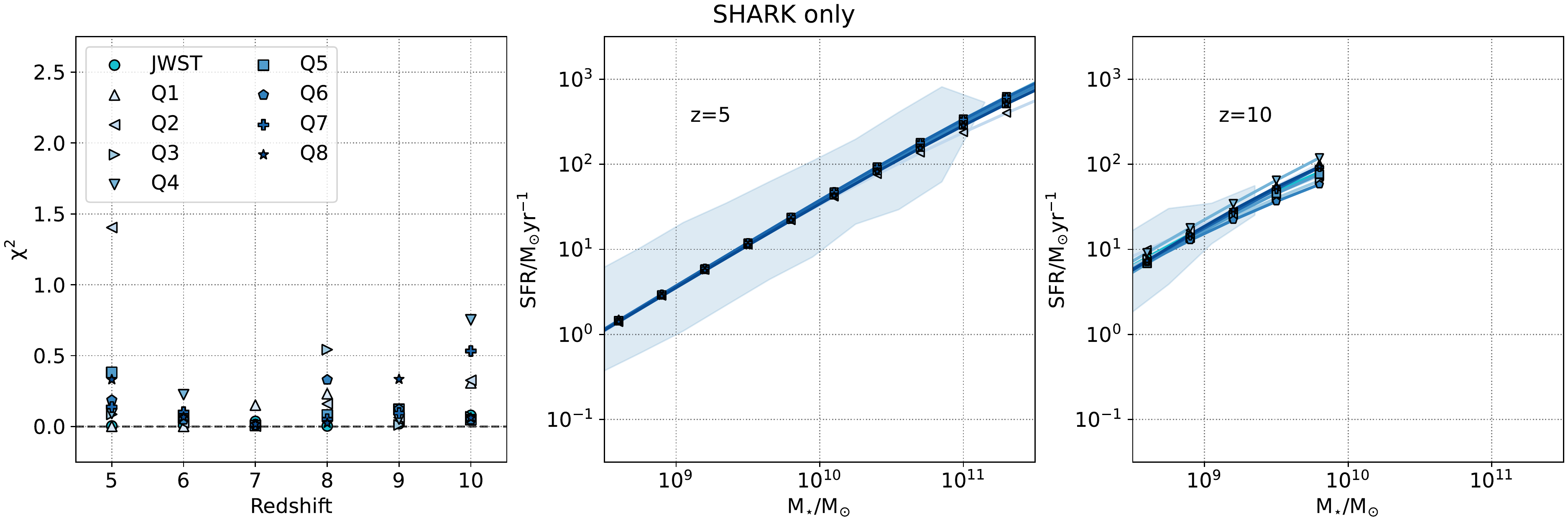}
    \caption{\textit{Left: } $\mathrm{\chi^{2}}$ values (as defined in \Cref{fig:resampled}), for the SFS fit performed in each subvolumes of the total \textsc{Shark} box. Galaxies selected based on their F200W magnitude, are shown with filled circles. We show each set of $8$ subvolumes with different symbols, as labelled. \textit{Middle and right: } SFS plots at ${z=5}$ (middle) and ${z=10}$ (right). We use the same symbols and colour gradient as the left panels to show the SFS fits of each galaxy sample.} 
    \label{fig:cosvar}
\end{figure*}

\Cref{fig:cosvar} shows the resulting SFS of this exercise. Cosmic variance mildly impacts the ability to constrain the SFS at all redshifts. 
At ${z=5}$ cosmic variance negligibly affects the SFS fit. By ${z=10}$ cosmic variance affects the whole appearance of the SFS, with the effect on the constraint at high masses being significant.
% {\bf At $z=5$, cosmic variance mostly affects the SFS fit at high masses (i.e. constraining the presence of a bend), while at $z=10$, it affects the whole appearance of the SFS.}
%We also see how parameter degeneracies become important in fitting the SFS with a similar spread in $\mathrm{chi^{2}}$ values at $\mathrm{z=5}$ and $\mathrm{z=10}$ that lead to visually different shapes and levels of agreement with the true population, once more highlighting that the SFS is not trivial to constrain. 
Therefore, in addition to our advocacy of wider surveys we further advocate for surveys distributed over a range of fields and sight lines. 

% Need to talk about Casey2014 and dust paradigms

%% file: Sections/Conclusion.tex
\section{Conclusions}
\label{sect:conclusion}

We have presented predictions on the ability for the JWST to quantify the details of star formation and stellar mass assembly of galaxies between ${z=5 \to 10}$. By using two distinct simulations, \textsc{Flares} and \textsc{Shark}, we have provided a theoretical framework, unmarred by potential systematic biases contained in a single simulation, in which the emerging observations of the JWST can be tested to better understand the astrophysics of the early Universe.

The key results are:

\begin{itemize}
    \item \textbf{Stellar mass and SFR functions: } We predict that throughout redshifts ${z=5 \to 10}$, the JWST will be able to detect all intrinsically bright galaxies up to redshift $\rm z=10$, and thus will be able to parametrise the distributions of stellar mass and SFRs of the first galaxies in the Universe. Conversely, both {\sc Flares} and {\sc Shark} predict that very dust-obscured galaxies are completely absent from current HST surveys. This is expected to affect current measurements of the SMF and SFRF from HST observations \citep{Katsianis2017, McLure2010, Bouwens2015, Song2016, Gonzalez2011, Thorne2021, Stefanon2021}. We predict that the JWST observations are only limited by the detection sensitivity of the instrument and survey area, and will not miss highly obscured galaxies.  
    
    \item \textbf{Star forming main sequence:} We predict that the JWST will be able to parametrise the shape and the stellar mass dependent scatter of the SFS between ${z=5\to10}$ for the first time by obtaining observations over a sufficient wide range of stellar masses and SFRs. As the SFS encodes information about the astrophysics driving the shape and scatter, the JWST will prove to be an invaluable tool in improving our understanding of galaxy formation up to the onset of cosmic reionisation when the first galaxies formed, and on the rising of massive, AGN-quenched galaxies towards the end of reionisation.
    
    \item \textbf{Cosmic stellar and SFR histories:}
    Because we predict the JWST to be able to observe every intrinsically bright galaxy out to redshift ${z\sim10}$, these observations should be able to constraint the CSMH and CSFRH between ${z=5\to10}$ robustly. The main issue here will be the contribution of low-mass galaxies, $M_{\star}\lesssim 10^8\,\rm M_{\odot}$ that are fainter than F200W 29mag, and which is expected to be significant at $z\gtrsim 5$. This could be resolved if the SMF and SFRF are extrapolated.
    
    %The only limit of the instrument will be that of sensitivity and area. This is an improvement upon existing short-wavelength surveys that potentially miss the dustiest galaxies in the Universe. As such, the JWST should be able to constrain the CSMH and CSFRH between ${z=5\to10}$ robustly.
    
    \item \textbf{The dependence of predictions on the baryon physics models: }
    By using two independent simulations we have shown that each prediction on the ability of the JWST is model dependent. Both \textsc{Flares} and \textsc{Shark} show similar distributions of stellar mass and SFR indicating that their sub-grid recipes are similar, but not exact. An important delineator between the two simulations is how each models dust that is itself sensitive to how each simulation deals with star formation and chemical enrichment. As we predict that the JWST will be able to provide a sufficient census of galaxies out to redshift ${z\sim10}$ we suggest that the JWST observations, in combination with multi-wavelength follow up observations, will be able distinguish between the two models.
    %by uncoupling the intricacies between the sub-grid parameters.  

    \item \textbf{The effect of survey area and cosmic variance: }
    By sampling from our simulated galaxy populations to obtain particular survey samples we find that wide area surveys are necessary to constrain the shape of the star-forming main sequence from $z=5$ to $z=10$. Furthermore, by dividing the \textsc{Shark} box into 8 independent and contiguous subvolumes and refitting the SFS in each we show that cosmic variance is an important factor in influencing the shape of the SFS at ${z\sim10}$, but less so at lower redshifts. 
    
\end{itemize}

With continued observations, the JWST will open the door to the next milestone toward our understanding of galaxy formation. We have presented predictions on the ability of the JWST so that we are well poised to compare them against the JWST observations as they continue to arrive. A deeper study of the specific sub-grid recipes used in simulations will allow us to interpret the astrophysics that the telescope recovers. Although \textsc{Flares} and \textsc{Shark} broadly agree with each other, it is a key objective of the JWST to distinguish, and comprehend the origin of, the slight differences between the independent models.

%--conclude with jwst should find all intrinsically bright galaxies needed to provide statistical sample blah blah blah--

%% file: main.bbl
\newcommand{\noop}[1]{}
\begin{thebibliography}{}
\makeatletter
\relax
\def\mn@urlcharsother{\let\do\@makeother \do\$\do\&\do\#\do\^\do\_\do\%\do\~}
\def\mn@doi{\begingroup\mn@urlcharsother \@ifnextchar [ {\mn@doi@}
  {\mn@doi@[]}}
\def\mn@doi@[#1]#2{\def\@tempa{#1}\ifx\@tempa\@empty \href
  {http://dx.doi.org/#2} {doi:#2}\else \href {http://dx.doi.org/#2} {#1}\fi
  \endgroup}
\def\mn@eprint#1#2{\mn@eprint@#1:#2::\@nil}
\def\mn@eprint@arXiv#1{\href {http://arxiv.org/abs/#1} {{\tt arXiv:#1}}}
\def\mn@eprint@dblp#1{\href {http://dblp.uni-trier.de/rec/bibtex/#1.xml}
  {dblp:#1}}
\def\mn@eprint@#1:#2:#3:#4\@nil{\def\@tempa {#1}\def\@tempb {#2}\def\@tempc
  {#3}\ifx \@tempc \@empty \let \@tempc \@tempb \let \@tempb \@tempa \fi \ifx
  \@tempb \@empty \def\@tempb {arXiv}\fi \@ifundefined
  {mn@eprint@\@tempb}{\@tempb:\@tempc}{\expandafter \expandafter \csname
  mn@eprint@\@tempb\endcsname \expandafter{\@tempc}}}

\bibitem[\protect\citeauthoryear{{Abramson}, {Kelson}, {Dressler}, {Poggianti},
  {Gladders}, {Oemler}  \& {Vulcani}}{{Abramson} et~al.}{2014}]{Abramson2014}
{Abramson} L.~E.,  {Kelson} D.~D.,  {Dressler} A.,  {Poggianti} B.,  {Gladders}
  M.~D.,  {Oemler} Augustus J.,   {Vulcani} B.,  2014, \mn@doi [\apjl]
  {10.1088/2041-8205/785/2/L36}, \href
  {https://ui.adsabs.harvard.edu/abs/2014ApJ...785L..36A} {785, L36}

\bibitem[\protect\citeauthoryear{Bahé et~al.,}{Bahé et~al.}{2017a}]{Bahe2017}
Bahé Y.~M.,  et~al., 2017a, \mn@doi [\mnras] {10.1093/mnras/stx1403}, 470,
  4186–4208

\bibitem[\protect\citeauthoryear{Bahé et~al.,}{Bahé
  et~al.}{2017b}]{Bah__2017}
Bahé Y.~M.,  et~al., 2017b, \mn@doi [\mnras] {10.1093/mnras/stx1403}, 470,
  4186–4208

\bibitem[\protect\citeauthoryear{{Barnes} et~al.,}{{Barnes}
  et~al.}{2017}]{Barnes2017}
{Barnes} D.~J.,  et~al., 2017, \mn@doi [\mnras] {10.1093/mnras/stx1647}, \href
  {https://ui.adsabs.harvard.edu/abs/2017MNRAS.471.1088B} {471, 1088}

\bibitem[\protect\citeauthoryear{{Baugh}}{{Baugh}}{2006}]{Baugh2006}
{Baugh} C.~M.,  2006, \mn@doi [Reports on Progress in Physics]
  {10.1088/0034-4885/69/12/R02}, \href
  {https://ui.adsabs.harvard.edu/abs/2006RPPh...69.3101B} {69, 3101}

\bibitem[\protect\citeauthoryear{Beckwith et~al.,}{Beckwith
  et~al.}{2006a}]{beckwithHubbleUltraDeep2006a}
Beckwith S. V.~W.,  et~al., 2006a, \mn@doi [\aj] {10.1086/507302}, 132, 1729

\bibitem[\protect\citeauthoryear{{Beckwith} et~al.,}{{Beckwith}
  et~al.}{2006b}]{Beckwith2006}
{Beckwith} S. V.~W.,  et~al., 2006b, \mn@doi [\aj] {10.1086/507302}, \href
  {https://ui.adsabs.harvard.edu/abs/2006AJ....132.1729B} {132, 1729}

\bibitem[\protect\citeauthoryear{Bellstedt et~al.,}{Bellstedt
  et~al.}{2020}]{bellstedtGalaxyMassAssembly2020}
Bellstedt S.,  et~al., 2020, \mn@doi [Monthly Notices of the Royal Astronomical
  Society] {10.1093/mnras/staa2620}, 498, 5581

\bibitem[\protect\citeauthoryear{{Benson}}{{Benson}}{2010}]{Benson2010}
{Benson} A.~J.,  2010, \mn@doi [\physrep] {10.1016/j.physrep.2010.06.001},
  \href {https://ui.adsabs.harvard.edu/abs/2010PhR...495...33B} {495, 33}

\bibitem[\protect\citeauthoryear{{Bouwens} et~al.,}{{Bouwens}
  et~al.}{2012}]{Bouwens2012}
{Bouwens} R.~J.,  et~al., 2012, \mn@doi [\apjl] {10.1088/2041-8205/752/1/L5},
  \href {https://ui.adsabs.harvard.edu/abs/2012ApJ...752L...5B} {752, L5}

\bibitem[\protect\citeauthoryear{{Bouwens} et~al.,}{{Bouwens}
  et~al.}{2015}]{Bouwens2015}
{Bouwens} R.~J.,  et~al., 2015, \mn@doi [\apj] {10.1088/0004-637X/803/1/34},
  \href {https://ui.adsabs.harvard.edu/abs/2015ApJ...803...34B} {803, 34}

\bibitem[\protect\citeauthoryear{Bouwens et~al.,}{Bouwens
  et~al.}{2020}]{bouwensALMASpectroscopicSurvey2020}
Bouwens R.,  et~al., 2020, \mn@doi [The Astrophysical Journal]
  {10.3847/1538-4357/abb830}, 902, 112

\bibitem[\protect\citeauthoryear{Bouwens et~al.,}{Bouwens
  et~al.}{2021}]{Bouwens2021}
Bouwens R.~J.,  et~al., 2021, \mn@doi [The Astronomical Journal]
  {10.3847/1538-3881/abf83e}, 162, 47

\bibitem[\protect\citeauthoryear{{Bower}, {Schaye}, {Frenk}, {Theuns},
  {Schaller}, {Crain}  \& {McAlpine}}{{Bower} et~al.}{2017}]{Bower17}
{Bower} R.~G.,  {Schaye} J.,  {Frenk} C.~S.,  {Theuns} T.,  {Schaller} M.,
  {Crain} R.~A.,   {McAlpine} S.,  2017, \mn@doi [\mnras]
  {10.1093/mnras/stw2735}, \href
  {https://ui.adsabs.harvard.edu/abs/2017MNRAS.465...32B} {465, 32}

\bibitem[\protect\citeauthoryear{{Brinchmann}, {Charlot}, {White}, {Tremonti},
  {Kauffmann}, {Heckman}  \& {Brinkmann}}{{Brinchmann}
  et~al.}{2004}]{Brichmann2004}
{Brinchmann} J.,  {Charlot} S.,  {White} S.~D.~M.,  {Tremonti} C.,  {Kauffmann}
  G.,  {Heckman} T.,   {Brinkmann} J.,  2004, \mn@doi [\mnras]
  {10.1111/j.1365-2966.2004.07881.x}, \href
  {https://ui.adsabs.harvard.edu/abs/2004MNRAS.351.1151B} {351, 1151}

\bibitem[\protect\citeauthoryear{{Bruzual} \& {Charlot}}{{Bruzual} \&
  {Charlot}}{2003}]{Bruzual&Charlot2003}
{Bruzual} G.,  {Charlot} S.,  2003, \mn@doi [\mnras]
  {10.1046/j.1365-8711.2003.06897.x}, \href
  {https://ui.adsabs.harvard.edu/abs/2003MNRAS.344.1000B} {344, 1000}

\bibitem[\protect\citeauthoryear{{Ca{\~n}as}, {Elahi}, {Welker}, {del P Lagos},
  {Power}, {Dubois}  \& {Pichon}}{{Ca{\~n}as} et~al.}{2019}]{Canas2019}
{Ca{\~n}as} R.,  {Elahi} P.~J.,  {Welker} C.,  {del P Lagos} C.,  {Power} C.,
  {Dubois} Y.,   {Pichon} C.,  2019, \mn@doi [\mnras] {10.1093/mnras/sty2725},
  \href {https://ui.adsabs.harvard.edu/abs/2019MNRAS.482.2039C} {482, 2039}

\bibitem[\protect\citeauthoryear{Calzetti, Armus, Bohlin, Kinney, Koornneef  \&
  {Storchi-Bergmann}}{Calzetti et~al.}{2000}]{calzettiDustContentOpacity2000}
Calzetti D.,  Armus L.,  Bohlin R.~C.,  Kinney A.~L.,  Koornneef J.,
  {Storchi-Bergmann} T.,  2000, \mn@doi [\apj] {10.1086/308692}, 533, 682

\bibitem[\protect\citeauthoryear{{Camps} \& {Baes}}{{Camps} \&
  {Baes}}{2015}]{Camps2015}
{Camps} P.,  {Baes} M.,  2015, \mn@doi [Astronomy and Computing]
  {10.1016/j.ascom.2014.10.004}, \href
  {https://ui.adsabs.harvard.edu/abs/2015A&C.....9...20C} {9, 20}

\bibitem[\protect\citeauthoryear{Casey, Narayanan  \& Cooray}{Casey
  et~al.}{2014}]{Casey_2014}
Casey C.~M.,  Narayanan D.,   Cooray A.,  2014, \mn@doi [Physics Reports]
  {10.1016/j.physrep.2014.02.009}, 541, 45–161

\bibitem[\protect\citeauthoryear{Casey et~al.,}{Casey
  et~al.}{2018}]{Casey_2018}
Casey C.~M.,  et~al., 2018, \mn@doi [\apj] {10.3847/1538-4357/aac82d}, 862, 77

\bibitem[\protect\citeauthoryear{{Chabrier}}{{Chabrier}}{2003}]{Chabrier2003}
{Chabrier} G.,  2003, \mn@doi [\pasp] {10.1086/376392}, \href
  {https://ui.adsabs.harvard.edu/abs/2003PASP..115..763C} {115, 763}

\bibitem[\protect\citeauthoryear{{Charlot} \& {Fall}}{{Charlot} \&
  {Fall}}{2000}]{Charlot&Fall2000}
{Charlot} S.,  {Fall} S.~M.,  2000, \mn@doi [\apj] {10.1086/309250}, \href
  {https://ui.adsabs.harvard.edu/abs/2000ApJ...539..718C} {539, 718}

\bibitem[\protect\citeauthoryear{{Cole}, {Lacey}, {Baugh}  \& {Frenk}}{{Cole}
  et~al.}{2000}]{Cole2000}
{Cole} S.,  {Lacey} C.~G.,  {Baugh} C.~M.,   {Frenk} C.~S.,  2000, \mn@doi
  [\mnras] {10.1046/j.1365-8711.2000.03879.x}, \href
  {https://ui.adsabs.harvard.edu/abs/2000MNRAS.319..168C} {319, 168}

\bibitem[\protect\citeauthoryear{Cook, Cortese, Catinella  \& Robotham}{Cook
  et~al.}{2020}]{Cook2020}
Cook R. H.~W.,  Cortese L.,  Catinella B.,   Robotham A.,  2020, \mn@doi
  [\mnras] {10.1093/mnras/staa666}, 493, 5596

\bibitem[\protect\citeauthoryear{Crain et~al.,}{Crain
  et~al.}{2015}]{Crain_2015}
Crain R.~A.,  et~al., 2015, \mn@doi [\mnras] {10.1093/mnras/stv725}, 450,
  1937–1961

\bibitem[\protect\citeauthoryear{Croton}{Croton}{2013}]{Croton2013}
Croton D.~J.,  2013, \mn@doi [\pasa] {10.1017/pasa.2013.31}, 30

\bibitem[\protect\citeauthoryear{{Croton} et~al.,}{{Croton}
  et~al.}{2016}]{Croton2016}
{Croton} D.~J.,  et~al., 2016, \mn@doi [\apjs] {10.3847/0067-0049/222/2/22},
  \href {https://ui.adsabs.harvard.edu/abs/2016ApJS..222...22C} {222, 22}

\bibitem[\protect\citeauthoryear{{Curtis-Lake}, {Chevallard}, {Charlot}  \&
  {Sandles}}{{Curtis-Lake} et~al.}{2021}]{Curtis-Lake2021}
{Curtis-Lake} E.,  {Chevallard} J.,  {Charlot} S.,   {Sandles} L.,  2021,
  \mn@doi [\mnras] {10.1093/mnras/stab698}, \href
  {https://ui.adsabs.harvard.edu/abs/2021MNRAS.503.4855C} {503, 4855}

\bibitem[\protect\citeauthoryear{{Dalla Vecchia} \& {Schaye}}{{Dalla Vecchia}
  \& {Schaye}}{2012}]{DallaVecchiaSchaye2012}
{Dalla Vecchia} C.,  {Schaye} J.,  2012, \mn@doi [\mnras]
  {10.1111/j.1365-2966.2012.21704.x}, \href
  {https://ui.adsabs.harvard.edu/abs/2012MNRAS.426..140D} {426, 140}

\bibitem[\protect\citeauthoryear{{Davies} et~al.,}{{Davies}
  et~al.}{2016}]{Davies2016}
{Davies} L.~J.~M.,  et~al., 2016, \mn@doi [\mnras] {10.1093/mnras/stw1342},
  \href {https://ui.adsabs.harvard.edu/abs/2016MNRAS.461..458D} {461, 458}

\bibitem[\protect\citeauthoryear{{Davies} et~al.,}{{Davies}
  et~al.}{2018}]{Davies2018}
{Davies} L.~J.~M.,  et~al., 2018, \mn@doi [\mnras] {10.1093/mnras/sty1553},
  \href {https://ui.adsabs.harvard.edu/abs/2018MNRAS.480..768D} {480, 768}

\bibitem[\protect\citeauthoryear{{Davies} et~al.,}{{Davies}
  et~al.}{2019}]{Davies2019}
{Davies} L.~J.~M.,  et~al., 2019, \mn@doi [\mnras] {10.1093/mnras/sty2957},
  \href {https://ui.adsabs.harvard.edu/abs/2019MNRAS.483.1881D} {483, 1881}

\bibitem[\protect\citeauthoryear{{Davies} et~al.,}{{Davies}
  et~al.}{2021}]{Davies2021b}
{Davies} L.~J.~M.,  et~al., 2021, \mn@doi [\mnras] {10.1093/mnras/stab1601},
  \href {https://ui.adsabs.harvard.edu/abs/2021MNRAS.506..256D} {506, 256}

\bibitem[\protect\citeauthoryear{{Davies} et~al.,}{{Davies}
  et~al.}{2022}]{Davies2021}
{Davies} L.~J.~M.,  et~al., 2022, \mn@doi [\mnras] {10.1093/mnras/stab3145},
  \href {https://ui.adsabs.harvard.edu/abs/2022MNRAS.509.4392D} {509, 4392}

\bibitem[\protect\citeauthoryear{{Davis}, {Efstathiou}, {Frenk}  \&
  {White}}{{Davis} et~al.}{1985}]{davis1985}
{Davis} M.,  {Efstathiou} G.,  {Frenk} C.~S.,   {White} S.~D.~M.,  1985,
  \mn@doi [\apj] {10.1086/163168}, \href
  {https://ui.adsabs.harvard.edu/abs/1985ApJ...292..371D} {292, 371}

\bibitem[\protect\citeauthoryear{De~Vis et~al.,}{De~Vis
  et~al.}{2019}]{DeVis2019}
De~Vis P.,  et~al., 2019, \mn@doi [Astronomy & Astrophysics]
  {10.1051/0004-6361/201834444}, 623, A5

\bibitem[\protect\citeauthoryear{{Draine}}{{Draine}}{2003}]{Draine2003}
{Draine} B.~T.,  2003, \mn@doi [\araa]
  {10.1146/annurev.astro.41.011802.094840}, \href
  {https://ui.adsabs.harvard.edu/abs/2003ARA&A..41..241D} {41, 241}

\bibitem[\protect\citeauthoryear{{Driver} \& {Robotham}}{{Driver} \&
  {Robotham}}{2010}]{DriverRobotham2010}
{Driver} S.~P.,  {Robotham} A. S.~G.,  2010, \mn@doi [\mnras]
  {10.1111/j.1365-2966.2010.17028.x}, \href
  {https://ui.adsabs.harvard.edu/abs/2010MNRAS.407.2131D} {407, 2131}

\bibitem[\protect\citeauthoryear{{Elahi}, {Welker}, {Power}, {Lagos},
  {Robotham}, {Ca{\~n}as}  \& {Poulton}}{{Elahi} et~al.}{2018}]{Elahi2018}
{Elahi} P.~J.,  {Welker} C.,  {Power} C.,  {Lagos} C. d.~P.,  {Robotham} A.
  S.~G.,  {Ca{\~n}as} R.,   {Poulton} R.,  2018, \mn@doi [\mnras]
  {10.1093/mnras/sty061}, \href
  {https://ui.adsabs.harvard.edu/abs/2018MNRAS.475.5338E} {475, 5338}

\bibitem[\protect\citeauthoryear{Elahi, Cañas, Poulton, Tobar, Willis, Lagos,
  Power  \& Robotham}{Elahi et~al.}{2019a}]{Elahi2019a}
Elahi P.~J.,  Cañas R.,  Poulton R. J.~J.,  Tobar R.~J.,  Willis J.~S.,  Lagos
  C. d.~P.,  Power C.,   Robotham A. S.~G.,  2019a, \mn@doi [\pasa]
  {10.1017/pasa.2019.12}, 36

\bibitem[\protect\citeauthoryear{{Elahi}, {Poulton}, {Tobar}, {Ca{\~n}as},
  {Lagos}, {Power}  \& {Robotham}}{{Elahi} et~al.}{2019b}]{Elahi2019b}
{Elahi} P.~J.,  {Poulton} R. J.~J.,  {Tobar} R.~J.,  {Ca{\~n}as} R.,  {Lagos}
  C. d.~P.,  {Power} C.,   {Robotham} A. S.~G.,  2019b, \mn@doi [\pasa]
  {10.1017/pasa.2019.18}, \href
  {https://ui.adsabs.harvard.edu/abs/2019PASA...36...28E} {36, e028}

\bibitem[\protect\citeauthoryear{{Eldridge}, {Stanway}, {Xiao}, {McClelland},
  {Taylor}, {Ng}, {Greis}  \& {Bray}}{{Eldridge} et~al.}{2017}]{Eldridge2017}
{Eldridge} J.~J.,  {Stanway} E.~R.,  {Xiao} L.,  {McClelland} L.~A.~S.,
  {Taylor} G.,  {Ng} M.,  {Greis} S.~M.~L.,   {Bray} J.~C.,  2017, \mn@doi
  [\pasa] {10.1017/pasa.2017.51}, \href
  {https://ui.adsabs.harvard.edu/abs/2017PASA...34...58E} {34, e058}

\bibitem[\protect\citeauthoryear{{Finkelstein} et~al.,}{{Finkelstein}
  et~al.}{2015}]{Finkelstein2015}
{Finkelstein} S.~L.,  et~al., 2015, \mn@doi [\apj]
  {10.1088/0004-637X/810/1/71}, \href
  {https://ui.adsabs.harvard.edu/abs/2015ApJ...810...71F} {810, 71}

\bibitem[\protect\citeauthoryear{{Foreman-Mackey}, {Hogg}, {Lang}  \&
  {Goodman}}{{Foreman-Mackey} et~al.}{2013}]{emcee}
{Foreman-Mackey} D.,  {Hogg} D.~W.,  {Lang} D.,   {Goodman} J.,  2013, \mn@doi
  [PASP] {10.1086/670067}, 125, 306

\bibitem[\protect\citeauthoryear{Fudamoto et~al.,}{Fudamoto
  et~al.}{2021}]{Fudamoto_2021}
Fudamoto Y.,  et~al., 2021, \mn@doi [Nature] {10.1038/s41586-021-03846-z}, 597,
  489–492

\bibitem[\protect\citeauthoryear{{Furlong} et~al.,}{{Furlong}
  et~al.}{2015}]{Furlong2015}
{Furlong} M.,  et~al., 2015, \mn@doi [\mnras] {10.1093/mnras/stv852}, \href
  {https://ui.adsabs.harvard.edu/abs/2015MNRAS.450.4486F} {450, 4486}

\bibitem[\protect\citeauthoryear{{Gardner} et~al.,}{{Gardner}
  et~al.}{2006}]{Gardner2006}
{Gardner} J.~P.,  et~al., 2006, \mn@doi [\ssr] {10.1007/s11214-006-8315-7},
  \href {https://ui.adsabs.harvard.edu/abs/2006SSRv..123..485G} {123, 485}

\bibitem[\protect\citeauthoryear{{Gonz{\'a}lez}, {Labb{\'e}}, {Bouwens},
  {Illingworth}, {Franx}  \& {Kriek}}{{Gonz{\'a}lez}
  et~al.}{2011}]{Gonzalez2011}
{Gonz{\'a}lez} V.,  {Labb{\'e}} I.,  {Bouwens} R.~J.,  {Illingworth} G.,
  {Franx} M.,   {Kriek} M.,  2011, \mn@doi [\apjl]
  {10.1088/2041-8205/735/2/L34}, \href
  {https://ui.adsabs.harvard.edu/abs/2011ApJ...735L..34G} {735, L34}

\bibitem[\protect\citeauthoryear{{Grogin} et~al.,}{{Grogin}
  et~al.}{2011}]{Grogin2011}
{Grogin} N.~A.,  et~al., 2011, \mn@doi [\apjs] {10.1088/0067-0049/197/2/35},
  \href {https://ui.adsabs.harvard.edu/abs/2011ApJS..197...35G} {197, 35}

\bibitem[\protect\citeauthoryear{{Hao}, {Kennicutt}, {Johnson}, {Calzetti},
  {Dale}  \& {Moustakas}}{{Hao} et~al.}{2011}]{Hao2011}
{Hao} C.-N.,  {Kennicutt} R.~C.,  {Johnson} B.~D.,  {Calzetti} D.,  {Dale}
  D.~A.,   {Moustakas} J.,  2011, \mn@doi [\apj] {10.1088/0004-637X/741/2/124},
  \href {https://ui.adsabs.harvard.edu/abs/2011ApJ...741..124H} {741, 124}

\bibitem[\protect\citeauthoryear{{Katsianis} et~al.,}{{Katsianis}
  et~al.}{2017}]{Katsianis2017}
{Katsianis} A.,  et~al., 2017, \mn@doi [\mnras] {10.1093/mnras/stx2020}, \href
  {https://ui.adsabs.harvard.edu/abs/2017MNRAS.472..919K} {472, 919}

\bibitem[\protect\citeauthoryear{{Katsianis} et~al.,}{{Katsianis}
  et~al.}{2019}]{Katsianis2019}
{Katsianis} A.,  et~al., 2019, \mn@doi [\apj] {10.3847/1538-4357/ab1f8d}, \href
  {https://ui.adsabs.harvard.edu/abs/2019ApJ...879...11K} {879, 11}

\bibitem[\protect\citeauthoryear{{Koekemoer} et~al.,}{{Koekemoer}
  et~al.}{2007}]{Koekemoer2007}
{Koekemoer} A.~M.,  et~al., 2007, \mn@doi [\apjs] {10.1086/520086}, \href
  {https://ui.adsabs.harvard.edu/abs/2007ApJS..172..196K} {172, 196}

\bibitem[\protect\citeauthoryear{Koushan et~al.,}{Koushan
  et~al.}{2021}]{koushanGAMADEVILSConstraining2021}
Koushan S.,  et~al., 2021, \mn@doi [Monthly Notices of the Royal Astronomical
  Society] {10.1093/mnras/stab540}, 503, 2033

\bibitem[\protect\citeauthoryear{{Kregel}, {van der Kruit}  \& {de
  Grijs}}{{Kregel} et~al.}{2002}]{Kregel2002}
{Kregel} M.,  {van der Kruit} P.~C.,   {de Grijs} R.,  2002, \mn@doi [\mnras]
  {10.1046/j.1365-8711.2002.05556.x}, \href
  {https://ui.adsabs.harvard.edu/abs/2002MNRAS.334..646K} {334, 646}

\bibitem[\protect\citeauthoryear{Krumholz, McKee  \& Tumlinson}{Krumholz
  et~al.}{2009}]{Krumholz_2009}
Krumholz M.~R.,  McKee C.~F.,   Tumlinson J.,  2009, \mn@doi [\apj]
  {10.1088/0004-637x/699/1/850}, 699, 850–856

\bibitem[\protect\citeauthoryear{{Labb{\'e}} et~al.,}{{Labb{\'e}}
  et~al.}{2013}]{Labbe2013}
{Labb{\'e}} I.,  et~al., 2013, \mn@doi [\apjl] {10.1088/2041-8205/777/2/L19},
  \href {https://ui.adsabs.harvard.edu/abs/2013ApJ...777L..19L} {777, L19}

\bibitem[\protect\citeauthoryear{{Lagos}, {Lacey}, {Baugh}, {Bower}  \&
  {Benson}}{{Lagos} et~al.}{2011}]{Lagos2011}
{Lagos} C. D.~P.,  {Lacey} C.~G.,  {Baugh} C.~M.,  {Bower} R.~G.,   {Benson}
  A.~J.,  2011, \mn@doi [\mnras] {10.1111/j.1365-2966.2011.19160.x}, \href
  {https://ui.adsabs.harvard.edu/abs/2011MNRAS.416.1566L} {416, 1566}

\bibitem[\protect\citeauthoryear{{Lagos}, {Baugh}, {Zwaan}, {Lacey},
  {Gonzalez-Perez}, {Power}, {Swinbank}  \& {van Kampen}}{{Lagos}
  et~al.}{2014}]{Lagos2014}
{Lagos} C.~D.~P.,  {Baugh} C.~M.,  {Zwaan} M.~A.,  {Lacey} C.~G.,
  {Gonzalez-Perez} V.,  {Power} C.,  {Swinbank} A.~M.,   {van Kampen} E.,
  2014, \mn@doi [\mnras] {10.1093/mnras/stu266}, \href
  {https://ui.adsabs.harvard.edu/abs/2014MNRAS.440..920L} {440, 920}

\bibitem[\protect\citeauthoryear{{Lagos}, {Tobar}, {Robotham}, {Obreschkow},
  {Mitchell}, {Power}  \& {Elahi}}{{Lagos} et~al.}{2018}]{Lagos2018}
{Lagos} C. d.~P.,  {Tobar} R.~J.,  {Robotham} A. S.~G.,  {Obreschkow} D.,
  {Mitchell} P.~D.,  {Power} C.,   {Elahi} P.~J.,  2018, \mn@doi [\mnras]
  {10.1093/mnras/sty2440}, \href
  {https://ui.adsabs.harvard.edu/abs/2018MNRAS.481.3573L} {481, 3573}

\bibitem[\protect\citeauthoryear{{Lagos} et~al.,}{{Lagos}
  et~al.}{2019}]{Lagos2019}
{Lagos} C. d.~P.,  et~al., 2019, \mn@doi [\mnras] {10.1093/mnras/stz2427},
  \href {https://ui.adsabs.harvard.edu/abs/2019MNRAS.489.4196L} {489, 4196}

\bibitem[\protect\citeauthoryear{{Lee} et~al.,}{{Lee} et~al.}{2012}]{Lee2012}
{Lee} K.-S.,  et~al., 2012, \mn@doi [\apj] {10.1088/0004-637X/752/1/66}, \href
  {https://ui.adsabs.harvard.edu/abs/2012ApJ...752...66L} {752, 66}

\bibitem[\protect\citeauthoryear{{Lee} et~al.,}{{Lee} et~al.}{2015}]{Lee2015}
{Lee} N.,  et~al., 2015, \mn@doi [\apj] {10.1088/0004-637X/801/2/80}, \href
  {https://ui.adsabs.harvard.edu/abs/2015ApJ...801...80L} {801, 80}

\bibitem[\protect\citeauthoryear{{Leslie} et~al.,}{{Leslie}
  et~al.}{2020}]{Leslie2020}
{Leslie} S.~K.,  et~al., 2020, \mn@doi [\apj] {10.3847/1538-4357/aba044}, \href
  {https://ui.adsabs.harvard.edu/abs/2020ApJ...899...58L} {899, 58}

\bibitem[\protect\citeauthoryear{{Lilly}, {Carollo}, {Pipino}, {Renzini}  \&
  {Peng}}{{Lilly} et~al.}{2013}]{Lilly2013}
{Lilly} S.~J.,  {Carollo} C.~M.,  {Pipino} A.,  {Renzini} A.,   {Peng} Y.,
  2013, \mn@doi [\apj] {10.1088/0004-637X/772/2/119}, \href
  {https://ui.adsabs.harvard.edu/abs/2013ApJ...772..119L} {772, 119}

\bibitem[\protect\citeauthoryear{{Lovell}, {Vijayan}, {Thomas}, {Wilkins},
  {Barnes}, {Irodotou}  \& {Roper}}{{Lovell} et~al.}{2021}]{Lovell2021}
{Lovell} C.~C.,  {Vijayan} A.~P.,  {Thomas} P.~A.,  {Wilkins} S.~M.,  {Barnes}
  D.~J.,  {Irodotou} D.,   {Roper} W.,  2021, \mn@doi [\mnras]
  {10.1093/mnras/staa3360}, \href
  {https://ui.adsabs.harvard.edu/abs/2021MNRAS.500.2127L} {500, 2127}

\bibitem[\protect\citeauthoryear{{Madau} \& {Dickinson}}{{Madau} \&
  {Dickinson}}{2014}]{Madau&Dickinson2014}
{Madau} P.,  {Dickinson} M.,  2014, \mn@doi [\araa]
  {10.1146/annurev-astro-081811-125615}, \href
  {https://ui.adsabs.harvard.edu/abs/2014ARA&A..52..415M} {52, 415}

\bibitem[\protect\citeauthoryear{{Matthee} \& {Schaye}}{{Matthee} \&
  {Schaye}}{2019}]{Matthee2019}
{Matthee} J.,  {Schaye} J.,  2019, \mn@doi [\mnras] {10.1093/mnras/stz030},
  \href {https://ui.adsabs.harvard.edu/abs/2019MNRAS.484..915M} {484, 915}

\bibitem[\protect\citeauthoryear{{McAlpine} et~al.,}{{McAlpine}
  et~al.}{2016}]{McAlpine2016}
{McAlpine} S.,  et~al., 2016, \mn@doi [Astronomy and Computing]
  {10.1016/j.ascom.2016.02.004}, \href
  {https://ui.adsabs.harvard.edu/abs/2016A&C....15...72M} {15, 72}

\bibitem[\protect\citeauthoryear{{McLure}, {Dunlop}, {Cirasuolo}, {Koekemoer},
  {Sabbi}, {Stark}, {Targett}  \& {Ellis}}{{McLure} et~al.}{2010}]{McLure2010}
{McLure} R.~J.,  {Dunlop} J.~S.,  {Cirasuolo} M.,  {Koekemoer} A.~M.,  {Sabbi}
  E.,  {Stark} D.~P.,  {Targett} T.~A.,   {Ellis} R.~S.,  2010, \mn@doi
  [\mnras] {10.1111/j.1365-2966.2009.16176.x}, \href
  {https://ui.adsabs.harvard.edu/abs/2010MNRAS.403..960M} {403, 960}

\bibitem[\protect\citeauthoryear{{Meurer}, {Heckman}  \& {Calzetti}}{{Meurer}
  et~al.}{1999}]{Meurer1999}
{Meurer} G.~R.,  {Heckman} T.~M.,   {Calzetti} D.,  1999, \mn@doi [\apj]
  {10.1086/307523}, \href
  {https://ui.adsabs.harvard.edu/abs/1999ApJ...521...64M} {521, 64}

\bibitem[\protect\citeauthoryear{{Micha{\l}owski}, {Hayward}, {Dunlop},
  {Bruce}, {Cirasuolo}, {Cullen}  \& {Hernquist}}{{Micha{\l}owski}
  et~al.}{2014}]{Michalowski2014}
{Micha{\l}owski} M.~J.,  {Hayward} C.~C.,  {Dunlop} J.~S.,  {Bruce} V.~A.,
  {Cirasuolo} M.,  {Cullen} F.,   {Hernquist} L.,  2014, \mn@doi [\aap]
  {10.1051/0004-6361/201424174}, \href
  {https://ui.adsabs.harvard.edu/abs/2014A&A...571A..75M} {571, A75}

\bibitem[\protect\citeauthoryear{Navarro, Frenk  \& White}{Navarro
  et~al.}{1997}]{NFW1997}
Navarro J.~F.,  Frenk C.~S.,   White S. D.~M.,  1997, \mn@doi [\apj]
  {10.1086/304888}, 490, 493–508

\bibitem[\protect\citeauthoryear{{Noeske} et~al.,}{{Noeske}
  et~al.}{2007a}]{Noeske2007b}
{Noeske} K.~G.,  et~al., 2007a, \mn@doi [\apjl] {10.1086/517926}, \href
  {https://ui.adsabs.harvard.edu/abs/2007ApJ...660L..43N} {660, L43}

\bibitem[\protect\citeauthoryear{{Noeske} et~al.,}{{Noeske}
  et~al.}{2007b}]{Noeske2007}
{Noeske} K.~G.,  et~al., 2007b, \mn@doi [\apjl] {10.1086/517927}, \href
  {https://ui.adsabs.harvard.edu/abs/2007ApJ...660L..47N} {660, L47}

\bibitem[\protect\citeauthoryear{Oesch, Bouwens, Illingworth, Labb{\'{e}}  \&
  Stefanon}{Oesch et~al.}{2018}]{Oesch_2018}
Oesch P.~A.,  Bouwens R.~J.,  Illingworth G.~D.,  Labb{\'{e}} I.,   Stefanon
  M.,  2018, \mn@doi [\apj] {10.3847/1538-4357/aab03f}, 855, 105

\bibitem[\protect\citeauthoryear{{Pearson} et~al.,}{{Pearson}
  et~al.}{2018}]{Pearson2018}
{Pearson} W.~J.,  et~al., 2018, \mn@doi [\aap] {10.1051/0004-6361/201832821},
  \href {https://ui.adsabs.harvard.edu/abs/2018A&A...615A.146P} {615, A146}

\bibitem[\protect\citeauthoryear{{Planck Collaboration} et~al.}{{Planck
  Collaboration} et~al.}{2014}]{Planck2014year1}
{Planck Collaboration} et~al., 2014, \mn@doi [\aap]
  {10.1051/0004-6361/201321529}, \href
  {https://ui.adsabs.harvard.edu/abs/2014A&A...571A...1P} {571, A1}

\bibitem[\protect\citeauthoryear{{Planck Collaboration} et~al.}{{Planck
  Collaboration} et~al.}{2016}]{Planck2025XIII}
{Planck Collaboration} et~al., 2016, \mn@doi [\aap]
  {10.1051/0004-6361/201525830}, \href
  {https://ui.adsabs.harvard.edu/abs/2016A&A...594A..13P} {594, A13}

\bibitem[\protect\citeauthoryear{{R{\'e}my-Ruyer} et~al.,}{{R{\'e}my-Ruyer}
  et~al.}{2014}]{Remy-Ruyer2014}
{R{\'e}my-Ruyer} A.,  et~al., 2014, \mn@doi [\aap]
  {10.1051/0004-6361/201322803}, \href
  {https://ui.adsabs.harvard.edu/abs/2014A&A...563A..31R} {563, A31}

\bibitem[\protect\citeauthoryear{Rigby et~al.,}{Rigby
  et~al.}{2022}]{rigbyCharacterizationJWSTScience2022a}
Rigby J.,  et~al., 2022, Characterization of {JWST} science performance from
  commissioning, \url {http://arxiv.org/abs/2207.05632}

\bibitem[\protect\citeauthoryear{{Robotham}, {Bellstedt}, {Lagos}, {Thorne},
  {Davies}, {Driver}  \& {Bravo}}{{Robotham} et~al.}{2020}]{Robotham2020}
{Robotham} A.~S.~G.,  {Bellstedt} S.,  {Lagos} C. d.~P.,  {Thorne} J.~E.,
  {Davies} L.~J.,  {Driver} S.~P.,   {Bravo} M.,  2020, \mn@doi [\mnras]
  {10.1093/mnras/staa1116}, \href
  {https://ui.adsabs.harvard.edu/abs/2020MNRAS.495..905R} {495, 905}

\bibitem[\protect\citeauthoryear{{Salim} et~al.,}{{Salim}
  et~al.}{2007}]{Salim2007}
{Salim} S.,  et~al., 2007, \mn@doi [\apjs] {10.1086/519218}, \href
  {https://ui.adsabs.harvard.edu/abs/2007ApJS..173..267S} {173, 267}

\bibitem[\protect\citeauthoryear{Schaller, Dalla~Vecchia, Schaye, Bower,
  Theuns, Crain, Furlong  \& McCarthy}{Schaller et~al.}{2015}]{Schaller2015}
Schaller M.,  Dalla~Vecchia C.,  Schaye J.,  Bower R.~G.,  Theuns T.,  Crain
  R.~A.,  Furlong M.,   McCarthy I.~G.,  2015, \mn@doi [\mnras]
  {10.1093/mnras/stv2169}, 454, 2277

\bibitem[\protect\citeauthoryear{{Schaye} et~al.,}{{Schaye}
  et~al.}{2010}]{Schaye2010}
{Schaye} J.,  et~al., 2010, \mn@doi [\mnras]
  {10.1111/j.1365-2966.2009.16029.x}, \href
  {https://ui.adsabs.harvard.edu/abs/2010MNRAS.402.1536S} {402, 1536}

\bibitem[\protect\citeauthoryear{{Schaye} et~al.,}{{Schaye}
  et~al.}{2015}]{Schaye2015}
{Schaye} J.,  et~al., 2015, \mn@doi [\mnras] {10.1093/mnras/stu2058}, \href
  {https://ui.adsabs.harvard.edu/abs/2015MNRAS.446..521S} {446, 521}

\bibitem[\protect\citeauthoryear{{Schenker} et~al.,}{{Schenker}
  et~al.}{2013}]{Schenker2013}
{Schenker} M.~A.,  et~al., 2013, \mn@doi [\apj] {10.1088/0004-637X/768/2/196},
  \href {https://ui.adsabs.harvard.edu/abs/2013ApJ...768..196S} {768, 196}

\bibitem[\protect\citeauthoryear{{Shen}, {Vogelsberger}, {Nelson}, {Tacchella},
  {Hernquist}, {Springel}, {Marinacci}  \& {Torrey}}{{Shen}
  et~al.}{2021}]{Shen2021}
{Shen} X.,  {Vogelsberger} M.,  {Nelson} D.,  {Tacchella} S.,  {Hernquist} L.,
  {Springel} V.,  {Marinacci} F.,   {Torrey} P.,  2021, arXiv e-prints, \href
  {https://ui.adsabs.harvard.edu/abs/2021arXiv210412788S} {p. arXiv:2104.12788}

\bibitem[\protect\citeauthoryear{Shivaei et~al.,}{Shivaei
  et~al.}{2020}]{shivaeiMOSDEFSurveyVariation2020b}
Shivaei I.,  et~al., 2020, \mn@doi [\apj] {10.3847/1538-4357/aba35e}, 899, 117

\bibitem[\protect\citeauthoryear{{Somerville} \& {Dav{\'e}}}{{Somerville} \&
  {Dav{\'e}}}{2015}]{Somerville&Dave2015}
{Somerville} R.~S.,  {Dav{\'e}} R.,  2015, \mn@doi [\araa]
  {10.1146/annurev-astro-082812-140951}, \href
  {https://ui.adsabs.harvard.edu/abs/2015ARA&A..53...51S} {53, 51}

\bibitem[\protect\citeauthoryear{Song et~al.,}{Song et~al.}{2016}]{Song2016}
Song M.,  et~al., 2016, \mn@doi [\apj] {10.3847/0004-637x/825/1/5}, 825, 5

\bibitem[\protect\citeauthoryear{{Speagle}, {Steinhardt}, {Capak}  \&
  {Silverman}}{{Speagle} et~al.}{2014}]{Speagle2014}
{Speagle} J.~S.,  {Steinhardt} C.~L.,  {Capak} P.~L.,   {Silverman} J.~D.,
  2014, \mn@doi [\apjs] {10.1088/0067-0049/214/2/15}, \href
  {https://ui.adsabs.harvard.edu/abs/2014ApJS..214...15S} {214, 15}

\bibitem[\protect\citeauthoryear{Springel}{Springel}{2005}]{Springel_2005}
Springel V.,  2005, \mn@doi [\mnras] {10.1111/j.1365-2966.2005.09655.x}, 364,
  1105–1134

\bibitem[\protect\citeauthoryear{{Springel}, {White}, {Tormen}  \&
  {Kauffmann}}{{Springel} et~al.}{2001}]{springel2001}
{Springel} V.,  {White} S. D.~M.,  {Tormen} G.,   {Kauffmann} G.,  2001,
  \mn@doi [\mnras] {10.1046/j.1365-8711.2001.04912.x}, \href
  {https://ui.adsabs.harvard.edu/abs/2001MNRAS.328..726S} {328, 726}

\bibitem[\protect\citeauthoryear{{Springel} et~al.,}{{Springel}
  et~al.}{2005}]{Springel2005}
{Springel} V.,  et~al., 2005, \mn@doi [\nat] {10.1038/nature03597}, \href
  {https://ui.adsabs.harvard.edu/abs/2005Natur.435..629S} {435, 629}

\bibitem[\protect\citeauthoryear{Stanway \& Eldridge}{Stanway \&
  Eldridge}{2018}]{stanwayReevaluatingOldStellar2018}
Stanway E.~R.,  Eldridge J.~J.,  2018, \mn@doi [Monthly Notices of the Royal
  Astronomical Society] {10.1093/mnras/sty1353}, 479, 75

\bibitem[\protect\citeauthoryear{Stefanon, Bouwens, Labbé, Illingworth,
  Gonzalez  \& Oesch}{Stefanon et~al.}{2021}]{Stefanon2021}
Stefanon M.,  Bouwens R.~J.,  Labbé I.,  Illingworth G.~D.,  Gonzalez V.,
  Oesch P.~A.,  2021, \mn@doi [\apj] {10.3847/1538-4357/ac1bb6}, 922, 29

\bibitem[\protect\citeauthoryear{{Tacchella}, {Dekel}, {Carollo}, {Ceverino},
  {DeGraf}, {Lapiner}, {Mandelker}  \& {Primack Joel}}{{Tacchella}
  et~al.}{2016}]{Tacchella2016}
{Tacchella} S.,  {Dekel} A.,  {Carollo} C.~M.,  {Ceverino} D.,  {DeGraf} C.,
  {Lapiner} S.,  {Mandelker} N.,   {Primack Joel} R.,  2016, \mn@doi [\mnras]
  {10.1093/mnras/stw131}, \href
  {https://ui.adsabs.harvard.edu/abs/2016MNRAS.457.2790T} {457, 2790}

\bibitem[\protect\citeauthoryear{{Terrazas} et~al.,}{{Terrazas}
  et~al.}{2020}]{Terrazas20}
{Terrazas} B.~A.,  et~al., 2020, \mn@doi [\mnras] {10.1093/mnras/staa374},
  \href {https://ui.adsabs.harvard.edu/abs/2020MNRAS.493.1888T} {493, 1888}

\bibitem[\protect\citeauthoryear{{Thorne} et~al.,}{{Thorne}
  et~al.}{2021}]{Thorne2021}
{Thorne} J.~E.,  et~al., 2021, \mn@doi [\mnras] {10.1093/mnras/stab1294}, \href
  {https://ui.adsabs.harvard.edu/abs/2021MNRAS.505..540T} {505, 540}

\bibitem[\protect\citeauthoryear{{Trayford}, {Lagos}, {Robotham}  \&
  {Obreschkow}}{{Trayford} et~al.}{2020}]{Trayford2020}
{Trayford} J.~W.,  {Lagos} C. d.~P.,  {Robotham} A. S.~G.,   {Obreschkow} D.,
  2020, \mn@doi [\mnras] {10.1093/mnras/stz3234}, \href
  {https://ui.adsabs.harvard.edu/abs/2020MNRAS.491.3937T} {491, 3937}

\bibitem[\protect\citeauthoryear{{Vazdekis}, {Koleva}, {Ricciardelli},
  {R{\"o}ck}  \& {Falc{\'o}n-Barroso}}{{Vazdekis} et~al.}{2016}]{Vazdekis2016}
{Vazdekis} A.,  {Koleva} M.,  {Ricciardelli} E.,  {R{\"o}ck} B.,
  {Falc{\'o}n-Barroso} J.,  2016, \mn@doi [\mnras] {10.1093/mnras/stw2231},
  \href {https://ui.adsabs.harvard.edu/abs/2016MNRAS.463.3409V} {463, 3409}

\bibitem[\protect\citeauthoryear{Vijayan, Clay, Thomas, Yates, Wilkins  \&
  Henriques}{Vijayan et~al.}{2019}]{Vijayan2019}
Vijayan A.~P.,  Clay S.~J.,  Thomas P.~A.,  Yates R.~M.,  Wilkins S.~M.,
  Henriques B.~M.,  2019, \mn@doi [\mnras] {10.1093/mnras/stz1948}, 489, 4072

\bibitem[\protect\citeauthoryear{{Vijayan}, {Lovell}, {Wilkins}, {Thomas},
  {Barnes}, {Irodotou}, {Kuusisto}  \& {Roper}}{{Vijayan}
  et~al.}{2021}]{Vijayan2021}
{Vijayan} A.~P.,  {Lovell} C.~C.,  {Wilkins} S.~M.,  {Thomas} P.~A.,  {Barnes}
  D.~J.,  {Irodotou} D.,  {Kuusisto} J.,   {Roper} W.~J.,  2021, \mn@doi
  [\mnras] {10.1093/mnras/staa3715}, \href
  {https://ui.adsabs.harvard.edu/abs/2021MNRAS.501.3289V} {501, 3289}

\bibitem[\protect\citeauthoryear{{Vijayan} et~al.,}{{Vijayan}
  et~al.}{2022}]{vijayan2021light}
{Vijayan} A.~P.,  et~al., 2022, \mn@doi [\mnras] {10.1093/mnras/stac338}, \href
  {https://ui.adsabs.harvard.edu/abs/2022MNRAS.511.4999V} {511, 4999}

\bibitem[\protect\citeauthoryear{{Vogelsberger} et~al.,}{{Vogelsberger}
  et~al.}{2014}]{Vogelsberger2014}
{Vogelsberger} M.,  et~al., 2014, \mn@doi [\mnras] {10.1093/mnras/stu1536},
  \href {https://ui.adsabs.harvard.edu/abs/2014MNRAS.444.1518V} {444, 1518}

\bibitem[\protect\citeauthoryear{{Vogelsberger} et~al.,}{{Vogelsberger}
  et~al.}{2020}]{Vogelsberger2020}
{Vogelsberger} M.,  et~al., 2020, \mn@doi [\mnras] {10.1093/mnras/staa137},
  \href {https://ui.adsabs.harvard.edu/abs/2020MNRAS.492.5167V} {492, 5167}

\bibitem[\protect\citeauthoryear{{Weigel}, {Schawinski}  \&
  {Bruderer}}{{Weigel} et~al.}{2016}]{Weigel2016}
{Weigel} A.~K.,  {Schawinski} K.,   {Bruderer} C.,  2016, \mn@doi [\mnras]
  {10.1093/mnras/stw756}, \href
  {https://ui.adsabs.harvard.edu/abs/2016MNRAS.459.2150W} {459, 2150}

\bibitem[\protect\citeauthoryear{Whitaker, van Dokkum, Brammer  \&
  Franx}{Whitaker et~al.}{2012a}]{Whitaker_2012}
Whitaker K.~E.,  van Dokkum P.~G.,  Brammer G.,   Franx M.,  2012a, \mn@doi
  [\apj] {10.1088/2041-8205/754/2/l29}, 754, L29

\bibitem[\protect\citeauthoryear{{Whitaker}, {van Dokkum}, {Brammer}  \&
  {Franx}}{{Whitaker} et~al.}{2012b}]{Whitaker2012}
{Whitaker} K.~E.,  {van Dokkum} P.~G.,  {Brammer} G.,   {Franx} M.,  2012b,
  \mn@doi [\apjl] {10.1088/2041-8205/754/2/L29}, \href
  {https://ui.adsabs.harvard.edu/abs/2012ApJ...754L..29W} {754, L29}

\bibitem[\protect\citeauthoryear{Wilkins et~al.,}{Wilkins
  et~al.}{2022a}]{wilkinsFirstLightReionisation2022}
Wilkins S.~M.,  et~al., 2022a, First {Light} {And} {Reionisation} {Epoch}
  {Simulations} ({FLARES}) {V}: {The} redshift frontier, \url
  {http://arxiv.org/abs/2204.09431}

\bibitem[\protect\citeauthoryear{Wilkins et~al.,}{Wilkins
  et~al.}{2022c}]{wilkinsFirstLightReionisation2022b}
Wilkins S.~M.,  et~al., 2022c, First {Light} {And} {Reionisation} {Epoch}
  {Simulations} ({FLARES}) {VI}: {The} colour evolution of galaxies \$z=5-15\$,
  \url {http://arxiv.org/abs/2207.10920}

\bibitem[\protect\citeauthoryear{Wilkins et~al.,}{Wilkins
  et~al.}{2022b}]{wilkinsFirstLightReionisation2022a}
Wilkins S.~M.,  et~al., 2022b, First {Light} {And} {Reionisation} {Epoch}
  {Simulations} ({FLARES}) {VII}: {The} {Star} {Formation} and {Metal}
  {Enrichment} {Histories} of {Galaxies} in the early {Universe}, \url
  {http://arxiv.org/abs/2208.00976}

\bibitem[\protect\citeauthoryear{Wuyts et~al.,}{Wuyts
  et~al.}{2011}]{Wuyts_2011}
Wuyts S.,  et~al., 2011, \mn@doi [\apj] {10.1088/0004-637x/742/2/96}, 742, 96

\bibitem[\protect\citeauthoryear{{Yabe}, {Ohta}, {Iwata}, {Sawicki}, {Tamura},
  {Akiyama}  \& {Aoki}}{{Yabe} et~al.}{2009}]{Yabe2009}
{Yabe} K.,  {Ohta} K.,  {Iwata} I.,  {Sawicki} M.,  {Tamura} N.,  {Akiyama} M.,
    {Aoki} K.,  2009, \mn@doi [\apj] {10.1088/0004-637X/693/1/507}, \href
  {https://ui.adsabs.harvard.edu/abs/2009ApJ...693..507Y} {693, 507}

\makeatother
\end{thebibliography}
